\documentclass{aa}  
\usepackage[dvipsnames]{xcolor}

\usepackage[normalem]{ulem}
\usepackage{float}
\usepackage[bottom,norule,flushmargin]{footmisc}
\usepackage{twoopt}
\usepackage{natbib}
\usepackage{graphicx}
\usepackage{mathtools}
\usepackage{array}
\usepackage{multirow}
\usepackage{amsmath}
\usepackage{amssymb}
\usepackage{mathrsfs}
\usepackage{bigints}
\usepackage{subfigure}
\usepackage{caption}
\usepackage{subcaption}
\usepackage{comment}
\usepackage{hyperref}
\hypersetup{colorlinks=true,
linkcolor=blue,
citecolor=blue,
urlcolor=blue,
bookmarksnumbered=true,%
bookmarksopen=true
}
\usepackage{enumitem,booktabs,cfr-lm}
\usepackage[referable]{threeparttablex}
\renewcommand{\arraystretch}{1.2}

\usepackage{txfonts}

\begin{document}

   \title{The dynamical lineage of ultra-diffuse galaxies from TNG50-1}
   \titlerunning{The dynamical lineage of ultra-diffuse galaxies from TNG50-1}
   \authorrunning{Nandi \& Banerjee}

        \author{Nilanjana Nandi\thanks{nandi.nilanjana154@gmail.com}
            \and
          Arunima Banerjee\thanks{arunima@iisertirupati.ac.in}
          }

   \institute{Indian Institute of Science Education and Research, Tirupati 517619, India
   }

   \date{Received June 27, 2025; Accepted December 12, 2025}

  \abstract
   {The formation and evolution of the ultra-diffuse galaxies (UDGs) continues to be a puzzle. The variety of formation scenarios can be broadly classified into two categories: a massive yet failed L$_*$-type and a dwarf-like origin. The similarities and the differences in the morphological and kinematical properties of the UDGs with their possible precursors may provide important constraints on their origin and evolutionary history.}
   {We compared and contrasted structural, orbital and kinematical properties of the UDGs with other galaxy populations - namely, the low-surface brightness (LSBs), L$_*$-type or high-surface brightness (HSBs), and the dwarf galaxies.}
   {We selected a sample of UDG, LSB, HSB and dwarf galaxies from the TNG50-1 box of the IllustrisTNG simulation. We first obtained a few possible scaling relations involving their mass properties and conducted Spearman's rank correlation tests to analyse if the regression fits for UDGs are in compliance with those of the other galaxy samples. Then, we studied the cutouts of the individual galaxies to investigate the intrinsic shapes of their dark matter (DM) and stellar components. We also investigated their orbital and kinematical properties by evaluating a few parameters composed of velocity dispersion components. Finally, we constructed the mock integral field-spectroscopic data using the publicly-available software \texttt{SimSpin} to extract the kinematic moment maps of the line-of-sight velocity distribution and probe the stellar kinematic properties of our galaxy samples. In all the cases, we divided the samples in two subpopulations: isolated and tidally bound to study the effect of the local environment.}
   {We observe that the UDGs and the dwarf galaxies have nearly similar regression fits in the following parameter spaces: (a) stellar-to-gas mass ratio versus gas mass, (b) stellar-to-gas mass ratio versus total dynamical mass, and (c) total baryonic mass versus total dynamical mass. Further, we may infer that the isolated UDGs can be classified as prolate, while the tidally bound UDGs can exhibit both prolate and oblate shapes. The DM and stellar velocity anisotropy of the UDGs suggest that they reside in a cored, low-mass halo and may be classified as early-type galaxies. Finally, their stellar kinematic properties suggest that the UDGs are slow-rotators exhibiting low to nearly no-rotation.}
   {The UDGs and the dwarf galaxies share similarities as far as the aforementioned possible scaling relations are concerned. Both the isolated UDGs and dwarf population can be characterised by prolate shapes unlike the other galaxy populations. However, the tidally bound UDGs exhibit both prolate as well as oblate shapes. The velocity anisotropy of the UDGs and the dwarfs hint at the fact that they may have originated in a dwarf-like halo, as opposed to the LSBs or the HSBs. Moreover, the UDGs and the dwarfs can be classified as early-type slow-rotating galaxies in contrast to the late-type, disc-dominated, fast-rotating LSBs and the HSBs. Therefore, we conclude that the UDGs and the dwarfs possibly have a common dynamical lineage.}


   \maketitle

\section{Introduction}\label{sec:intro}
The ultra-diffuse galaxies (UDGs) have attracted considerable attention since their discovery in the Coma cluster, and then subsequently in almost all cosmological environments in the nearby universe, owing to their remarkably faint yet extended appearance.(\citealt{2015vanDokkum_UDG_formation_a,2015vanDokkum_UDG_formation_b}; \citealt{2015Koda_UDGENV}; \citealt{2015Mihos_UDGENV}; \citealt{2015Munoz_UDGENV}; \citealt{2016Yagi_UDG_morphology}; \citealt{2016MartinezDelgado_UDGENV}; \citealt{2017vanderBurg_UDGENV}; \citealt{2017RomanTrujillo_UDG}; \citealt{2017Venhola_UDGENV}; \citealt{2017Janssens_UDGENV,2019Janssens_UDGENV}; \citealt{2018Muller_UDGENV}; \citealt{2019Forbes_UDGENV}; \citealt{2021Prole_UDGENV}; \citealt{2019Zaritsky_UDGENV}; \citealt{2019Roman_UDGENV}; \citealt{2020Barbosa_UDGENV}). 
In fact, UDGs are defined as galaxies with their $g$-band central surface brightness, $\mu_{\rm g,0} >$ 24 mag arcsec$^{-2}$ and effective radii, R$_e > 1.5$ kpc.
In general, the UDGs are DM-dominated (\citealt{2017DiCintio_UDGFormation}; \citealt{2018Toloba_UDGDM}; \citealt{2019vanDokkum_UDG_LOSVD}; \citealt{2021Forbes_UDGDM}; \citealt{2021Gannon_UDGDM}; \citealt{2022Kong_UDG}; \citealt{2023Benavides_UDG_TNG}); nevertheless, there exists some UDGs which are reportedly DM-free (\citealt{2022Pina_UDGDM}; \citealt{2022vanDokkum}). Therefore, these galaxies are ideal testing-beds to constrain models of formation and evolution of galaxies in different cosmological paradigms.
\\
\indent A variety of physical mechanisms have been proposed in the literature to explain the formation and evolution of the UDGs. Since UDGs consist of older stellar populations, they are primarily understood to originate from {\textquotesingle}failed{\textquotesingle} L$_*$-type galaxies which ceased star formation after gas-loss driven by ram pressure stripping or harassment, and as a result, consist of older stellar populations (\citealt{2015vanDokkum_UDG_formation_a,2015vanDokkum_UDG_formation_b}; \citealt{2016Peng_UDG_formation}; \citealt{2018Toloba_UDGDM}; \citealt{2022Janssens_UDGENV}). 
Alternatively, UDGs are believed to be formed from dwarf galaxies via transformation through several intrinsic or/and extrinsic physical mechanisms, such as: (1) formation in dwarf-like yet high-spin DM halos \citep{2016AmoriscoLoeb_UDGformation}, (2) expansion of DM and stellar halos due to feedback-driven gas outflows \citep{2017DiCintio_UDGFormation}; (3) mass-loss or mass-redistribution due to the group/cluster tidal field \citep{2019Jiang_UDG_ShapeTensor,2019Carleton_UDG_formation,2020Tremmel_UDG_formation}; (4) accretion or infall into a cluster environment \citep{2020Sales_UDG_formation}; (5) backsplash dwarf galaxies \citep{2021Benavides_UDG_formation}; (6) fall-outs of early mergers \citep{2021Wright_UDG_formation}; (7) fraction of star-forming gas redistributed to outer region while undergoing galactic fountain therefore triggering star-formation at extended radii \citep{2025_ZhengEAGLE}. Interestingly, while this alternative formation scenarios appear to be more viable, they can hardly explain the presence of surprisingly larger number of globular clusters (GCs) in some UDGs (\citealt{2023Toloba_UDG_formation}; \citealt{2025Haacke_UDG_formation}). 
Moreover, recent studies substantiate the validity of both the contrasting formation scenarios based on the number of GCs found in the UDGs (\citealt{2020Forbes_UDGENVb}, \citealt{2022Buzzo_UDG_formation,2024Buzzo_UDG_formation}). Clearly, a general agreement is still lacking in the literature involving the formation of the UDGs.
\\
\indent To identify the fundamental formation mechanism of the UDGs, several observational and simulation studies probed their intrinsic stellar morphology which emerged as another subject of debate.
While the Coma UDGs were observed to exhibit prolate morphologies \citep{2017Burkert_UDG_ShapeTensor}, 
an extensive sample of UDGs, including the ones existing in the field as well as those located in several low-to-intermediate redshift galaxy clusters, exhibit stellar shapes which are well-described by an oblate-triaxial model \citep{2020Rong_UDG_ShapeTensor,2021KadoFong_UDG_ShapeTensor}.
On the other hand, the field UDGs found in the idealised zoom-in simulations were observed to display both prolate as well as oblate shapes \citep{2019Jiang_UDG_ShapeTensor,2020CardonaBarrero_UDG_ShapeTensor}. In parallel, some studies suggested that the UDGs display oblate shapes when they are in the field and prolate when in denser environments \citep{2019Liang_UDG_ShapeTensor,2022vanNest_UDG_morphology}; some other studies observe no significant difference between the cluster and field UDGs as far as their intrinsic shapes are concerned \citep{2021Kadowaki_UDG,2023Benavides_UDG_TNG}.   
Although the intrinsic shapes of the DM halo may impact the stellar orbital distribution by modifying the underlying halo potential, this aspect has not received much attention in the literature in the context of UDGs. Furthermore, the stellar orbital and kinematic features may impose important constraints on the formation and evolutionary history of these galaxies.
A number of field and cluster UDGs show relatively lower circular velocities or nearly no significant rotation, while some of the cluster UDGs exhibit major, intermediate or minor axis rotation (\citealt{2017Leisman_ALFALFA}; \citealt{2019Pina_UDG_LOSVD_offBTFR}; \citealt{2019Sengupta_UDG_LOSVD}; \citealt{2019Chilingarian_UDG_LOSVD}). The integral-field spectroscopic (IFS) observations of some cluster UDGs suggest the presence of both rotating (including NGC1052-DF2) and non-rotating (including DF44) UDGs (\citealt{2019Emsellem_UDG_LOSVD}; \citealt{2019vanDokkum_UDG_LOSVD}; \citealt{2023Iodice_UDG_LEWIS}; \citealt{2025Buttitta_UDG_LEWIS}).
Similarly, by analysing the synthetic 2D stellar kinematic map of the field UDGs in the NIHAO zoom-in simulation, both the rotation- and dispersion-supported UDGs were observed to be present in the isolated environment with nearly equal probability \citep{2020CardonaBarrero_UDG_ShapeTensor}. In fact, the higher-order Gauss-Hermite (GH) moments of the line-of-sight velocity distribution (LOSVD) extracted from the IFS data may describe the stellar kinematic properties of these galaxies in a greater detail \citep{1993vanderMarel_Franx_LOSVD,1993Gerhard_LOSVD}. 
However, due to their faintness, the observations require relatively longer exposure time to achieve a significant signal-to-noise ratio for a robust GH parametrisation. Till date, DF44 is the only UDG for which the higher-order GH coefficients were derived from the IFS data gathered via a series of VLT/MUSE IFS observations over a span of a few months amounting to an overall exposure time of 25.3 hr \citep{2019vanDokkum_UDG_LOSVD}. One way to overcome this limitation is by simulating the mock IFS spectra of the UDGs by using galaxy cutouts from state-of-the-art large-scale cosmological simulations as initial conditions. This will also help in studying and analysing a relatively larger sample of galaxies which is necessary from a statistical viewpoint to identify the possible trends, if any, present in their structural and morphological properties.
\\
\indent In our previous study, we observed that the isolated, HI-rich UDGs and the dwarf irregular galaxies share similarities in their structural and kinematic properties and concluded a possible dynamical lineage between these two galaxy populations as opposed to the field late-type low-surface brightness galaxies (LSBs) \citep{2025NandiPaper1}. Nonetheless, the number of galaxy samples considered previously were limited due to availability of the mass-modelling data in the literature. As an extension of the previous work, we aim to analyse the structural and kinematical properties of the UDGs and contrast with those of the other galaxy populations which are possible precursors to the UDGs - namely, the LSBs, dwarf galaxies (or dwarfs) and in addition, the high-surface brightness (HSBs) galaxies. We conduct this study by incorporating the galaxy samples selected from TNG50-1, the highest-resolution sub-box of the IllustrisTNG cosmological simulation (\citealt{2019TNG50_Nelson_b}; \citealt{2019TNG50_Pillepich}). To begin with, we studied some possible scaling relations considering a few pairs of basic properties directly obtained from the TNG database. Further, we analysed the galaxy cutouts to derive a few morphological and kinematical properties of the stellar and DM components of our galaxy samples. In this context, we note that \cite{2023Benavides_UDG_TNG} studied the stellar morphology of the UDGs from the TNG50 simulation box based on their projected axial ratios. In comparison, our methodology to study the intrinsic morphology is quite different and slightly more robust. Finally, we simulated the IFS data cubes of these galaxies and constructed the 2D-maps of their higher-order GH moments (up to fourth order) using the publicly-available software \texttt{SimSpin} (\citealt{2020SimSpin_v1,2023SimSpin_v2}). In all the cases, the galaxy samples were divided into two subsamples of isolated and tidally bound galaxies based on the local environment to investigate the effect of the environment on their morphology and evolution.
\\
\indent The organisation of the paper is as follows: we introduce the samples of galaxies selected from TNG50-1 and the analysis techniques in Sect. \ref{sec:data} and Sect. \ref{sec:method}, respectively. In Sect. \ref{sec:results}, we present and discuss the results. Finally, we derive the conclusions in Sect. \ref{sec:conclusions}.
\section{Data}\label{sec:data}
We selected the galaxy samples in our study from the TNG50-1 box of the IllustrisTNG  simulation (\citealt{2018TNG_Pillepich_a, 2018TNG_Pillepich_b}; \citealt{2018TNG_Naiman}; \citealt{2018TNG_Springel}; \citealt{2018TNG_Marinacci}; \citealt{2018TNG_Nelson, 2019TNG_Nelson_a}). The IllustrisTNG Project is a suite of gravo-magnetohydrodynamical simulation run separately for three different cosmological volumes of comoving box-length 51.7 Mpc, 110.7 Mpc and 302.6 Mpc (referred to as TNG50, TNG100 and TNG300, respectively) implementing the moving mesh code \texttt{AREPO}, assuming the cosmological parameters consistent with the Planck 2015 results ($h$ = 0.6774, $\Omega_m$ = 0.3089, $\Omega_\Lambda$ = 0.6911, $\Omega_b$ = 0.0486, $n_s$ = 0.9677, $\sigma_8$ = 0.8159) and invoking prescriptions to account for a realistic baryonic physics (\citealt{2001Springel_AREPO_GADGET}; \citealt{2016PlanckColab}; \citealt{2017Weinberger_TNGFeedbackProcess}; \citealt{2018TNG_Pillepich_b}). Each simulation run has 100 snapshots corresponding to redshifts ($z$) equally-spaced between $z \sim$ 20 to $z$ = 0. At each snapshot, the \texttt{Friends-of-Friends} (\texttt{FoF}) (\citealt{1985DavisFoF}) algorithm and, then, the \texttt{SUBFIND} subroutine (\citealt{2001SpringelSUBFIND}; \citealt{2009DolagSUBFIND}) were implemented to identify individual galaxies (referred as subhalos). The basic subhalo properties have already been derived and catalogued in the TNG database that can be directly accessed using their web-based interface\footnote{\href{https://www.tng-project.org/data/}{https://www.tng-project.org/data/}}.
\\
\indent Among all the simulation boxes, TNG50-1 (\citealt{2019TNG50_Nelson_b}; \citealt{2019TNG50_Pillepich}) has the highest resolution with 2160$^3$ DM particles of mass 4.5 $\times$ 10$^5$ M$_\odot$ and baryonic particles of mass 8.5 $\times$ 10$^4$ M$_\odot$ originating from an initial number of 2160$^3$ Voronoi gas cells. The significantly higher particle resolution of the simulation box makes it best-suited for studying the intrinsic properties of high as well as low-mass galaxies at different cosmological settings. The individual subhalo cutouts of interest can be directly downloaded for further analysis. We chose our galaxy samples from the snapshot corresponding to $z$ = 0. The process of sample selection is discussed below.
\\

\noindent Sample selection: We selected a sample of UDGs, LSBs, HSBs, and the dwarf galaxies from TNG50-1 by implementing their well-accepted definitions in the literature:
\begin{itemize}
    \item[$\circ$] UDGs: $g$-band absolute magnitude, M$_g$ $>$ -17 and effective radius, R$_e$ $>$ 1.5 kpc \citep{2015vanDokkum_UDG_formation_a,2015vanDokkum_UDG_formation_b};
    \item[$\circ$] LSBs: maximum circular velocity, V$_{\text{max}}$ within the range 100 - 200 km s$^{-1}$ and $B$-band absolute magnitude, M$_B$ $>$ -18 \citep{1996deBlok_SVE_LSB,1995AMcGaugh_SVE_LSB}; 
    \item[$\circ$] HSBs: V$_{\text{max}}$ within the range 170 - 250 km s$^{-1}$ and stellar mass M$_*$ $>$ 10$^{9.7}$ M$_{\odot}$ \citep{2005Cooray_HSB_Selection,2005Pizzella_HSB_Selection};
    \item[$\circ$] Dwarfs: stellar mass, M$_*$ within the range 10$^{6}$ - 10$^{9}$ M$_{\odot}$, M$_g$ $>$ -17 and R$_e$ $<$ 1.5 kpc \citep{2018Revaz_Dwarf_Selection,2022Poulain_UDG+Dwarf_Selection}.
\end{itemize}
\begin{figure*}
    \centering
    \includegraphics[width=0.33\textwidth]{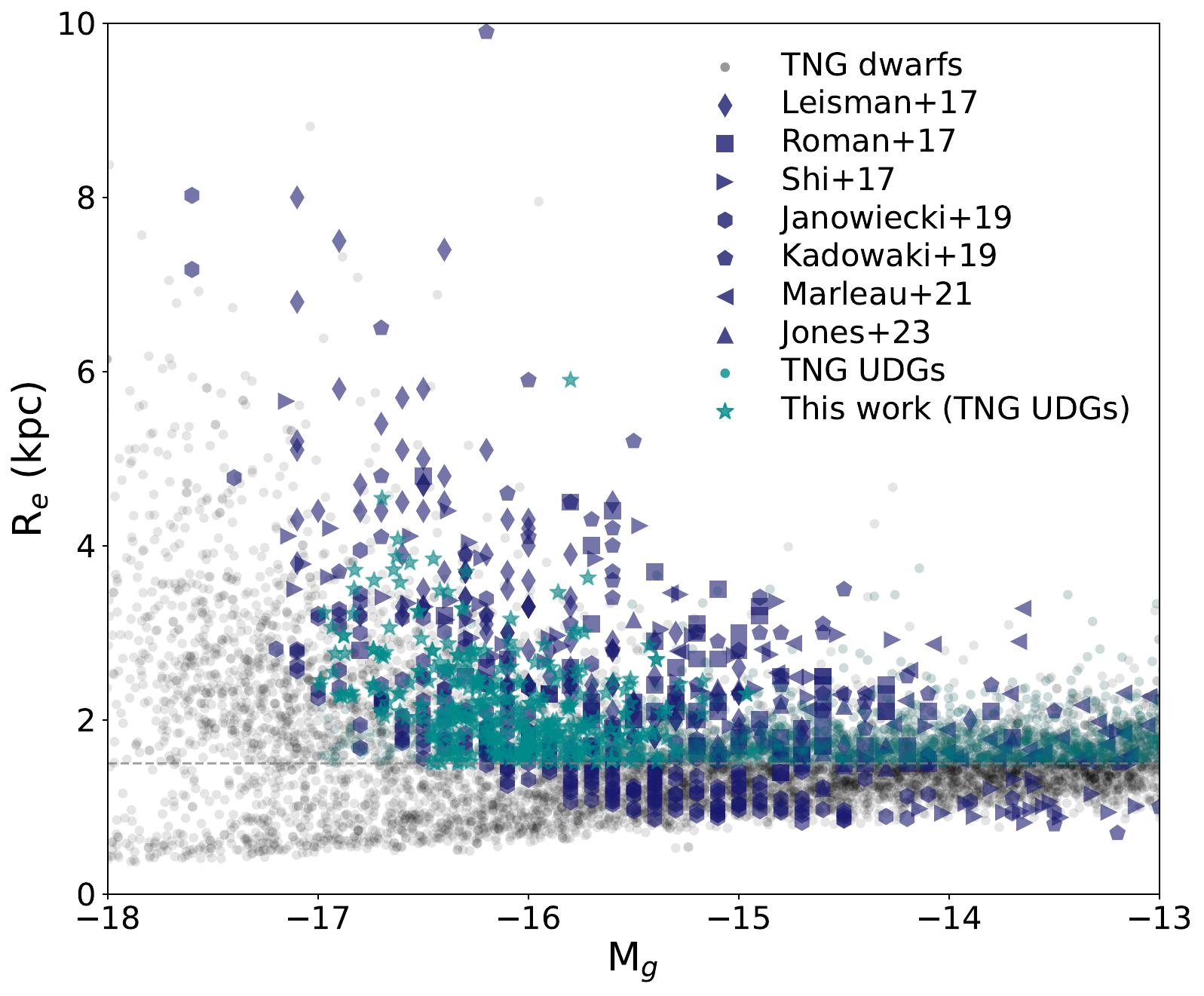}
    \includegraphics[width=0.333\textwidth]{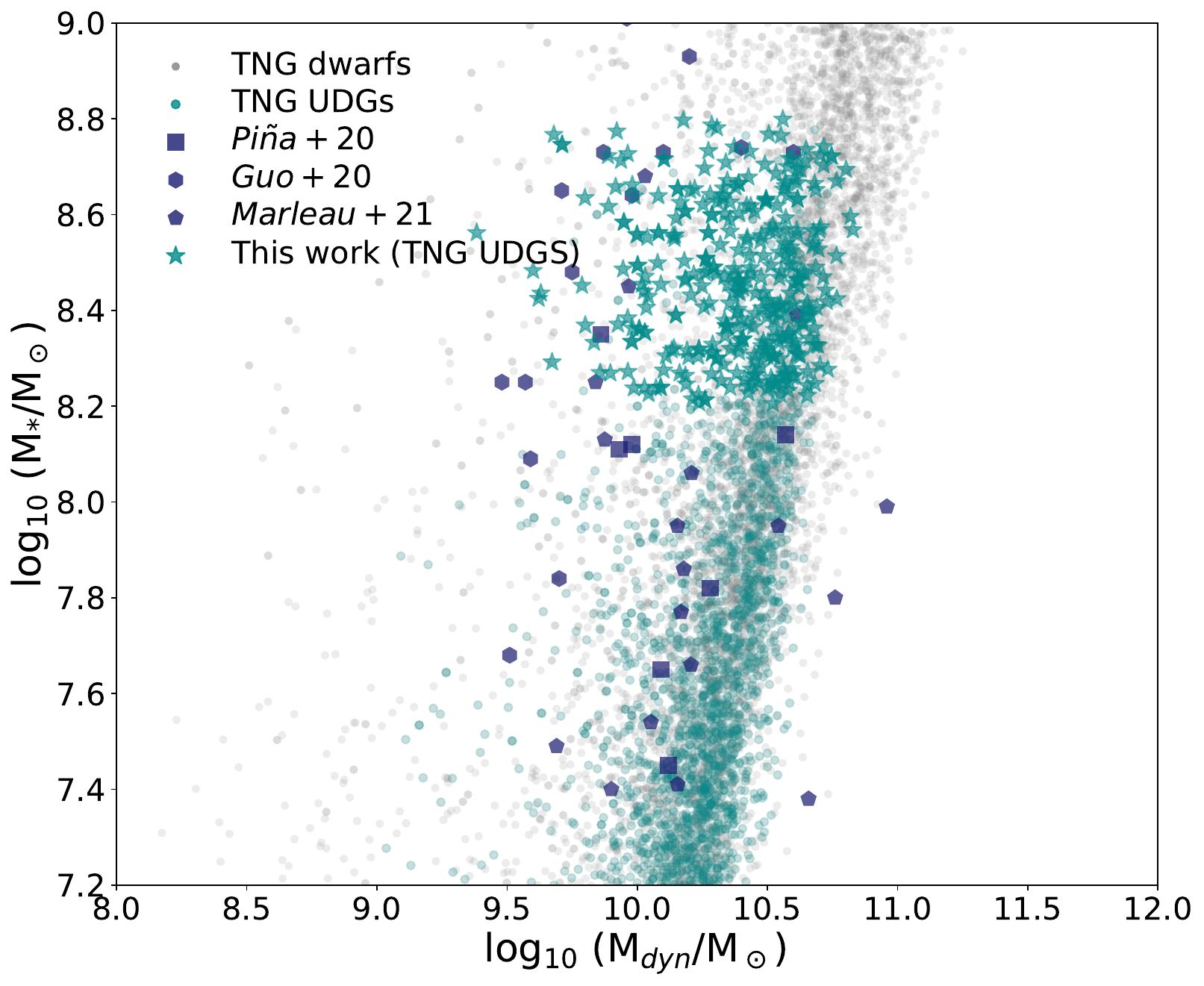}
    \includegraphics[width=0.32\textwidth]{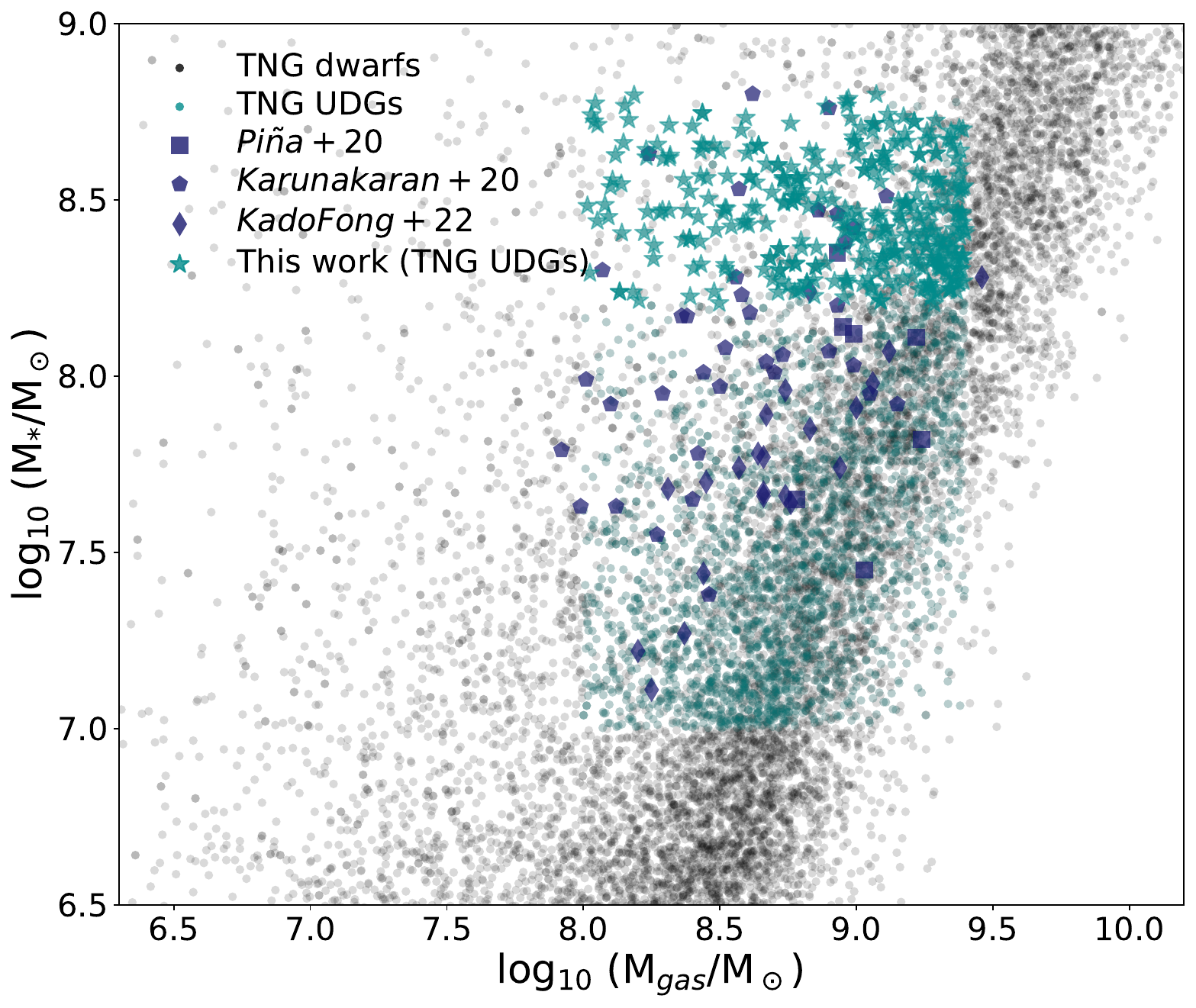}
    \caption{Distribution of the TNG50-1 UDGs selected in our sample in the (left) M$_g$-R$_e$, (middle) M$_*$-M$_\text{dyn}$, and (right) M$_*$-M$_\text{gas}$ space marked in teal stars. To compare with the observed populations, the UDGs obtained from \cite{2017Leisman_ALFALFA}, \cite{2017RomanTrujillo_UDG}, \cite{2017Shi_UDG}, \cite{2019Janowiecki_UDG}, \cite{2021Marleau_UDG}, \cite{2021Kadowaki_UDG}, \cite{2022Poulain_UDG+Dwarf_Selection}, and \cite{2023Jones_UDG} are superposed on the left panel. Similarly, on the middle and the right panel, the stellar, dynamical, and gas masses of the UDGs taken from \cite{2020Guo_UDG}, \cite{2020PinaHI_UDG}, \cite{2020KarunakaranHI_UDG}, \cite{2021Marleau_UDG}, \cite{2022Kong_UDG}, and \cite{2022KadoFong_UDG} are shown for comparison. The teal dots represent the remaining UDGs identified in TNG50-1 while the grey dots denote the whole sample space from which the UDGs were selected.}
    \label{fig:UDG_Selection}
\end{figure*}
\noindent UDGs: We searched for all the galaxies in the $z$ = 0 snapshot of TNG50-1 having stellar masses within 10$^{6}$ - 10$^{9}$ M$_{\odot}$, DM mass, M$_{\text{DM}}$ $>$ 10$^{8}$ M$_{\odot}$ and non-zero gas cells. The latter two criteria were to eliminate the objects with non-cosmological origin and imposed on other galaxy populations as well.
Further, to distinguish the UDGs from 15,714 dwarf galaxies identified in this process, we restricted the dynamical (M$_{\text{dyn}}$) and the gas mass (M$_{\text{gas}}$) within ranges that match the observed UDGs: 10$^{9}$ $<$ M$_{\text{dyn}}$ $<$ 10$^{11.2}$ M$_{\odot}$ and 10$^{8}$ $<$ M$_{\text{gas}}$ $<$ 10$^{9.4}$ M$_{\odot}$ (\citealt{2020Guo_UDG}; \citealt{2020PinaHI_UDG}; \citealt{2020KarunakaranHI_UDG}; \citealt{2021Marleau_UDG}; \citealt{2022Kong_UDG}; \citealt{2022KadoFong_UDG}). The dynamical mass was obtained by summing a galaxy's stellar, gas and DM masses. Next, to select our sample UDGs, we mapped the whole sample of galaxies in the $g$-band absolute magnitude (M$_g$) - effective radius (R$_e$) plane such that they closely and proportionately match the distribution of the observed UDGs (for example, \citealt{2017Leisman_ALFALFA}; \citealt{2017RomanTrujillo_UDG}; \citealt{2017Shi_UDG}; \citealt{2019Janowiecki_UDG}; \citealt{2021Marleau_UDG}; \citealt{2021Kadowaki_UDG}; \citealt{2022Poulain_UDG+Dwarf_Selection}; \citealt{2023Jones_UDG}). Towards this end, we listed the M$_g$ and stellar half-mass radius (R$_{1/2}$) values from the TNG data archive and converted R$_{1/2}$ to $g$-band optical R$_{e}$ by using the following power-law: 
\begin{equation}\label{eq:Re_conversion}
\frac{\text{R}_{e}}{\text{kpc}} = C_g \times 0.999 \bigg(\frac{\text{R}_{1/2}}{\text{kpc}}\bigg)^{0.922},
\end{equation}
where $C_g$ $\sim$ 1.58 \citep{2024BeasTNGSKIRT_wavelengthRe}\footnote{In this context we note that \cite{2024BeasTNGSKIRT_wavelengthRe} found no significant dependence of optical-to-stellar effective radii ratio on stellar mass for galaxies with M$_*$ $\sim$ 10$^{9.8-12}$ M$_{\odot}$. Therefore, we assumed the same conversion relation for all the galaxy samples in our study regardless of their stellar mass.}. 
We implemented a criteria of M$_g$ $>$ -17 and R$_e$ $\gtrsim$ 1.5 kpc and identified 2,845 UDGs. Moreover, we limited the UDGs to consist of at least 3,000 stellar particles to maintain a minimum resolution and thus selected 305 galaxies in our final UDG sample. In Figure \ref{fig:UDG_Selection}, we compare our final sample of UDGs with those available in the literature. On the left panel of Figure \ref{fig:UDG_Selection}, the M$_g$-R$_e$ distribution of our final UDG samples are shown in teal stars, while the remaining UDGs and the dwarf galaxies are shown in teal and grey dots, respectively. The observed UDG samples from \cite{2017Leisman_ALFALFA}, \cite{2017RomanTrujillo_UDG}, \cite{2017Shi_UDG}, \cite{2019Janowiecki_UDG}, \cite{2021Marleau_UDG}, \cite{2021Kadowaki_UDG} and \cite{2023Jones_UDG} are superposed on this plane for comparison. On the middle and right panels of Figure \ref{fig:UDG_Selection}, the distributions the stellar, gas, and the dynamical masses are compared with the observations (\citealt{2020Guo_UDG}; \citealt{2020PinaHI_UDG}; \citealt{2020KarunakaranHI_UDG}; \citealt{2021Marleau_UDG}; \citealt{2022Kong_UDG}; \citealt{2022KadoFong_UDG}). We observe that the UDGs identified in this process represent the original UDG population fairly well. However, we note that our final UDG samples lie near the higher mass end of the observed M$_*$-range and we couldn't probe the fainter, low-mass UDGs due to the resolution criterion.
\\
\noindent LSBs: Our LSB sample consists of the non-dwarf subset of the low-surface brightness galaxies which are characterised by their relatively higher asymptotic velocities (V$_{\text{max}}$) \citep{1996deBlok_SVE_LSB,1995AMcGaugh_SVE_LSB}. The UDGs and the LSBs are often considered and studied on an equal footing in the literature (see for example \citealt{2019Martin}; \citealt{2021KadoFong_UDG_ShapeTensor}; \citealt{2024PerezMontano}) since the UDG formation scenario proposed by \cite{2016AmoriscoLoeb_UDGformation} is analogous to that of the LSBs (\citealt{1996deBlok_SVE_LSB}; \citealt{1995AMcGaugh_SVE_LSB}; \citealt{2019JadhavBanerjee}). In general, the LSBs are defined to be a class of galaxies with characteristically low stellar surface densities regardless of their masses, sizes and morphologies. While an overlap may exist between the two groups, the classical late-type LSBs are distinct from the classical dwarf galaxies
(\citealt{2000vanHoek}; \citealt{2001Mathews_LSB_Selection}; \citealt{2004O'Neil}; \citealt{2013KimLee}). This is the main reason to incorporate the LSBs in our classification study.
To choose the LSBs in our sample, we limited the V$_{\text{max}}$ within the range 100 - 200 km s$^{-1}$ and M$_B$ $>$ -18.5 (\citealt{1996Impey_LSB_Selection}; \citealt{1996deBlok_SVE_LSB}; \citealt{2001Mathews_LSB_Selection}; \citealt{2004Kniazev_LSB_Selection}). In addition, we restricted the galaxies to have M$_*$ < 10$^{9.6}$ M$_\odot$ following their observed stellar mass range. In this way, we selected the 251 galaxies in our final LSB sample. 
\\
\noindent HSBs: We termed the Milky Way-like L$_*$-type galaxies as the HSBs in our study and identified them based on their M$_*$ and V$_{\text{max}}$. We selected 300 galaxies in our final HSB sample which satisfied the following selection criteria: M$_*$ $>$ 10$^{9.7}$ M$_\odot$ and 170 $<$ V$_{\text{max}}$ $<$ 250 km s$^{-1}$ (\citealt{2005Cooray_HSB_Selection}; \citealt{2005Pizzella_HSB_Selection}; \citealt{2013Robotham_HSB_Selection}).
\\
\noindent Dwarfs: Finally, we chose our dwarf sample from the set of low-luminosity galaxies with M$_{g}$ $>$ -17 and M$_*$ $\sim$ 10$^{6-9}$ M$_\odot$ (\citealt{2018Revaz_Dwarf_Selection}; \citealt{2022Poulain_UDG+Dwarf_Selection}). To avoid overlap with the UDGs, our dwarf samples were chosen to have R$_e$ $<$ 1.5 kpc. Similar to the case of the UDGs, we imposed a resolution criterion of 3000 stellar particles, therefore ending up with 538 dwarf galaxies in our final sample. Here, in all cases, M$_*$, M$_{\rm gas}$, and M$_{\rm DM}$ represent the total stellar, gas, and DM masses of a galaxy. Therefore, to list the mass properties of our galaxy samples from the TNG data archive, we chose those definitions which consider all the (star or DM) particles/(gas) cells bound to a galaxy subhalo. A similar argument holds for M$_g$ and M$_B$ as well.
\begin{figure}
    \centering
    \hspace{-1cm}
    \includegraphics[width=1\linewidth]{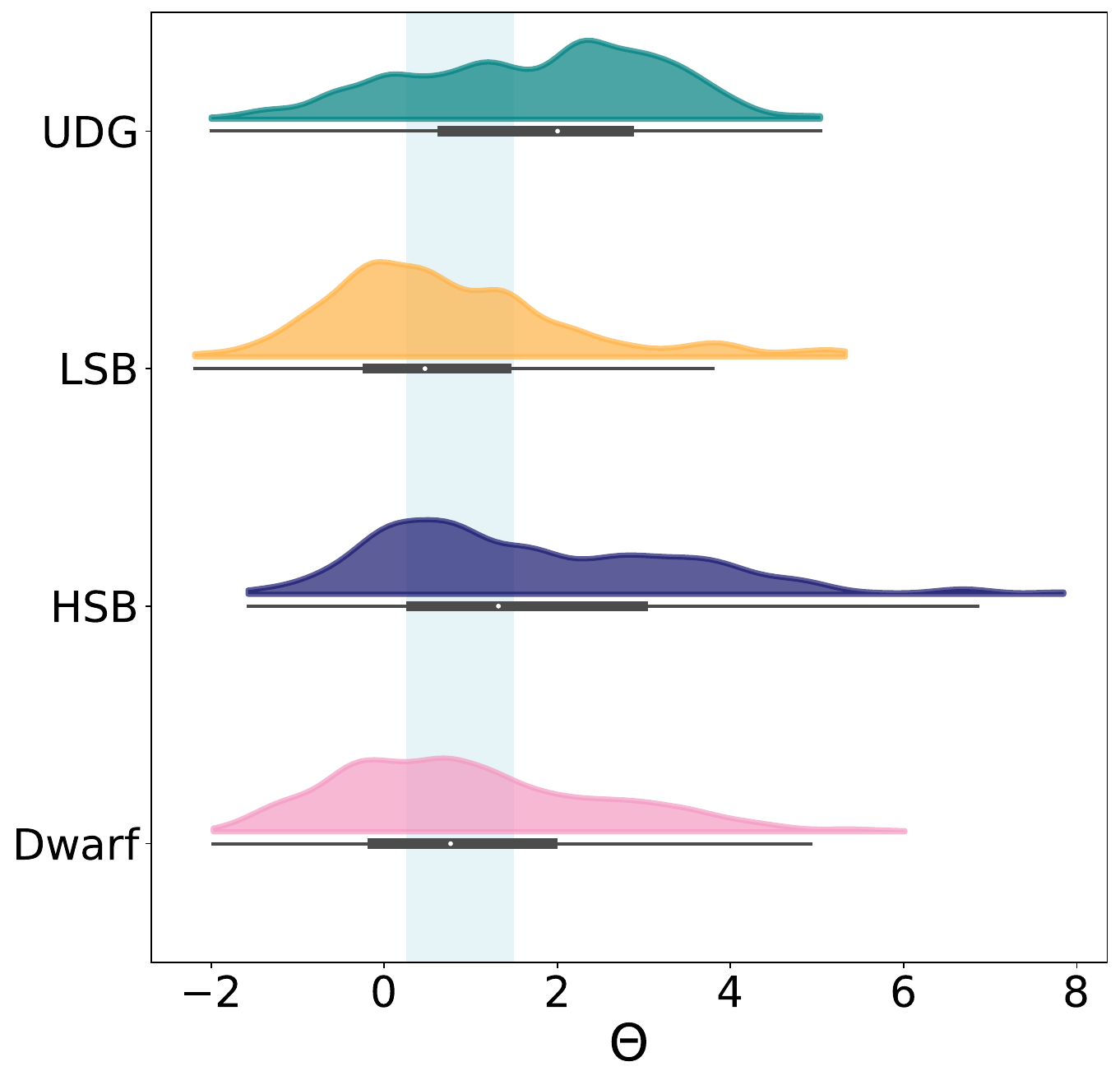}
    \caption{Distribution of the tidal index, $\Theta$, for our galaxy samples: UDGs in teal, LSBs in yellow, HSBs in blue and the dwarfs in pink,  along with their box plots. The blue region shows the buffer region which we rejected from our study.}
    \label{fig:theta}
\end{figure}
\begin{figure*}
    \centering
    \includegraphics[width=1.\linewidth]{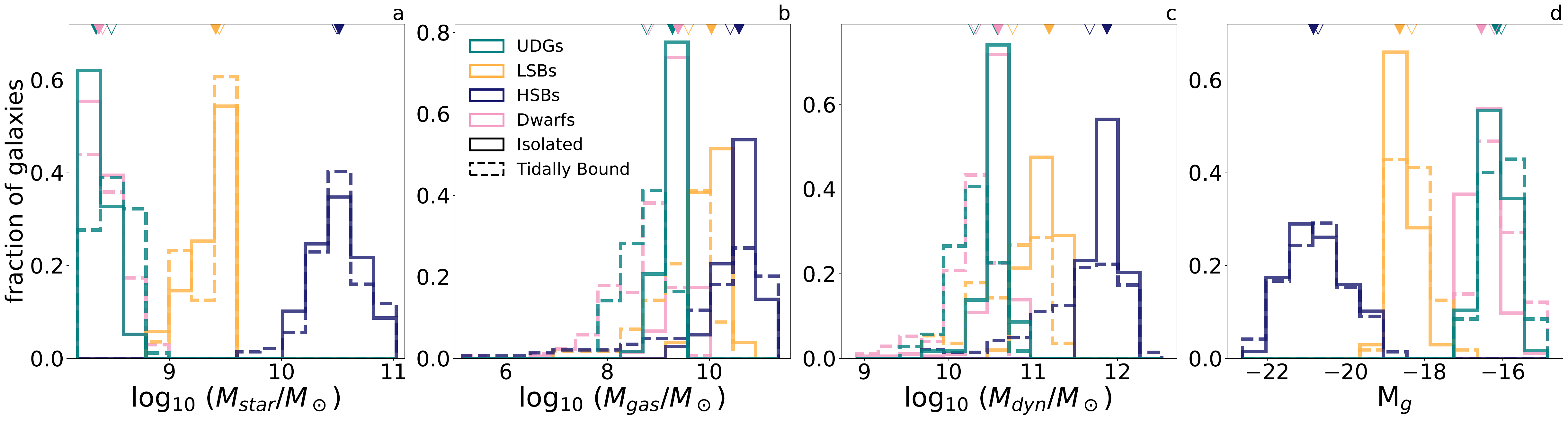}
    \caption{Histograms of (a) M$_*$, (b) M$_{\rm gas}$, (c) M$_{\rm dyn}$ and (d) M$_g$ of the UDGs, LSBs, HSBs, and the dwarfs considered in our study presented in teal, yellow, blue and pink colour, respectively. The solid histograms represent the isolated galaxies and the dotted histograms denote the tidally bound subsamples. The medians of the isolated and tidally bound subsamples are denoted with, respectively, solid and hollow triangle on the top following the same colour scheme.}
    \label{fig:basic_properties}
\end{figure*}
\section{Methods}\label{sec:method}
\subsection{Analysis of the local environment}\label{sec:tidal_index}
In TNG50, the usual convention to distinguish the field galaxies from the satellites relies on the \texttt{FOF} and \texttt{SUBFIND} algorithm. Although it provides a qualitative classification of galaxies as either ‘central’ or ‘satellite’\footnote{See Appendix \ref{appendix1}.}, it may not be sufficient to quantify the tidal influence of the local environment on a galaxy. For a robust classification of the environment, we employed an observationally motivated approach, i.e., by evaluating the dimensionless tidal index, $\Theta$, for each galaxy (\citealt{2004Karachentsev_TidalIndex, 2013AKarachentsev_TidalIndex, 2018Karachentsev_TidalIndex}; \citealt{2018Besla_TidalIndex}; \citealt{2024Mutlu-Pakdil_TidalIndex}; \citealt{2025Joy_LocalOverdensity}). $\Theta_i$ determines the influence of the local tidal field on the $i$-th galaxy by finding out the maximum gravitational force ($F_{i,n}$) exerted on it by one of its 5 massive, nearest neighbours with a dynamical mass $M_{\text{dyn},n}$ located at a distance $d_{i,n}$ ($F_{i,n}$ $\sim$ $M_n/d^3_{i,n}$). $\Theta_{i}$ can be expressed as follows:
\begin{equation}\label{eq:tidal_index}
    \Theta_{i} = \text{max}\Biggl\{ \text{log}\Bigg( \frac{\text{M}_{\text{dyn} ,n}/\text{M}_\odot}{(d_{i,n}/\text{Mpc})^3} \Bigg) \Biggl\}_{n = 1,\; 2,\; \dots,\; 5} + \; C.
\end{equation}
The galaxy with the strongest gravitational influence is termed as the main disturber (MD). The constant C = -10.96 is chosen such that the Keplerian period of a galaxy about its MD becomes equal to the Hubble time and the galaxy remains within a \textquotesingle zero velocity sphere\textquotesingle. $\Theta_i$ = 0 can also be regarded as the density enhancement about the $i$-th galaxy ($\Delta \rho_{i,n}$ $\sim$ $M_n/d^3_{i,n}$) caused by its MD which reduces to the critical density of the universe for $\Theta_i$ = 0 (\citealt{1999Karachentsev_TidalIndex, 2013AKarachentsev_TidalIndex}). The galaxies with $\Theta_i$ $<$ 0 can be termed as isolated and undisturbed by the environment and $\Theta_i$ $>$ 0 can be considered as tidally bound to the MD; larger the $\Theta$, stronger the local tidal influence. 
\\
\indent To calculate $\Theta$, we listed the 3D positions and the total dynamical masses of 18,850 galaxies with M$_*$ $>$ 10$^{7}$ M$_\odot$ and imposed the \texttt{scipy.spatial.cKDTree} algorithm to find the nearest neighbours with M$_*$ $>$ 10$^{9.5}$ M$_\odot$. To account for the periodic boundary conditions, the masses and positions of the galaxies within the TNG50 box were mirrored along all directions, such that the original box at the centre was surrounded by 26 replicas (eight in the same plane as the original box, nine above, and nine below). Both the original and mirrored galaxies were considered when identifying the nearest massive neighbours to evaluate $\Theta$ using Eq. \ref{eq:tidal_index}. Since the goal was to determine the nearest neighbours, the $\Theta_i$ values of most galaxies remain unaffected by this mirroring process, except for the relevant ones located near the boundaries. Furthermore, we labelled galaxies with $\Theta < 0.25$ as isolated and those with $\Theta > 1.5$ as tidally bound, while rejecting galaxies with $\Theta$ in the range 0.25 – 1.5. The rationale behind this selection criteria is two-fold: (1) to differentiate the isolated galaxies from the tidally bound ones, and (2) to eliminate galaxies possibly in the process of transitioning from isolated to tidally influenced scenarios. It is important to highlight that $\Theta = 0$ is not a rigid boundary; galaxies with positive yet close-to-zero $\Theta$ values may not differ significantly from those with negative $\Theta$ values tending toward zero. A comparable argument holds for tidally bound galaxies as well: galaxies with relatively smaller $\Theta$, which may be termed as transition populations, may be significantly different from highly tidally influenced galaxies (with higher $\Theta$ values). Hence, the aforementioned selection criteria in $\Theta$ allow the division of each sample into two fairly distinct subsamples of isolated and tidally bound galaxies, thereby eliminating the transition populations. With this, the numbers of galaxies in our UDG, LSB, HSB, and dwarf samples are finalised as: 235 (58 isolated, 177 tidally bound), 159 (103 isolated, 56 tidally bound), 213 (69 isolated, 144 tidally bound), and 368 (195 isolated, 173 tidally bound), respectively. The distributions of the tidal indices for our galaxy samples, along with their box plots (obtained using the python module \texttt{PtitPrince}) are presented in Figure \ref{fig:theta} \citep{2018PtitPrince_HalfViolin}. The light blue region represents the galaxies that were excluded from our samples. 
The overall distribution of our galaxy samples within the TNG50-1 box are presented in Figure \ref{fig:environment}. Here, it is worth noting that the UDGs are more prevalent in the tidally bound region compared to the other galaxy samples, suggesting that the tidal field may be an important factor in their origin. Finally, for a basic comparison among the chosen galaxies in our sample, we present the histograms of M$_*$, M$_{\text{gas}}$, M$_{\text{dyn}}$ and M$_g$ in Figure \ref{fig:basic_properties}. The colour scheme of the histograms are as follows: UDGs in teal, LSBs in yellow, HSBs in blue, and dwarfs in pink. Furthermore, the solid and the dotted lines represent the isolated and the tidally bound subsamples, respectively. The filled lower triangles at the top of each panel indicate the medians of the distributions for the isolated galaxy subsamples, while the empty ones represent the medians for the tidally bound subsamples. The same colour code and line styles are used throughout the paper.
\subsection{Intrinsic morphology}\label{sec:morphology}
We studied the intrinsic morphology of the DM and stellar components of our galaxy samples by solving eigenvalues and eigenvectors of the second moment of the mass distribution, the matrix known as the shape tensor in the literature (\citealt{2016Tomassetti_ShapeTensor}; \citealt{2019Chua_ShapeTensor,2022Chua_ShapeTensor}; \citealt{2021Cataldi_ShapeTensor}). The $ij$-component of the shape tensor is defined as $\frac{1}{\text{M}} \sum_{n = 1}^N m_n \; r_{n,i} \; r_{n,j}$, where $m_n$ is the mass of the $n$-th particle, M = $\sum_{n=1}^N m_n$ is the total mass, and $r_{n,i}$ and $r_{n,j}$ are the $i$-th and $j$-th component of its position vector ($i,j$ = $x$, $y$, $z$). By construction, the shape tensor can be dominated by the particles located in the outer region, particularly in systems with relatively smaller numbers of stellar and dark-matter particles (as in case of our UDG and dwarf samples). Therefore, we employed the reduced shape tensor,
\begin{equation}\label{eq:shape_tensor}
    S_{i,j} = \frac{1}{\text{M}} \sum_{n = 1}^N m_n \; \frac{r_{n,i} \; r_{n,j}}{|r|^2},
\end{equation}
where each of $r_{n,i}$ and $r_{n,j}$ is weighted by the radial distance of the particle, $|r| = \sqrt{x^2 + y^2 + z^2}$. 
Theoretically, the eigenvalues of the shape tensor are proportional to the squares of the semi-major ($a$), semi-intermediate ($b$) and semi-minor ($c$) axes such that $a > b > c$ and the corresponding eigenvectors align with the principal axes of the underlying ellipsoidal shape distribution. The shape tensor is solved iteratively in order to adopt the intrinsic shape of the matter distribution with each step.
\\
\indent For the shape estimation, we considered the DM (stellar) particles lying within a spherical volume of 2 $\times$ DM (stellar) half-mass radius, $R_{h,\text{DM}}$ ($R_{h,*}$, estimated based on the 3D particle distribution in a left-handed coordinate system centred at the galactic centre of mass. Here, we note that $R_{h}$ and R$_{1/2}$ (mentioned in Section \ref{sec:data}) are different in the sense that R$_{1/2}$ was directly noted from the TNG data archive; on the other hand, we calculated $R_h$ based on the distribution of particles in the individual galaxy cutouts. Following \cite{2016Tomassetti_ShapeTensor}, first we computed eigenvalues and eigenvectors of the shape tensor of a spherical volume of radius $R$ = 2$R_h$, fixing $a$ = $b$ = $c$ = $R$. Then at each iteration, (a) the coordinate system was rotated such that it aligned along the eigenvectors from the previous iteration, (b) the eigenvalues were normalised with respect to the largest one ($a$) while fixing $a = R$, and (c) the particles within the ellipsoidal volume $\frac{x^2}{a^2} + \frac{y^2}{b^2}+ \frac{z^2}{c^2}= 1 $ were considered for the next iteration. The loop continued until the percentage error in $b/a$ and $c/a$ between two successive iterations decreased below 1\% for the DM and 8\% for the stellar component. In some cases, the shape tensors may fail to converge if two of the axes attain nearly equal lengths or when the convergence radius is not implemented (\citealt{2006Allgood_ShapeTensor}; \citealt{2019Chua_ShapeTensor}); also relaxing the convergence criterion may also aid in achieving convergence.
\\
\indent We obtained the elongation ($e$), flattening ($f$), and the triaxiality parameter ($T$) of the DM and stellar morphology by using the following formulae:
\begin{equation}\label{eq:elongation_flattening}
    e = \sqrt{1 - \bigg(\frac{b}{a}\bigg)^2},   \;\;\;  f = \sqrt{1 - \bigg(\frac{c}{b}\bigg)^2},  \;\;\;  T = \frac{1 - (\frac{b}{a})^2}{1 - (\frac{c}{a})^2}.
\end{equation}
All of these parameter can have values ranging between 0 to 1. Larger values of $e$ and $f$ represents highly elongated and flattened shapes, respectively. Further, T $<$ 1/3, 1/3 $<$ T $<$ 2/3 and T $>$ 2/3 corresponds to the oblate, triaxial and prolate shape. 
\subsection{Kinematical properties involving velocity dispersion}\label{sec:dispersion}
\subsubsection{Velocity anisotropy parameter:}\label{sec:anisotropy}
To probe the orbital structure of the DM and stellar components of the galaxies in our sample, we studied their velocity anisotropy parameter ($\beta$) as a function of galactocentric radii following the definition \citep{1980Binney_anisotropy}, 
\begin{equation}\label{eq:anisotropy}
    \beta  = 1 - \frac{\sigma_\theta^2 \;+\; \sigma_\phi^2}{2\sigma_r^2}.
\end{equation}
Here, $\sigma_r$, $\sigma_\theta$ and $\sigma_\phi$ represent the diagonal components of velocity dispersion tensor in the spherical rest-frame. We started with a coordinate transformation from Cartesian to a new left-handed orthogonal coordinate system where the $z$-axis aligned with the direction of angular momentum and the $x$-$y$ lying arbitrarily in the plane perpendicular to $z$ (essentially the galactic plane). Next, the Cartesian velocity components ($v_x$, $v_y$ and $v_z$) were converted to those in the spherical coordinates ($v_r$, $v_\theta$ and $v_\phi$). Further, we used the following formula to calculate $\sigma_r$, $\sigma_\theta$ and $\sigma_\phi$ at 15 radial points equally-spaced between 0 to 2R$_h$ by considering the particles within each spherical shell:
\begin{equation}\label{eq:velocity_dispersion}
    \sigma_{i} = \sqrt{\langle v_i^2 \rangle \; - \; \langle v_i \rangle^2}.
\end{equation}
From Eq. \ref{eq:anisotropy}, it is evident that the $\beta$ parameter is defined in such a manner that it can attain values between -$\infty$ to 1. $\beta$ = -$\infty$ and $\beta$ = 1 represent, purely-tangential and purely-radial orbits, respectively, while $\beta$ = 0 describes an isotropic orbital structure. 
\subsubsection{Stellar velocity ellipsoid:}\label{sec:sve}
We further studied the components of the stellar velocity dispersion in the cylindrical rest-frame ($\sigma_r$, $\sigma_\theta$, $\sigma_z$) whose ratios define the shape of the stellar velocity ellipsoid (SVE) (\citealt{1907Schwarzschild}; \citealt{1997Gerseen_SVE}; \citealt{2003Shapiro_SVE}; \citealt{2012Gerssen_SVE}). In a similar manner as discussed in \ref{sec:anisotropy}, we evaluated $\sigma_r$, $\sigma_\theta$ and $\sigma_z$ in the cylindrical coordinates and evaluated their ratios - $\sigma_z$/$\sigma_r$, $\sigma_\theta$/$\sigma_r$ - as a function of galactocentric radius up to 2.5R$_h$. $\sigma_z$/$\sigma_r$ $>$ 0.5 and $<$ 0.5 are suggestive of, respectively, dynamically hot and cold stellar components revealing key information about the underlying kinematic heating mechanisms. The effects of vertical disc heating agents (for example, galactic bars) increases the $\sigma_z/\sigma_r$, on the other hand, non-axisymmetric features such as spirals can cause decrease in $\sigma_z/\sigma_r$ (\citealt{2018Pinna_SVE} and the references therein). In general, $\sigma_\theta$/$\sigma_r$ only validates the epicyclic approximation projected to the plane of the galactic disc. However, $\sigma_\theta$/$\sigma_r$ may be crucial in understanding the variations in stellar orbital trajectories, therefore, inferring the differences in the underlying halo potential \citep{2002Combes,2008BinneyTremaine}. It may also indicate the presence of bulge or bar-like structures in a galaxy \citep{2018Pinna_SVE, 2018Chemin_SVE, 2022WaloMartin_SVE}.
\subsection{Kinematical properties extracted from mock IFS data cubes}\label{sec:LOSVD}
In our study, we employed the publicly-available \texttt{R}-software framework \texttt{SimSpin} to simulate mock integral field spectroscopic (IFS) data using the TNG50-1 galaxy cutouts as initial conditions (\citealt{2020SimSpin_v1, 2023SimSpin_v2}). The line-of-sight velocity distribution (LOSVD) $f(v)$ of a galaxy reveals the distribution of stellar motions projected along the line-of-sight and can be extracted from its IFS data cube. The LOSVD of a galaxy can be parametrised with a Gaussian and higher-order (> 2) Gauss-Hermite functions (\citealt{1993Gerhard_LOSVD}; \citealt{1993vanderMarel_Franx_LOSVD}) as:
\begin{eqnarray}
    &&f(v) = I_0 e^{-y^2/2} (1 + h_3 \mathscr{H}_3(y) + h_4 \mathscr{H}_4(y) + \dots) \label{eq:losvd}, \\
    \text{where,} \label{eq:dummy1} \nonumber 
    &&\mathscr{H}_3(y) = (2\sqrt{2} y^3 - 3\sqrt{2}y)/\sqrt{6} \;\;\; {\rm and} \nonumber\\
    &&\mathscr{H}_4(y) = (4y^4 - 12y^2 + 3)/\sqrt{24} \;. \nonumber
\end{eqnarray}
Here, $I_0$ is a normalization constant; $v_{\text{fit}}$ and $\sigma_{\text{fit}}$ are the measured radial velocity and velocity dispersion, respectively, while $v$ is the mean radial velocity with $y = \frac{(v_{\text{fit}} - v)}{\sigma_{\text{fit}}}$. $\mathscr{H}_3$ and $\mathscr{H}_4$ represent the 3rd and 4th order Hermite polynomials. 
The coefficients associated with $\mathscr{H}_3$ ($h_3$) and $\mathscr{H}_4$ ($h_4$) in Eq. \ref{eq:losvd} quantify the deviation of the LOSVD from a standard Gaussian velocity distribution. In general, the IFS data cubes are binned by implementing the Voronoi 2D-binning algorithm \texttt{vorbin} \citep{2003Vorbin} and fitted using \texttt{ppxf} scheme \citep{2004ppxf} to extract the pixel-by-pixel velocity moment maps. However, \texttt{SimSpin} offers a choice to generate the line-of-sight velocity ($V$), dispersion ($\sigma$), $h_3$, and $h_4$ moments directly (only up to the fourth order).
\\
\indent \texttt{SimSpin} generates the mock IFS spectra by implementing three main functions - (1) \texttt{telescope} that can be set to mimic the instrumental specifications of some well-known IFS survey like MUSE, CALIFA, MANGA, SAMI; (2) \texttt{observing\_strategy} that defines the distance, inclination etc. of a galaxy; and (3) \texttt{build\_datacube} that generates the spectral data cubes or kinematic moment maps specified by \texttt{method} for a given telescope specifications and observational conditions. 
In this work, we set the telescope \texttt{type} to match the MUSE survey while applying $g$-band SDSS filter bandpass function (\texttt{filter} = {\textquotesingle}$g${\textquotesingle}) and adding some artificial {\textquotesingle}noise{\textquotesingle} such that \texttt{signal\_to\_noise} ratio $\sim$ 30. The field-of-view (\texttt{fov}) was chosen suitably for different galaxy populations based on their radial extent. Next, the galaxies were specified to be observed at a projected redshift-distance (\texttt{dist\_z}) of 0.05,  inclined at angles (\texttt{inc}) 10 and 50 degrees. Finally, the mass-weighted 2D-maps of the velocity moments were constructed by setting \texttt{method = velocity}, \texttt{mass\_flag = T} and turning on the particle-based Voronoi tessellation (\texttt{voronoi\_bin = T}). To minimise the effect of shot-noise on the mock IFS data cubes, we restricted each Voronoi pixel to contain a minimum (\texttt{voronoi\_limit}) of 200 particles \citep{2024SimSpin_VorbinLimit}. 
\\
\indent To avoid using effective radii obtained from the 3D stellar-particle distributions (as discussed in Sect. \ref{sec:data}), we used the 2D-flux distribution of the datacubes inclined at angle 10 to obtain the effective radii. Additionally,  the galaxy datacubes with 50 degree-inclination were employed to find the projected ellipticity. However, to extract the stellar kinematic properties, we used only the velocity moment map which were constructed for the 50 degree inclination. We fitted S\'{e}rsic model to the 2D-observed flux distribution of a galaxy at its face-on orientation while choosing the pixel with the highest mass as the centre of the 2D map. The R$^{1/n}$ model proposed by S\'{e}rsic \citep{1963Sersic, 1968Sersic} expresses the intensity of light as a function of galactocentric radius as,
\begin{equation}
    I({\rm R}) = I_e \; \text{exp}\bigg\{ -b_n \bigg[ \bigg( \frac{\rm R}{\text{R}_{\text{eff}}}\bigg)^{\frac{1}{n}} - 1 \bigg] \bigg\},
\end{equation}
where, $I_e$ represents the intensity at the effective radius, R$_{\text{eff}}$, $n$ is the S\'{e}rsic index, and the constant $b_n = 2n \; - \; 1/3$. Next, we constructed the inertia tensor of the 2D-flux distribution to estimate the alignment ($\vartheta$) of the principal axes with respect to the 2D Cartesian $x$-$y$ (following \citealt{1980Carter_Metcalfe_LOSVD_PA}). The formula for obtaining $\vartheta$ is given by, 
\begin{equation}\label{eq:alignment}
\vartheta = \frac{1}{2} \text{tan}^{-1}\bigg(\frac{2\langle xy \rangle}{\langle x^2\rangle - \langle y^2\rangle}\bigg) = \frac{1}{2} \text{tan}^{-1}\Bigg(\frac{2\sum_{i=1} F_ix_iy_i}{\sum_{i=1} F_ix^2_i - \sum_{i=1} F_iy^2_i} \Bigg),
\end{equation} 
Here, ($x_i$, $y_i$) and $F_i$ represent the coordinate and the flux of the $i$-th pixel on the 2D map considering all the valid pixels. Further, we calculated the projected ellipticity of the galaxy inclined at an angle of 50 degrees using the following expression \citep{2007Emsellem_LOSVD}:
\begin{equation}\label{eq:epsilon}
\epsilon = 1- \sqrt{\frac{\langle y^2 \rangle}{\langle x^2 \rangle}} = 1- \Bigg(\frac{\sum_{i=1}^N F_i y^2_i}{\sum_{i=1}^N F_ix^2_i}\Bigg)^\frac{1}{2},
\end{equation}
keeping $x$ and $y$ aligned along the photometric major and minor-axis, respectively. Here, N represents the total number of pixels within 2.5R$_{\rm eff}$.
We further extracted the kinematic properties such as $\langle$V/$\sigma \rangle$, proxy of the spin parameter ($\lambda_{\rm R_e}$), $\langle h_3 \rangle$, and $\langle h_4 \rangle$. Here, $\langle$ $\rangle$ represent the flux-weighted mean evaluated within an elliptical region of semi-major axes R$_{\rm eff}$ and ellipticity $\epsilon$, inclined at an angle $\vartheta$ with respect to $x$-$y$, as follows \citep{2007Emsellem_LOSVD}:
\begin{eqnarray}\label{eq:average}
    \langle A \rangle &=& \frac{\sum_{i=1}^N F_i \; a_i}{\sum_{i=1}^N F_i}, \quad A \equiv h_3, h_4, \text{etc.}
\end{eqnarray}
Here, we note that the dwarfs in our sample are comparatively less extended than the other galaxy populations, therefore, we considered 2R$_{\text{eff}}$ for obtaining the flux-weighted averages for the dwarf sample. Further, the following formulae were implemented to calculate $V/\sigma$ and $\lambda_{\rm R_e}$ (\citealt{2005Binney,2007Emsellem_LOSVD,2007Cappellari_LOSVD}):
\begin{eqnarray} 
    \frac{V}{\sigma} &\equiv& \sqrt{\frac{\langle V^2 \rangle}{\langle \sigma^2 \rangle}} = \Bigg(\frac{\sum_{i=1}^N F_i V^2_i}{\sum_{i=1}^N F_i \sigma^2_i}\Bigg)^\frac{1}{2}, \label{eq:v-sigma} \\
    \text{\(\lambda\)}_{\rm R_e} &\equiv& \frac{R|V|}{R\sqrt{V^2 + \sigma^2}} = \frac{\sum_{i=1}^N F_i R_i |V_i|}{\sum_{i=1}^N F_i R_i \sqrt{V_i^2 + \sigma_i^2}}. \label{eq:spin_parameter_LOSVD}
\end{eqnarray}
Here, $R_i$ is the distance to the $i$-th pixel in the elliptical region. In the Eqs. \ref{eq:v-sigma} and \ref{eq:spin_parameter_LOSVD}, we used the inclination-corrected velocity by substituting $V$ with $V/\text{cos} \;50$ to incorporate the circular velocity in these expressions. Eq. \ref{eq:v-sigma} characterises the ratio of ordered to random motion: $V/\sigma$ $>$ 1 signifies the dominance of rotational motions observed in fast-rotating (FRs) galaxies and $V/\sigma$ $<$ 1 suggests the influence of dispersion suggestive of slow-rotating (SRs) galaxies. Moreover, $\lambda_{\rm R_e}$ is a measure for the galaxy's global rotation that can be employed to investigate the kinematic features of the galaxy. Similarly, $\lambda_{\rm R_e}$ $>$ and $<$ 0.1 roughly indicates the FRs and the SRs, respectively. Finally, the average of the coefficient $h_3$ measures the global asymmetry in the velocity distribution such that $\langle h_3\rangle$ > 0 (< 0) describes skewness towards velocities lower (higher) than the mean velocity of the galaxy. Parallelly, the average of $\langle h_4\rangle > 0$ corresponds to an overall heavy-tailed distribution with sharper peak, whereas, $\langle h_4\rangle < 0$ indicates an overall light-tailed blunt distribution compared to the standard Gaussian \citep{2001Halliday_LOSVD,2017vandeSande_LOSVD}. 
\section{Results and discussion}\label{sec:results}
We present our results in three main segments divided based on the analysis techniques. In Sect. \ref{results:regression}, we study possible scaling relations between a few pairs of mass properties which were directly noted from the TNG archive. In Sect. \ref{results:shape_DM} and Sect. \ref{results:orbits_n_kinematics}, we discuss the morphological, orbital and kinematical properties of our galaxy samples obtained by analysing the galaxy cutouts. Finally, we explore the kinematic properties extracted from the mock IFS data cubes in Sect. \ref{results:results_losvd}. In all the cases, we primarily focus on the intrinsic properties of the UDGs and observe their similarities and differences with other galaxy populations.
\begin{figure*}[ht!]
    \centering
    \setlength{\tabcolsep}{0.2pt} 
    \renewcommand{\arraystretch}{0} 
    \includegraphics[width=0.32\linewidth]{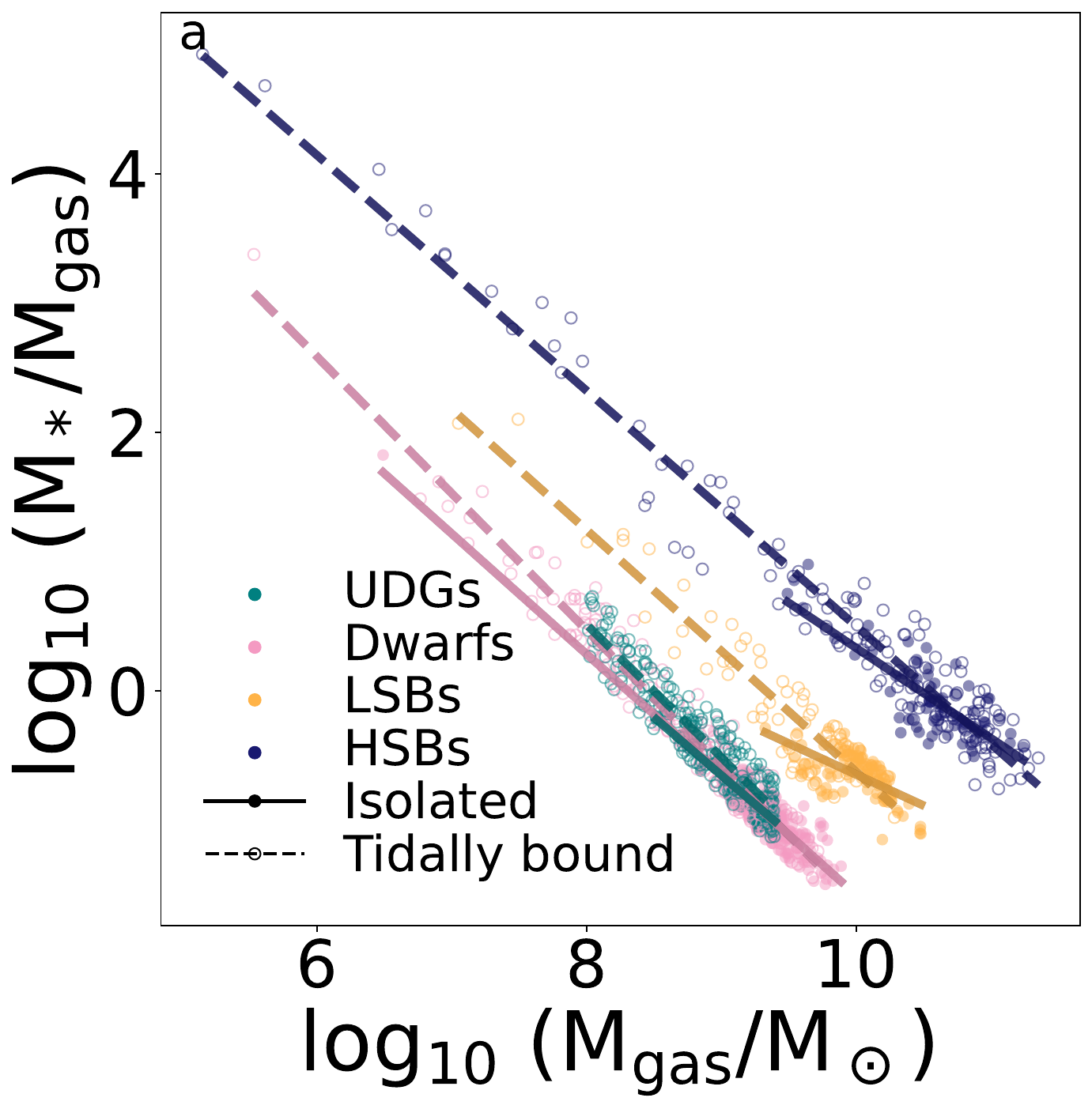} 
    \includegraphics[width=0.32\linewidth]{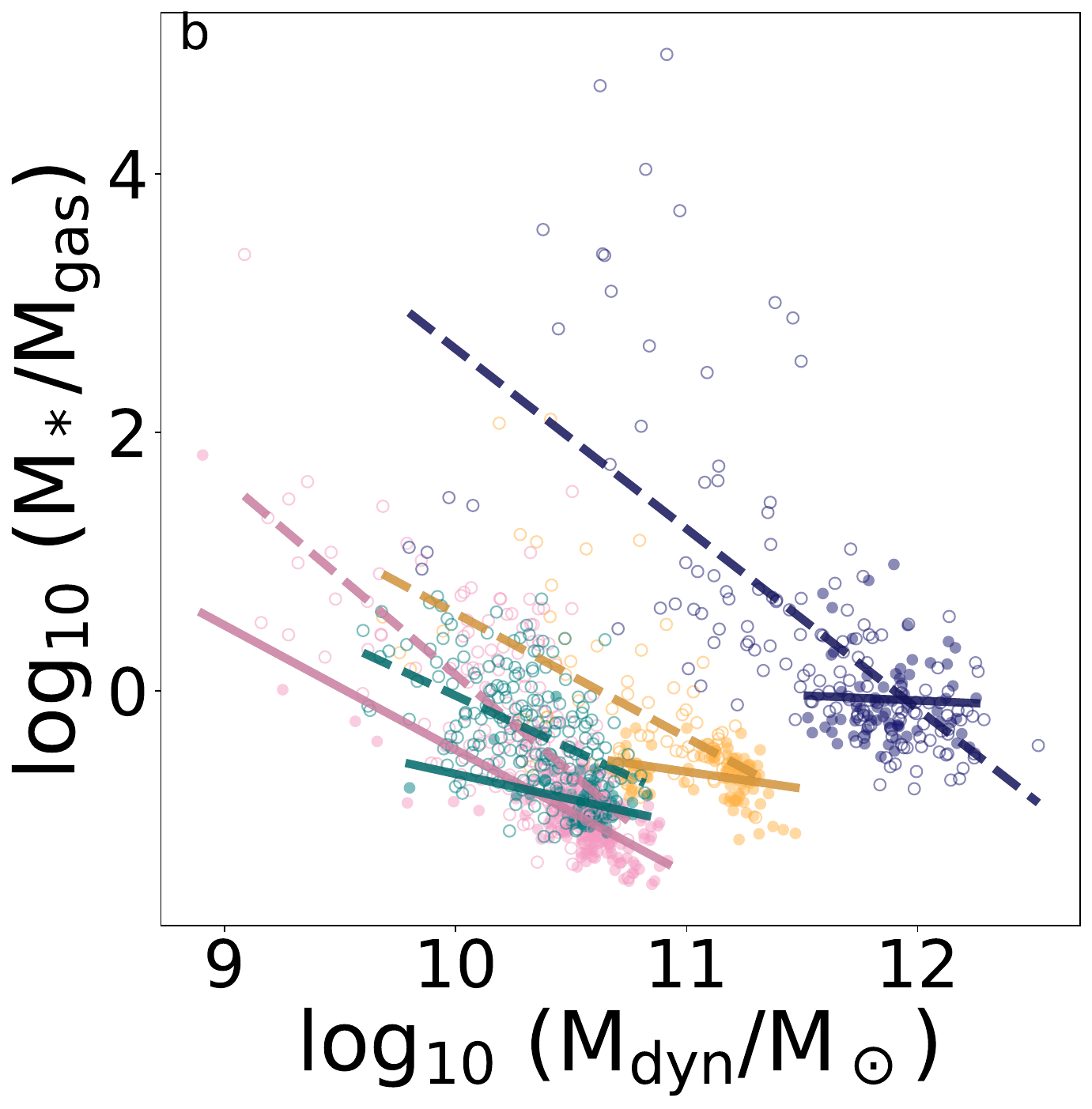}
    \includegraphics[width=0.33\linewidth]{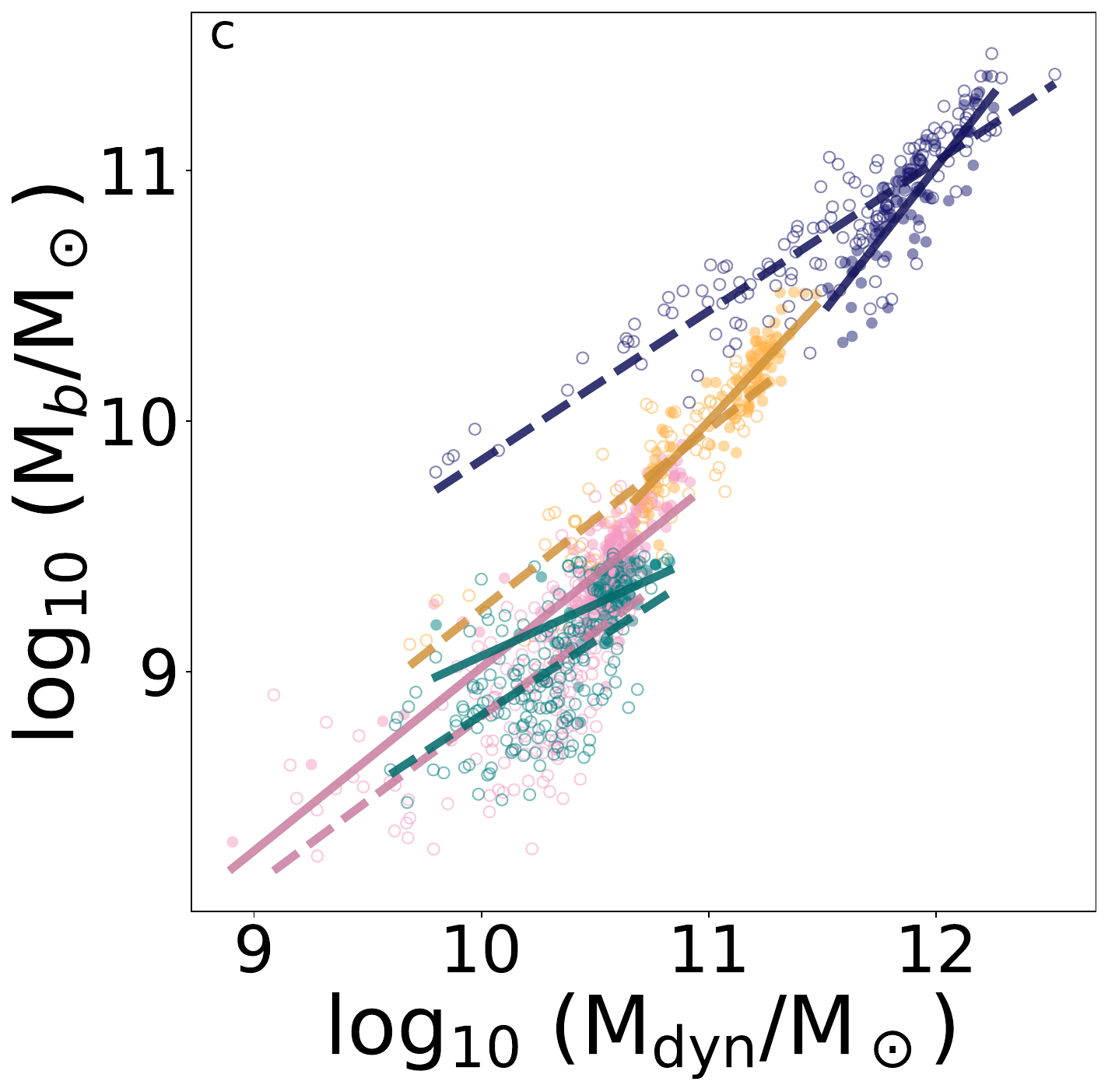} 
    \caption{Possible scaling relations of (a) stellar-to-gas mass ratio versus gas mass (log$_{10}$ M$_*$/M$_{\rm gas}$ vs. log$_{10}$ M$_{\rm gas}$), (b) stellar-to-gas mass ratio versus total dynamical mass (log$_{10}$ M$_*$/M$_{\rm gas}$ vs. log$_{10}$ M$_{\rm dyn}$), and (c) total baryonic mass versus total dynamical mass log$_{10}$ M$_{\rm b}$ vs. log$_{10}$ M$_{\rm dyn}$). The UDGs, LSBs, HSBs, and the dwarfs are shown in teal, yellow, blue and magenta colour, respectively. Each of the galaxy samples are divided in two subsamples - isolated (denoted with filled circles) and tidally bound (denoted with empty circles). The scaling relations are obtained for the two subsamples of each galaxy classes separately and plotted with solid and dashed lines, respectively.}
    \label{fig:basic_regression}
\end{figure*}
\subsection{Possible galaxy scaling relations}\label{results:regression}
Galaxies may appear similar by means of individual parameters, however, the differences in their origin are reflected on the galaxy scaling relations which are regulated by the underlying physical processes \citep{1996deCosta_Renzini}. In Figure \ref{fig:basic_regression}, we present a set of possible scaling relations obtained separately for the isolated and tidally bound subsets of our galaxy samples. On each panel, scatter distributions and regression fits for the isolated galaxies are plotted with solid circles and solid lines, respectively; those of the tidally bound galaxies are denoted with empty circles and dotted lines. We followed the same colour scheme for our galaxy populations as discussed in Sect. \ref{sec:tidal_index}. 
For each pair of parameters, we obtained the ordinary regression relation using the python module \texttt{statsmodels} and noted the coefficient of determination (R$^2$) and the corresponding $p$-value. The null hypothesis in this case states that there exists no correlation between the independent and the dependent variables. 
$p <$ 0.05 rejects the null hypothesis thereby suggesting a possible correlation, while, $p >$ 0.05 offers no conclusion. In parallel, we examined validity and strength of the correlation between each pair of parameters by conducting Spearman's rank correlation test by employing the python module \texttt{scipy}. Again, we wrote down Spearman's correlation coefficient (hereafter denoted by S) and the corresponding $p$ values. 
Similar to the previous case, $p <$ 0.05 rejects the null hypothesis revealing a monotonic correlation between two physical parameters. For a statistically significant correlation, positive and negative signs of the S-coefficient denote, respectively, correlations and anti-correlations; the larger the absolute value of S, the stronger the (anti-)correlation. We assumed a threshold of 0.5 for S above which a correlation was termed to be stronger. The slopes ($m$), intercepts ($c$), R$^2$ and the S-coefficients are listed in Table \ref{tab:regression}. The $p$-values are smaller than 0.05 in all cases, unless mentioned. We observed that in cases of mild (anti-)correlations where S $<$ 0.5, the R$^2$ values are significantly smaller. Therefore, we considered the S-coefficients as the indicators of correlations and use the $m$ and $c$-values to differentiate between the galaxy populations.
\\
\indent Panel $a$ of Figure \ref{fig:basic_regression} presents the correlation between stellar-to-gas mass ratio, M$_*$/M$_{\rm gas}$ and M$_{\rm gas}$ for our galaxy samples. The gas mass present in a galaxy constitutes the fuel for star formation and the stellar mass characterises the final outcome of the past star formation activities. The correlation between M$_*$/M$_{\rm gas}$ and M$_{\rm gas}$ may be studied to understand how efficiently the gas is converted to stars for a given M$_{\rm gas}$, therefore, M$_*$/M$_{\rm gas}$ may be considered as the proxy of star formation efficiency. We observe strong anti-correlations between these two parameters (S-values $\lesssim$ -0.6) for all the galaxy populations. Here, we note that the regression fits for the UDGs and the dwarfs are almost identical. It is also evident that the UDGs and the dwarfs have significantly lesser M$_*$/M$_{\rm gas}$ for a given value of M$_{\rm gas}$ compared to the HSBs similar to what has been suggested for the low-mass dwarf galaxies \citep{2018Catinella_PlotA,2024Dou_PlotA,2025Dou_plotD}; LSBs, on the other hand, seem to constitute an intermediate population. Hence, we may infer that the star formation activities in the UDGs and the dwarfs are reasonably different than that of the LSBs and the HSBs.
\\
\indent DM may have an important role in regulating the star formation activities in galaxies, especially in case the DM-dominated galaxy populations (\citealt{2017GargBanerjee_plotC}). Therefore, we studied how M$_*$/M$_{\rm gas}$ varies as a function of M$_{\rm dyn}$ in panel $b$ of Figure \ref{fig:basic_regression}. M$_*$/M$_{\rm gas}$ anti-correlates with M$_{\rm dyn}$ as suggested by their S-values. Here we observe that the $p$-values associated with R$^2$ and S for the isolated UDGs and the isolated HSBs are greater than 0.05, making the correlation statistically insignificant. Hence, based on their distributions in this parameter space, we note that the UDGs and dwarfs cluster in a similar region in contrast to the other galaxy populations. This suggests that the influence of DM on the star formation activities in the UDGs and the dwarfs are almost alike.
\\
\indent Finally, in panel $c$ of Figure \ref{fig:basic_regression}, we observe the correlation between the baryonic mass, M$_b$ and M$_{\rm dyn}$, analogous to the baryonic Tully-Fisher relation \citep{2000McGaugh_BTFR,2005McGaugh_BTFR}. Here, we obtained the total baryonic mass as M$_b$ = M$_*$ + M$_{\rm gas}$. As expected, all the galaxy populations show a strong correlation between these two parameters with correlation coefficients $>$ 0.6. The distribution of the UDGs and the dwarfs and their regression fits are nearly similar in this parameter space. Interestingly, the baryon content in the UDGs
seems to be {\textquotesingle}normal{\textquotesingle}, in contrary to the baryon-rich field UDGs studied by \cite{2019Pina_UDG_LOSVD_offBTFR}. 
\\
\indent Therefore, we observe that the UDGs and the dwarfs follow nearly similar scaling relations which are significantly different from that of the HSBs. This corroborates the idea that the UDGs and the dwarfs share a common dynamical origin, in contrast to the failed L$_*$-type galaxy formation scenario. Here, it is worth-noting that a clear bimodality is observable in the way the LSB galaxies are distributed in these spaces. Some of the LSBs lie closer to the regression fits for the UDGs and the dwarfs. These may represent the transition population between late-type classical LSBs and LSB-dwarfs. Thus, there is a possibility that UDGs are transformed from these type of LSBs.
\begin{figure*}
    \centering
    \setlength{\tabcolsep}{2pt} 
    \renewcommand{\arraystretch}{0} 
    \begin{tabular}{c}
        \includegraphics[width=1\textwidth]{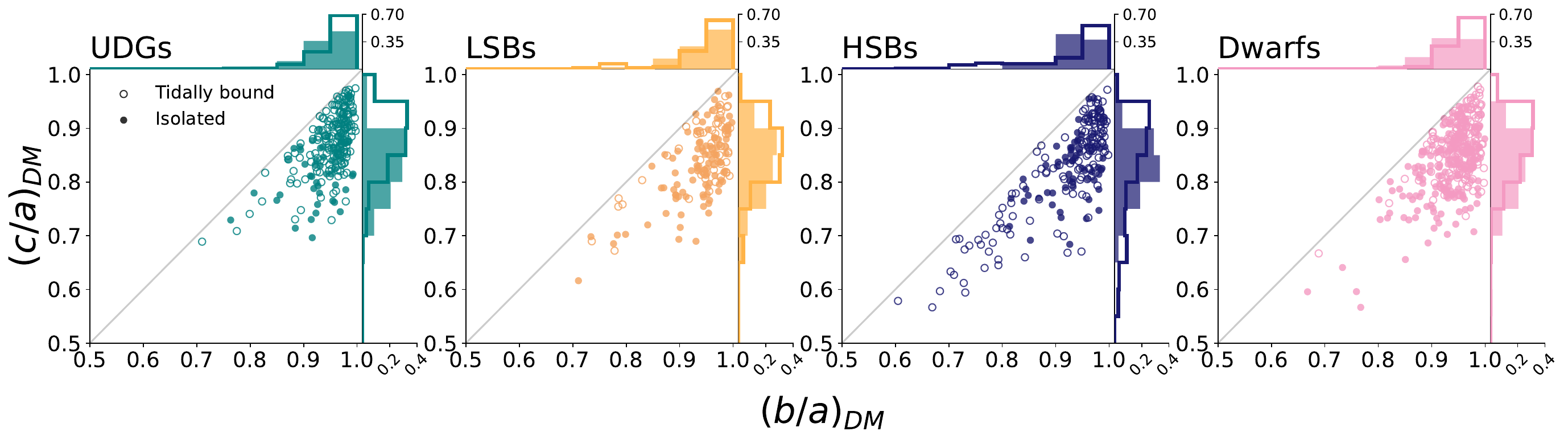} \\
        \includegraphics[width=1\textwidth, height=5cm]{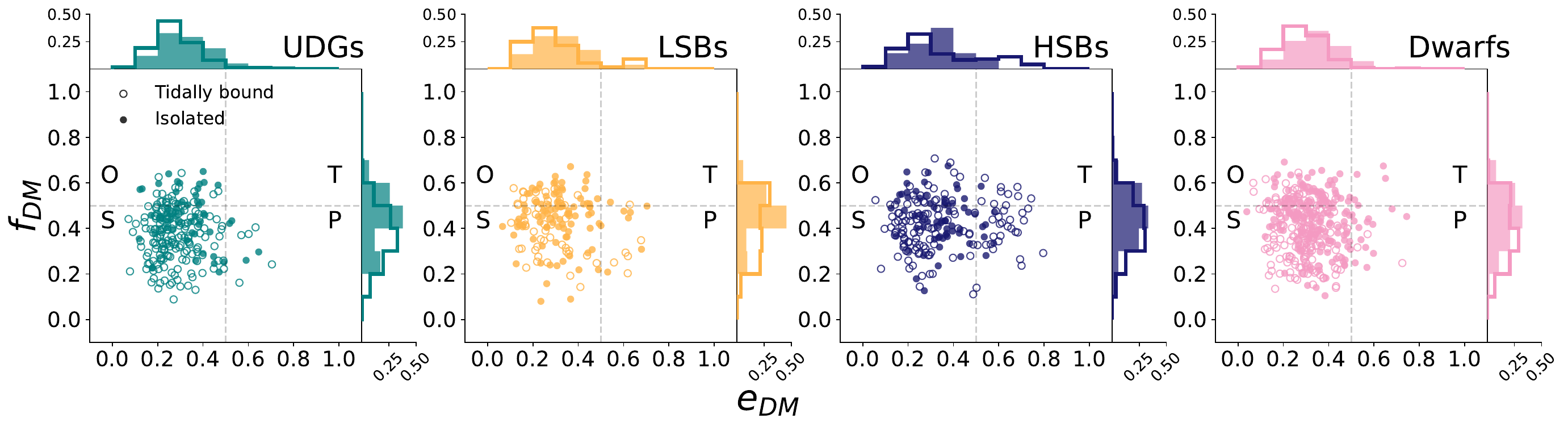}
    \end{tabular}
    \caption{Distribution of the (top) intermediate-to-major and minor-to-major ((b/a)$_{\rm DM}$-(c/a)$_{\rm DM}$) and (bottom) elongation and flattening ($e_{\rm DM}$-$f_{\rm DM}$) of the DM component of our galaxy samples. The histograms of the $x$ and $y$ data are plotted on the top and right sides of each panel, respectively, with equal bin widths, showing the fraction of galaxies in each bin. The isolated galaxies are shown with filled circles and filled histograms, while the empty circles and empty histograms represent the tidally bound subsamples. The same colour scheme as in Figure \ref{fig:basic_regression} is adopted.}
    \label{fig:dm_axes_ratios_elong_flat}
\end{figure*}
\subsection{Intrinsic morphology of the DM and stellar components}\label{results:shape_DM}
The intrinsic shape of the DM distribution reveals important insights to the halo and baryon assembly history, effects of the baryonic feedback mechanisms and the surrounding environment (see for example, \citealt{2021Cataldi_ShapeTensor}; \citealt{2023Orkney_haloShape}; \citealt{2024Rodriguez_haloalignment} and the references therein). The shape of the DM halo modifies the underlying halo potential and thus deeply impacts the stellar orbits, thus the stellar morphology of a galaxy (\citealt{2014Zotos_stellarOrbit_haloshape1,2014Zotos_stellarOrbit_haloshape2}). Therefore, it is an important parameter to constraint the picture of galaxy formation and evolution. 
In Figure \ref{fig:dm_axes_ratios_elong_flat}, we present the intrinsic shapes of the DM halos of the UDGs, LSBs, HSBs, and the dwarfs from left to right panels, respectively. The distribution of the intermediate-to-major (b/a)$_{\text{DM}}$ and minor-to-major (c/a)$_{\text{DM}}$ axes ratios of the DM distribution is shown on the top row. The colour code has been maintained as described in \ref{sec:tidal_index}. The histograms of $x$ and $y$-data on each panel are shown on the top and right of the panel. For a better comparison, the histograms are plotted with same bin widths for all the galaxy classes and represent the fraction of occurrences in each bin. More than 70\% of the UDGs, LSBs, HSBs, and the dwarfs (in both isolated and tidally bound conditions) have (b/a)$_{\text{DM}}$ values within the range 0.9 - 1 and (c/a)$_{\text{DM}}$ ranging between 0.75 - 0.95. For a clearer classification, the elongation ($e_{\text{DM}}$) and flattening ($f_{\text{DM}}$) of the DM distribution is shown in the bottom row of Figure \ref{fig:dm_axes_ratios_elong_flat}. The $e_{\text{DM}}$-$f_{\text{DM}}$ spaces are divided in four quadrants with the following approximate ranges: (i) spherical (S) for 0 $\lesssim$ $e_{\text{DM}}$ $\lesssim$ 0.5 \& 0 $\lesssim$ $f_{\text{DM}}$ $\lesssim$ 0.5; (ii) oblate (O) for 0 $\lesssim$ $e_{\text{DM}}$ $\lesssim$ 0.5 \& 0.5 $\lesssim$ $f_{\text{DM}}$ $\lesssim$ 1; (iii) prolate (P) for 0.5 $\lesssim$ $e_{\text{DM}}$ $\lesssim$ 1 \& 0 $\lesssim$ $f_{\text{DM}}$ $\lesssim$ 0.5 and (iv) triaxial (T) for 0.5 $\lesssim$ $e_{\text{DM}}$ $\lesssim$ 1 \& 0.5 $\lesssim$ $f_{\text{DM}}$ $\lesssim$ 1 \citep{2016Tomassetti_ShapeTensor}. Most of the galaxies in UDG, LSB, HSB and dwarf samples are observed to reside in spherical DM halos as suggested by their $e_{\text{DM}}$ and $f_{\text{DM}}$ values. In both cases, no striking differences can be identified based on their distributions in the (b/a)$_{\text{DM}}$-(c/a)$_{\text{DM}}$ and $e_{\text{DM}}$-$f_{\text{DM}}$ spaces.
\\
\begin{figure*}
    \centering
    \setlength{\tabcolsep}{2pt} 
    \renewcommand{\arraystretch}{0} 
    \begin{tabular}{c}
        \includegraphics[width=1\textwidth]{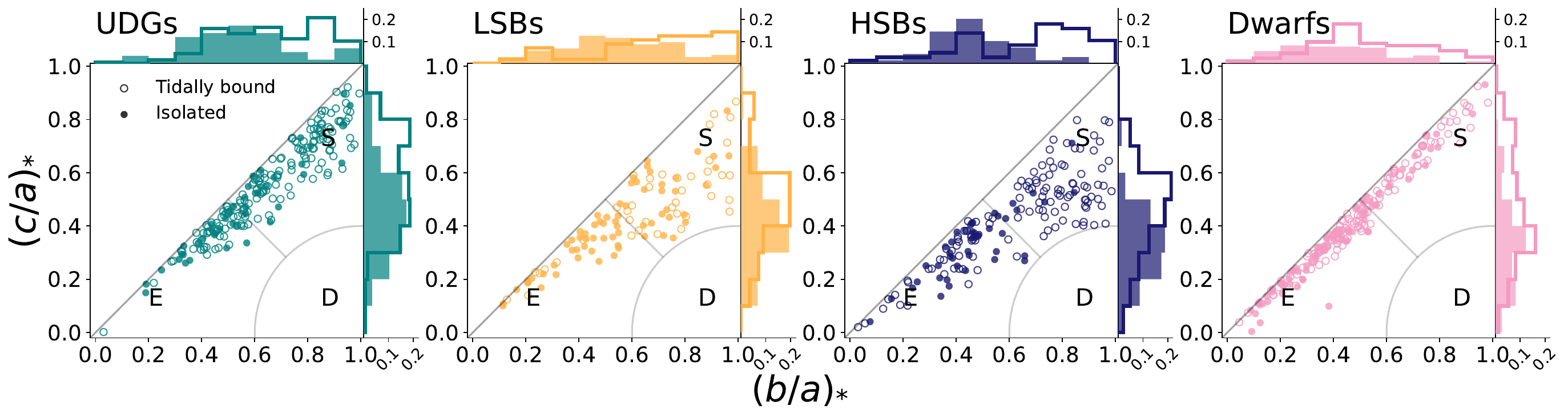} \\
        \includegraphics[width=1\textwidth]{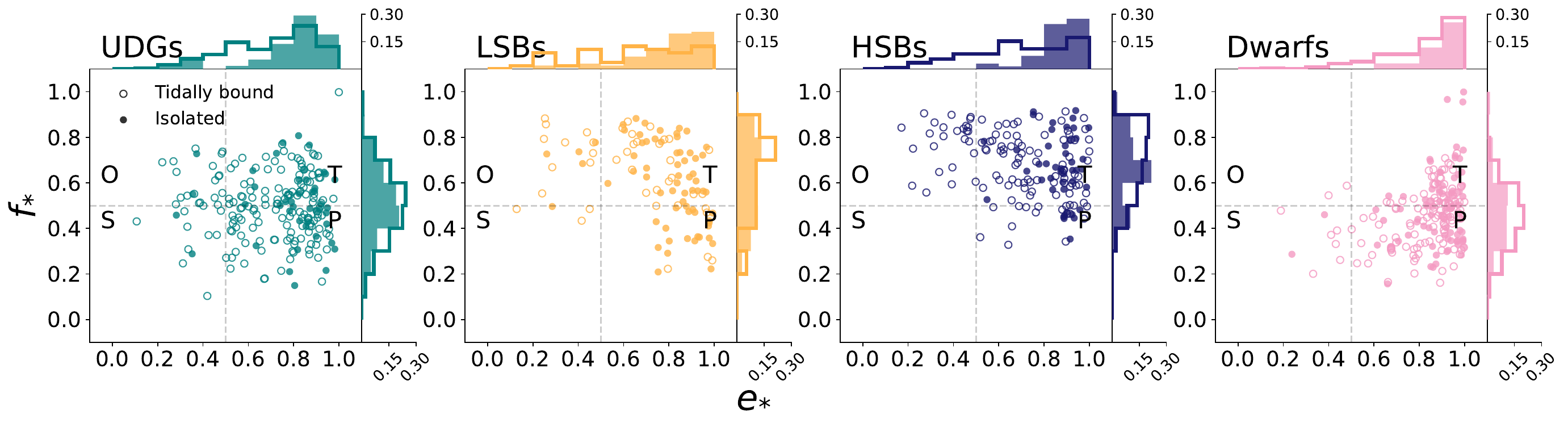}
    \end{tabular}
    \caption{Distribution of the stellar component of our galaxy samples in (top) (b/a)$_*$-(c/a)$_*$ and (bottom) $e_*$-$f_*$ spaces, similar to Figure \ref{fig:dm_axes_ratios_elong_flat}.}
    \label{fig:star_axes_ratios_elong_flat}
\end{figure*}
\begin{figure*}
    \centering
    \includegraphics[width=1.0\linewidth, height=5cm]{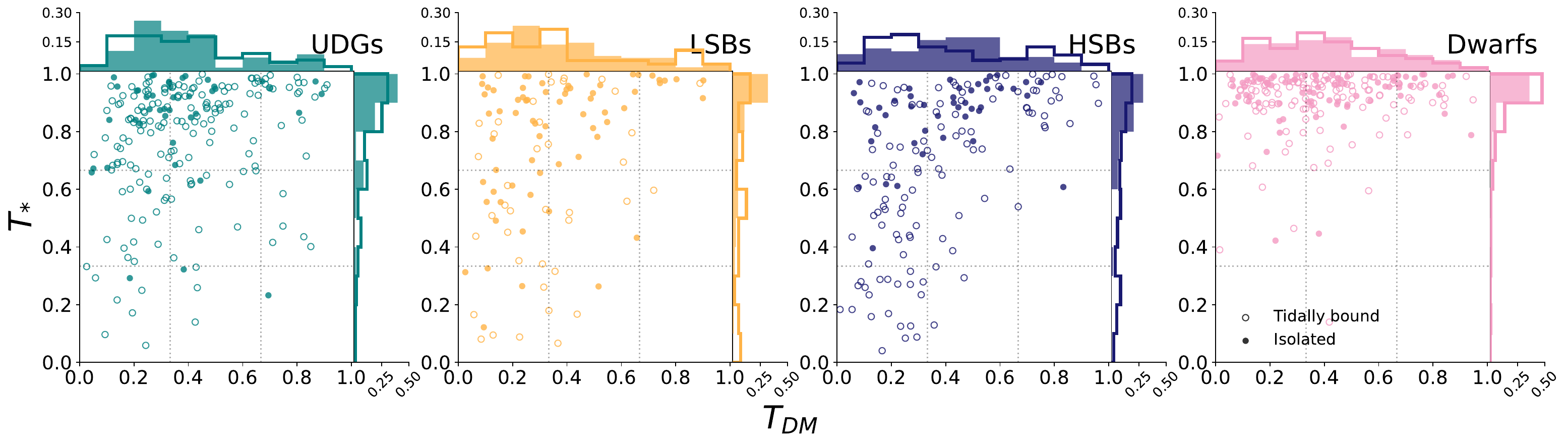}
    \caption{Distribution of the stellar component of our galaxy samples in $T_*$-$T_{\rm DM}$ space, similar to Figure \ref{fig:dm_axes_ratios_elong_flat}.}
    \label{fig:Tstar_Tdm}
\end{figure*}
\indent Stellar morphology of a galaxy preserves imprints of its formation mechanisms and reveals important information about the intrinsic and extrinsic physical mechanisms influencing its evolution (\citealt{2017RodriguezGomez_morphology}; \citealt{2021Emami_ShapeTensor_a}; \citealt{2022Zhang_ShapeTensor}; \citealt{2025Kolesnikov_morphology} and the references therein). Based on the existing morphological classifications, galaxies can be broadly classified in two categories: elliptical or dispersion-supported early-type galaxies (ETGs)  and disc or rotation-supported late-type galaxies (LTGs) - both being the relics of distinct formation mechanics (\citealt{1926Hubble}; \citealt{1936Hubble}; \citealt{1966Liller}; \citealt{1996KormendyBender}). In Figure \ref{fig:star_axes_ratios_elong_flat}, the stellar shape parameters for the UDG, LSB, HSB, and dwarf samples are shown, following the same colour scheme as earlier. As discussed in Sect. \ref{sec:morphology}, the stellar shape tensor did not converge for 23 out of 235 UDGs (14 isolated, 9 tidally bound), 52 out of 159 LSBs (36 isolated, 16 tidally bound), 52 out of 213 HSBs (25 isolated, 27 tidally bound), and 160 out of 368 dwarfs (108 isolated, 52 tidally bound). Therefore, the results are presented only for the galaxies for which the stellar shape tensor converged. The axes ratios of the stellar shapes ((b/a)$_{*}$ and (c/a)$_*$) are shown on the top row of Figure \ref{fig:star_axes_ratios_elong_flat}. On each panel, the (b/a)$_{*}$-(c/a)$_*$ space is divided in three sections relating to three different morphological types: (i) spherical (S) for $a_*$ $\sim$ $b_*$ $\sim$ $c_*$; (ii) discy/oblate (D) for $a_*$ $\sim$ $b_*$ $>$ $c_*$ and (iii) elongated/prolate (E) for $a_*$ $>$ $b_*$ $\sim$ $c_*$ \citep{2014vanderWel_ShapeTensor, 2022Zhang_ShapeTensor}. 
The bottom row of Figure \ref{fig:star_axes_ratios_elong_flat} shows the elongation and flattening of the stellar shape distributions ($e_*$-$f_*$). From the distributions of our galaxy samples in these two parameter spaces, we note that isolated UDGs predominantly exhibit prolate-to-triaxial shapes, whereas they display a wider range of morphologies under tidally bound conditions. In contrast, the stellar components of dwarfs are better described by prolate shapes. The LSBs and HSBs, on the other hand, are more likely to display triaxial shapes isolated and tidally bound environments. 
\\
For a comprehensive understanding of the overall picture, we show the distributions of the stellar ($T_*$) and dark matter triaxiality parameters ($T_{\rm DM}$) in Figure \ref{fig:Tstar_Tdm}. To examine the similarities in the intrinsic dark matter and stellar shapes between the UDGs and the other galaxy samples, we performed two-sample Anderson–Darling (AD) tests using $T_*$ and $T_{\rm DM}$. For each parameter: (1) we randomly sampled 500 values with replacement from each galaxy class; (2) conducted AD tests between pairs of classes, subdivided into isolated and tidally bound subsamples - (UDGs and LSBs), (UDGs and HSBs), and (UDGs and dwarfs) - and recorded the AD coefficients and corresponding $p$-values; and (3) repeated this process 5,000 times and computed the median AD coefficients and $p$-values. The null hypothesis in this statistical test states that the shape parameters for the UDGs and the comparison galaxy samples (LSBs, HSBs, or dwarfs) follow the same continuous distribution. The null hypothesis was rejected if the AD coefficient $\geq$ critical value (AD$_{\rm crit}$ = 2.718 at a 2.5\% significance level) with a $p$-value < 0.02, thereby indicating that the distributions of the corresponding parameters for the UDGs and the other galaxy samples are statistically distinct.
We find that the median AD coefficients exceed AD$_{\rm crit}$, and the $p$-values are below 0.02 for both $T_*$ and $T_{\rm DM}$ across all pairs, except for the isolated UDGs and dwarfs. For these former cases, the null hypotheses are rejected, indicating that the distributions of the respective parameters are statistically distinct; in contrast, the null hypothesis cannot be rejected for the latter case. It may suggest that the UDGs and the dwarfs originate in DM halos of similar intrinsic shapes.
\\
\indent Our results for the isolated UDGs seem to be in compliance with the inferences drawn by \cite{2019Jiang_UDG_ShapeTensor} who studied the field UDGs in the NIHAO simulation and concluded that multiple stellar feedback episodes may be responsible for their non-rotating prolate shapes. On the contrary, follow-up studies incorporating the isolated UDGs in the NIHAO simulation conducted by \cite{2020CardonaBarrero_UDG_ShapeTensor} suggest that the existence of dispersion-supported triaxial/prolate UDGs and rotation-supported oblate-discy UDGs are almost equally probable. \cite{2019Liang_UDG_ShapeTensor} also found that star-forming (blue) field UDGs in the Auriga simulation tend to have discy-shape similar to the classical low-surface brightness galaxies (LSBs), while the quenched (red) satellite UDGs acquire a prolate-shape possibly due to tidal effects. On a similar note, \cite{2022vanNest_UDG_morphology} studied field UDGs from the ROMULUS25 and ROMULUS C simulation and observed that the isolated UDGs exhibit oblate-triaxial shapes in contrast to their non-UDG counterparts. Our results obtained for the isolated UDGs seem to be in contrast with these studies. On the other hand, the shapes of the tidally bound UDGs in our study are in compliance with the cluster UDGs studied in the literature. By studying the projected axial ratios of the Coma UDGs, \cite{2017Burkert_UDG_ShapeTensor} observed that the UDGs exhibit more elongated, bar-like prolate shapes. \cite{2020Rong_UDG_ShapeTensor} observed that the low- and intermediate-redshift cluster UDGs display an oblate-triaxial shape and become puffier with (i) decreasing redshift, possibly indicating a {\textquotesingle}discy{\textquotesingle} origin and (ii) decreasing cluster-centric distance, suggesting a crucial role of tidal interactions on their evolution. A somewhat similar observation was made by \cite{2021KadoFong_UDG_ShapeTensor} that the UDGs are well-characterised by the oblate-spheroid shape, irrespective of their colour or environment. Studying the cluster UDGs from the ROMULUS25 and ROMULUS C simulation, \cite{2022vanNest_UDG_morphology} suggested that the tidal field make these galaxies relatively spherical. 
Therefore, we may infer that the UDGs have intrinsically prolate stellar distributions in isolation, possibly due to episodic stellar feedback. However, in a denser environment, the tidal field transforms their stellar distributions therefore giving rise to a range of intrinsic shapes. In this context, it is important to note that our results are different compared to the inferences made by \cite{2023Benavides_UDG_TNG} on the morphologies of the UDGs in the TNG50; they found no significant difference in the shapes of the UDGs in different environments.  
\begin{figure*}
    \centering
    \setlength{\tabcolsep}{2pt} 
    \renewcommand{\arraystretch}{0} 
    \begin{tabular}{c}
        \includegraphics[width=1\textwidth]{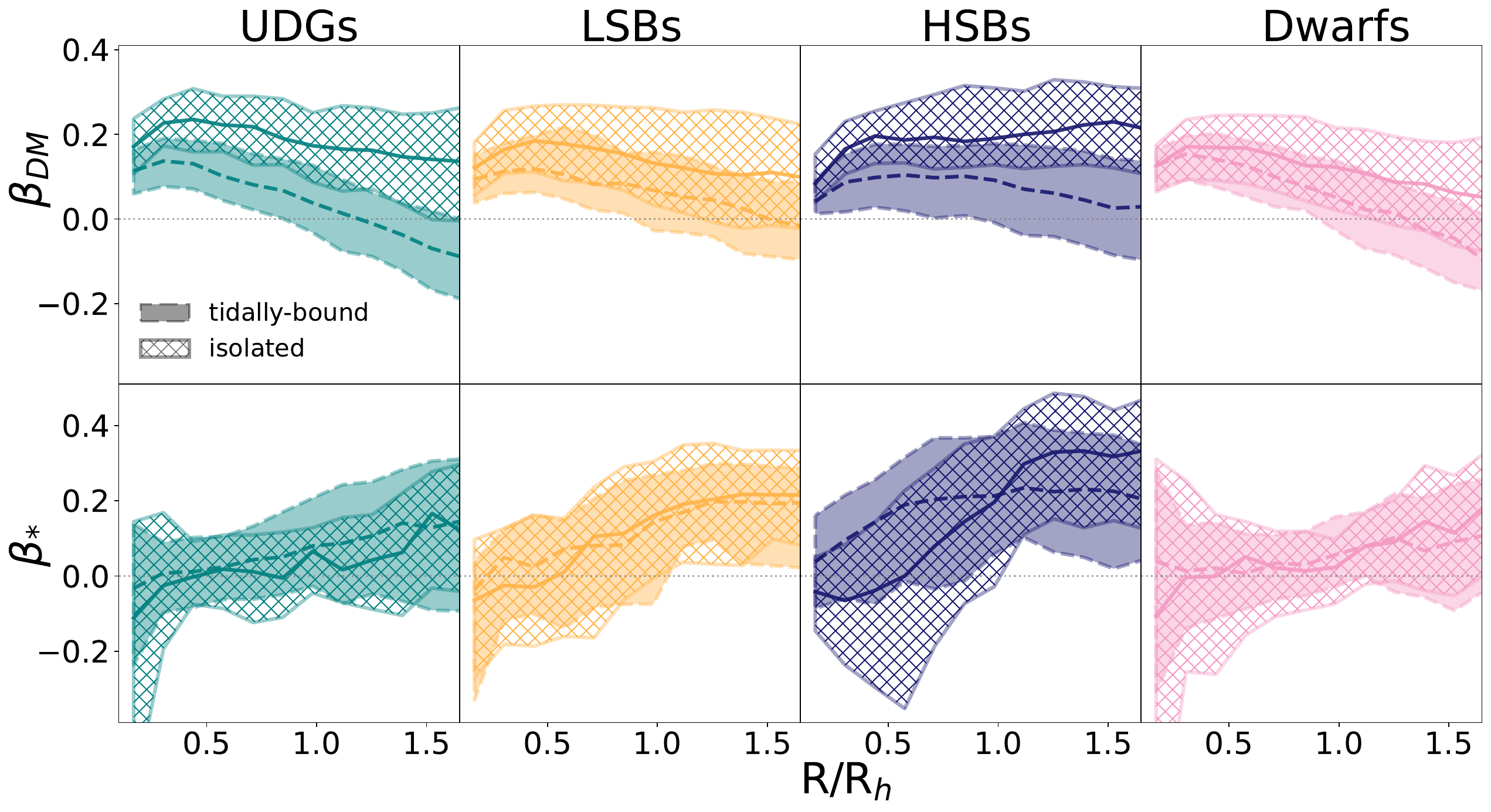} 
    \end{tabular}
    \caption{Variation of (top) DM ($\beta_{\rm DM}$) and (bottom) stellar ($\beta_*$) velocity anisotropy as a function of galactocentric radius normalised by the respective half-mass radius. Solid and dashed lines represent the median values of the isolated and tidally bound galaxy subsamples, respectively. The 25th and 75th quartile regions are shown with shaded region - solid-filled for the isolated subsamples and crossed diagonal lines for the tidally bound ones.}
    \label{fig:anisotropy}
\end{figure*}
\subsection{Orbital and kinematical properties from the galaxy cutouts}\label{results:orbits_n_kinematics}
Elements of the velocity dispersion tensor trace the dynamical status of a galaxy and, therefore, crucial to probe its formation and evolution. In this section, we study a few parameters constructed using the components of DM and stellar velocity dispersions which were evaluated from the individual galaxy cutouts.
\subsubsection{DM and stellar velocity anisotropy: an indicator of the underlying orbital structure}\label{results:anisotropy}
The velocity anisotropy parameter measures orbital structure of a galaxy's DM and stellar components. The variation in $\beta$ can be insightful to understand the concentration and radial distribution of DM halos (\citealt{2024He_anisotropy} and the references therein); at the same time, the stellar velocity anisotropy may be employed to distinguish the ETGs from the LTGs \citep{2014Agnello_anisotropy,2022Emami_VelocityDispersion}. In Figure \ref{fig:anisotropy}, we present the (top) DM anisotropy ($\beta_{\text{DM}}$) as a function of galactocentric radii normalised by the DM half-mass radius. We mainly focus on the inner and middle region (up to nearly 1.5R$_{\rm h}$) of the DM halos as the number of particles near the outer region become significantly smaller. The solid (dashed) lines represent medians of the isolated (tidally bound) galaxy populations and the region shaded (with crossed diagonal lines) denote the 25th to 75th quartile range. By studying $\beta_{\rm DM}$ as a function of R/R$_{\rm h}$, we observe that the DM orbits are closer to isotropic ($\beta_{\text{DM}} \sim$ 0) near the centre of the halo and become radial-biased ($\beta_{\text{DM}} >$ 0) with increasing radius for all the galaxy populations. In case of the UDGs, LSBs, and the dwarfs, the orbits become isotropic again while approaching the outer region. In contrast, the DM orbits continue to remain radially biased in case of the HSBs. Presence of more radial orbits near the outer halo of the HSBs may be indicative of the active radial infall of stars through accretion. On the other hand, the DM orbits of the tidally bound UDGs and dwarfs become rather tangentially-biased ($\beta_{\text{DM}} <$ 0) near the outer halo maybe in response to the tidal-interactions \citep{2014VeraCiro_ShapeTensor}. \cite{2024He_anisotropy} observed that the variation in $\beta_{\rm DM}$ depends on the total dynamical mass of the galaxies. Thus, we may conclude that the UDGs were possibly formed in lower-mass DM halos, consistent with their total dynamical mass.
Similar observation can be made from the radial variation of stellar velocity anisotropies ($\beta_*$) shown in the bottom row of Figure \ref{fig:anisotropy}. The $\beta_*$ profiles for the UDGs and the dwarf galaxies slowly increase with increasing radii, the LSBs displaying a slightly steeper rise. In contrast, the HSBs do not show such a gradual change: the increase in $\beta_*$ in their case seem to be quite steeper. The variation in $\beta_*$ depends on the concentration parameter ($c$) of the DM halo - galaxies with higher $c$ exhibit more radially-biased orbits for a given dynamical mass, therefore resulting in steeply rising radial profiles of $\beta_*$ - a possible signature of different assembly history or evolution \citep{2024He_anisotropy}. Therefore, we may speculate that the variations in the $\beta_*$ profiles of our galaxy samples could also be attributed to their DM $c$-values. However, a detailed analysis of the dark matter density distribution is necessary to confirm this hypothesis.
\begin{figure*}
    \centering
    \hspace{-2em}
    \includegraphics[width=\linewidth]{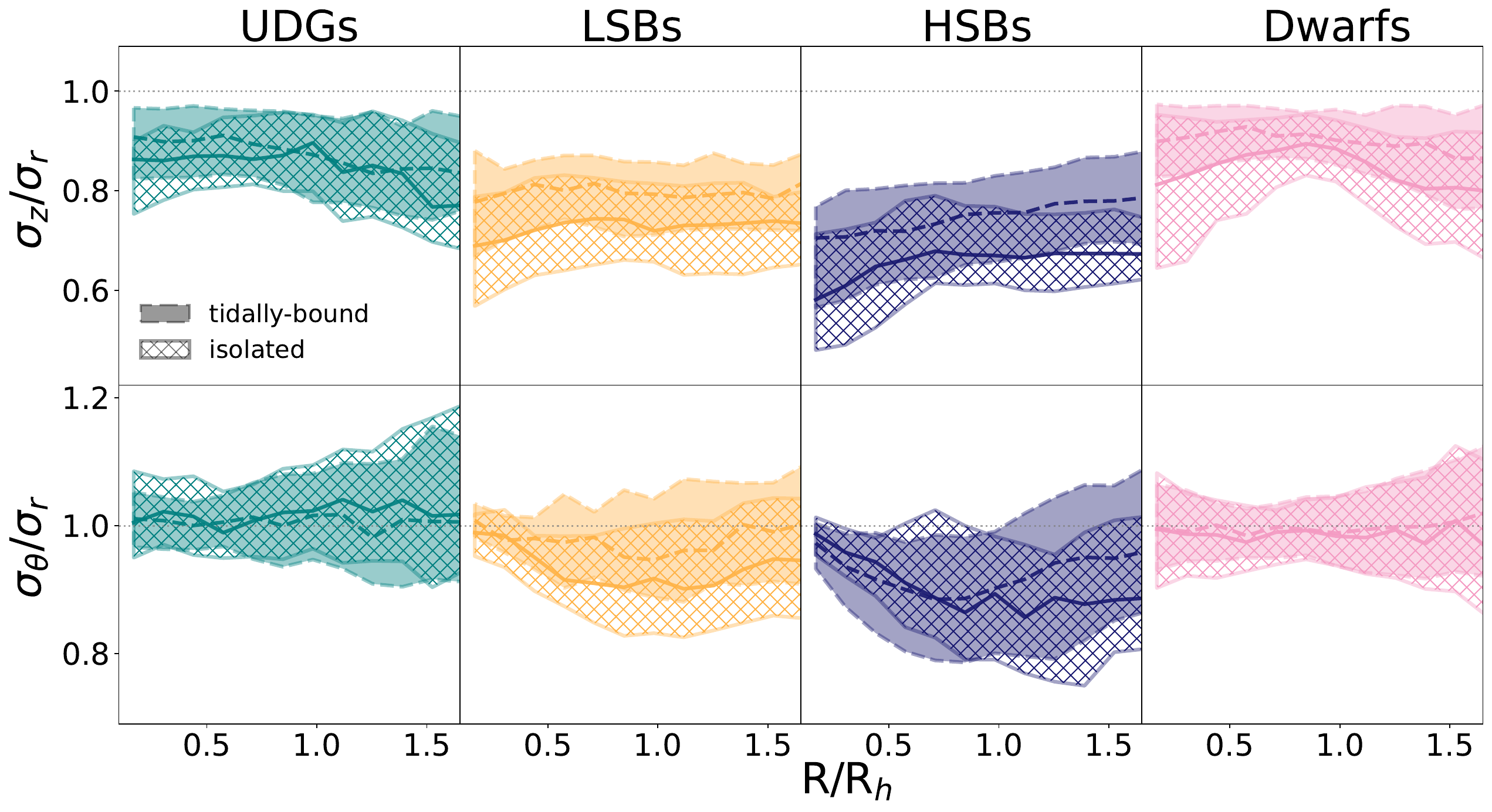}
    \caption{Variation of stellar (top) vertical-to-radial ($\sigma_z/\sigma_r$) and (bottom) tangential-to-radial ($\sigma_\theta/\sigma_r$) velocity dispersion ratios as a function of galactocentric radius normalised by the stellar half-mass radius. The plotting scheme is similar to Figure \ref{fig:anisotropy}.}
    \label{fig:sve}
\end{figure*}
\subsubsection{The components of the stellar velocity ellipsoid: insights to the stellar kinematics}\label{results:sve}
Stars in a galaxy are considered to be formed from a gas disc that settled down at the centre of a DM halo. The gas component has relatively lower velocity dispersion due to its collisional nature. However, the stars end up with gaining sufficient amount of velocity dispersion in response to the dynamical heating mechanisms. Such disc heating agents play an important role in the secular evolution of the galaxies and can be identified by the shape of the stellar velocity ellipsoid (\citealt{1997Gerseen_SVE}; \citealt{2003Shapiro_SVE}; \citealt{2012Gerssen_SVE}). In the top and bottom panel of Figure \ref{fig:sve}, we present the ratios of vertical-to-radial ($\sigma_z/\sigma_r$) and tangential-to-radial ($\sigma_\theta/\sigma_r$) velocity dispersions as a function of galactocentric radius normalised by the stellar half-mass radius. The UDGs and the dwarfs have significantly higher values of $\sigma_z/\sigma_r$ compared to the LSBs and the HSBs. The $\sigma_z/\sigma_r$ values suggest that the UDGs and the dwarfs are dispersion-dominated ETGs while the LSBs and the HSBs are LTGs supported by their rotational motions \citep{1996deBlok_SVE_LSB, 2012Gerssen_SVE, 2014Adams_SVE_Dwarfs}. A similar observation was made in our earlier study where the HI-rich, isolated UDGs and the dwarf irregular galaxies were observed to be earlier-types compared to the LSBs \citep{2025NandiPaper1}. We also note that range of $\sigma_z/\sigma_r$-values of the UDGs are quite similar to those of the Coma UDGs \citep{2019Chilingarian_UDG_LOSVD}. Further, the bottom row of Figure \ref{fig:sve} shows that the variation in $\sigma_\theta/\sigma_r$ as a function of radius. For the UDGs and the dwarfs, $\sigma_\theta/\sigma_r$-values remain close to 1 throughout their galactic plane, indicating isotropically circular orbits. In comparison, the LSBs and the HSBs exhibit isotropic orbits near the galactic centres and the orbits become more radially biased while approaching the outer disc. The presence of increasingly radial orbits may indicate the presence of the underlying bulge or bar-like structures in the LSBs and the HSBs which are rarer in case of the UDGs and the dwarfs \citep{2018Chemin_SVE, 2019Mogotsi_SVE, 2022WaloMartin_SVE}. Therefore, we conclude that the UDGs and the dwarfs exhibit similar kinematic properties as suggested by their SVE.
\begin{figure*}
    \centering
    \includegraphics[width=\linewidth]{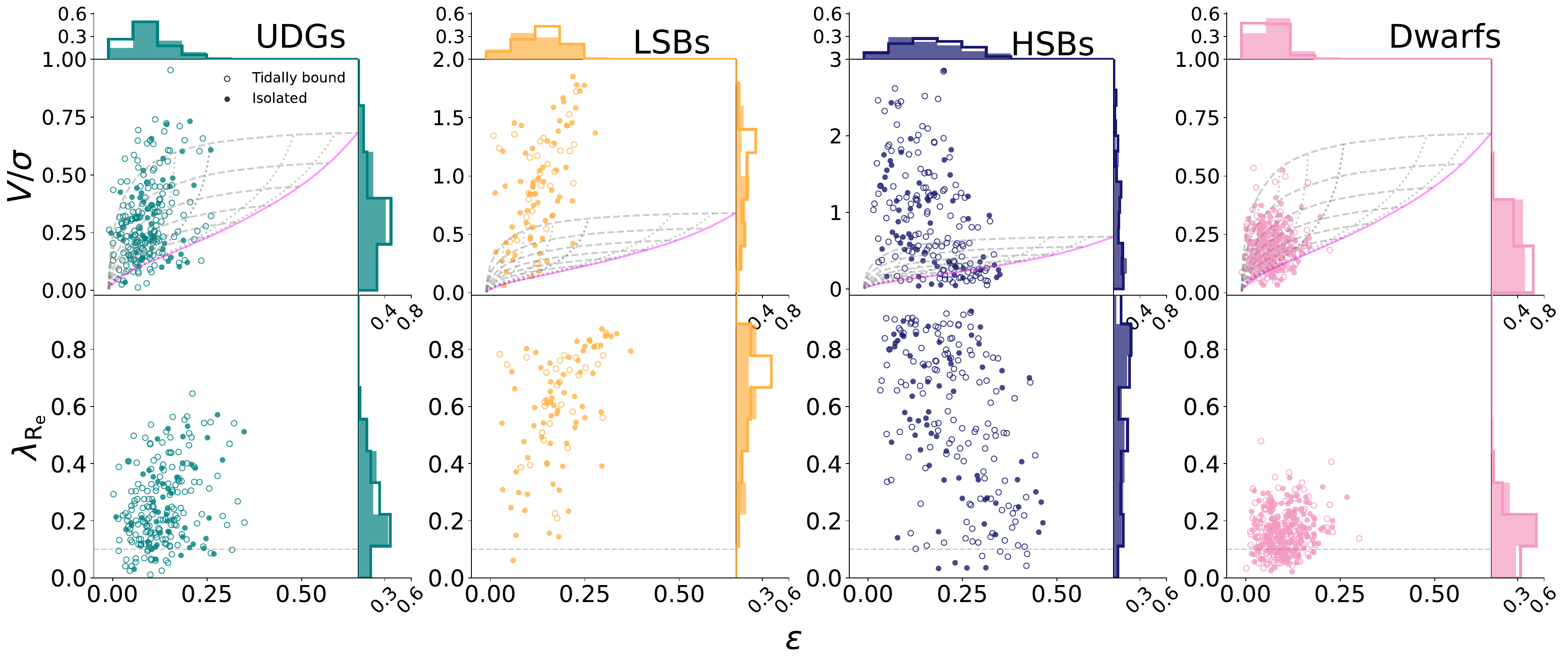} 
    \caption{Distribution of our galaxy samples in (top) $V/\sigma$ vs. $\epsilon$ and (bottom) $\lambda_{\rm R_e}$ vs. $\epsilon$ space following the same plotting scheme as Figure \ref{fig:dm_axes_ratios_elong_flat}. The solid magenta line predicts the theoretical distribution of the axisymmetric galaxies with anisotropy parameter $\beta_z$ = 0.7$\epsilon$ at an inclination of 90 degrees and dotted lines from right to left denote the theoretical distribution for decreasing inclinations, leftmost corresponding to the face-on orientation. Further, The grey dashed lines corresponds locations of galaxies at edge-on orientations with intrinsic ellipticities $\epsilon$ = 0.85 - 0.35 (see \citealt{2007Cappellari_LOSVD}; \citealt{2007Emsellem_LOSVD}). }
    \label{fig:vbys_spin_ellipticity} 
\end{figure*}
\subsection{Stellar kinematics from the mock IFS spectra}\label{results:results_losvd} 
The velocity ($V$), velocity dispersion ($\sigma$), and the higher order ($h_3$ \& $h_4$) moment maps constitute a set of parameter space which directly explores the stellar kinematics of a galaxy (\citealt{1993Gerhard_LOSVD}; \citealt{2001Halliday_LOSVD}; \citealt{2001Bacon_LOSVD}; \citealt{2002deZeeuw_LOSVD}; \citealt{2004Emsellem_LOSVD,2007Cappellari_LOSVD}; \citealt{2007Emsellem_LOSVD}; \citealt{2011Emsellem_vbys_ATLAS}; \citealt{2017Veale_LOSVD};  \citealt{2017FalconBarroso_kine,2019FalconBarroso_kine}; \citealt{2017vandeSande_LOSVD,2021vandeSande_VbyS_SAMI}; \citealt{2019Bernardi_MANGA}; \citealt{2021Pinna_VbyS_SAMI}). In this section, we analysed the properties of the stellar kinematic moment maps obtained from the mock IFS spectra. 
\subsubsection{Stellar kinematics from first and second moments}\label{results:v_by_sigma}
Based on the degree of ordered motion, galaxies can be broadly classified into two categories: rotation-supported FRs and pressure-supported SRs. Emergence of these two distinct classes may be explained by the differences in their assembly histories \citep{2006Cappellari_FR_SR_Formation, 2018vandeSande_LOSVD_VbyS}. The top panel of Figure \ref{fig:vbys_spin_ellipticity} shows the distribution of our galaxy samples in rotation velocity-to-velocity dispersion ($V/\sigma$) versus projected ellipticity ($\epsilon$) space. In all cases, the solid magenta and the dotted lines predict the theoretical distribution of the axisymmetric galaxies with anisotropy parameter $\beta_z$ = 0.7$\epsilon$ for different viewing angles - starting on the right with the solid magenta line for the edge-on galaxies to the left-most for the face-on viewing angles. The grey dashed lines, on the other hand, correspond to the theoretical distributions of galaxies with intrinsic ellipticities $\epsilon$ = 0.85 - 0.35 viewed in edge-on orientations (see \citealt{2007Cappellari_LOSVD}; \citealt{2007Emsellem_LOSVD}).  
\begin{figure*}[ht!]
    \centering
    \includegraphics[width=0.9\linewidth]{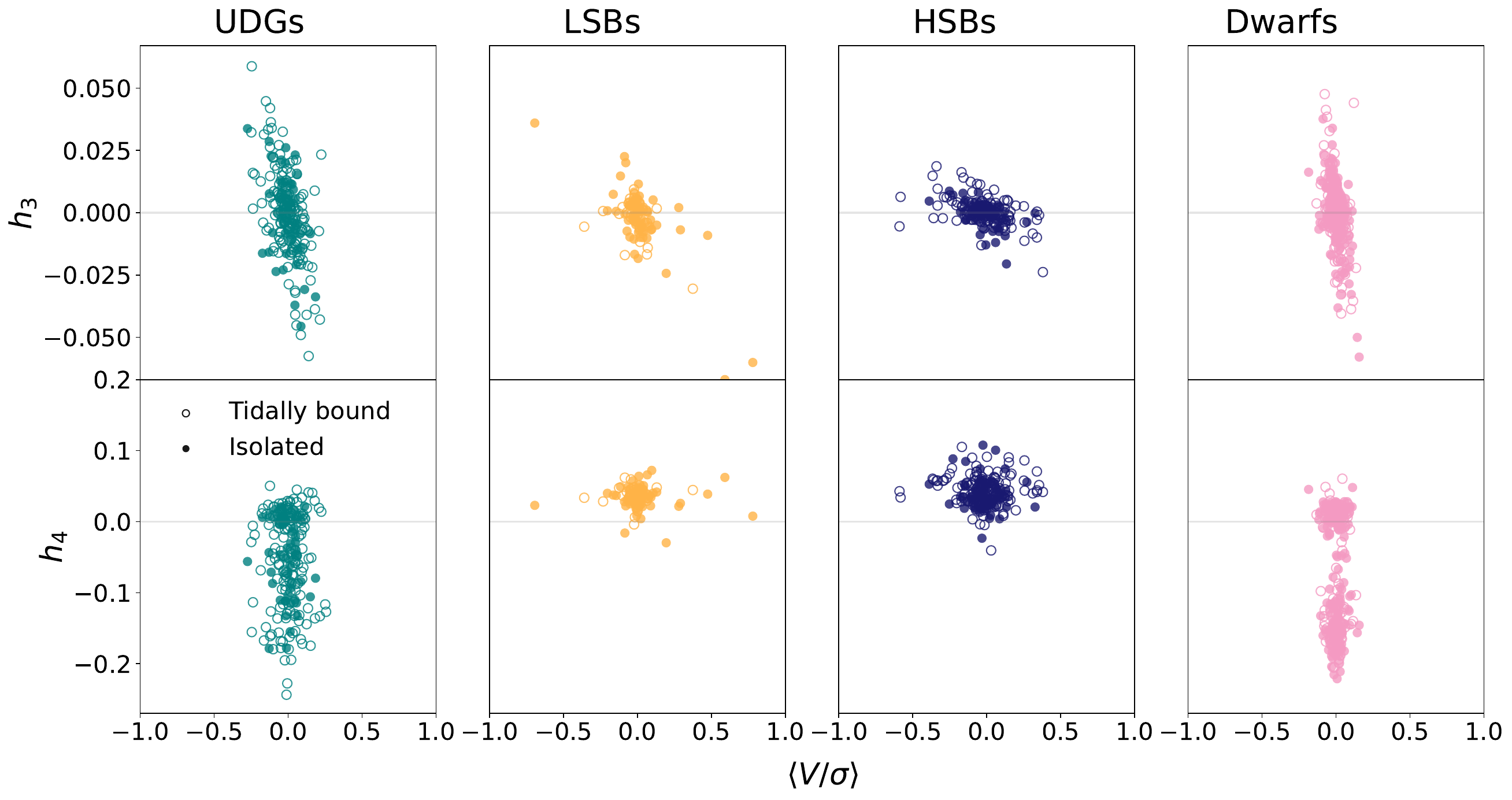} 
    \caption{ Distribution of our galaxy samples in the (top) $h_3$ and (bottom) $h_4$ versus ratio of flux-weighted average of velocity and velocity dispersion ($\langle V/\sigma\rangle$). A similar plotting scheme is adopted as that of the Figure \ref{fig:dm_axes_ratios_elong_flat}.}
    \label{fig:h3_h4_vbys}
\end{figure*}
Based on the distribution of the galaxies in the $V/\sigma$-$\epsilon$ space, we observe that the UDGs and the dwarfs are dispersion-supported systems with $V/\sigma$ $<$ 1, therefore, they can be termed as SRs. In comparison, the LSBs and HSBs seem to have both $V/\sigma$ $>$ 1 and $<$ 1 with nearly equal probability suggesting that they can be both FRs and SRs. Here it is notable that the HSBs are attaining higher values of $V/\sigma$ compared to the LSBs. Environment seems to have no significant impact on any of these galaxy populations as their distributions looks almost similar in case of isolated and tidally bound galaxies except for the isolated LSBs. In this context, we note that R$_{\rm eff}$ values for the UDGs obtained in Sect. \ref{sec:LOSVD} is consistent with their definition i.e. > 1.5 kpc with roughly 12\% of them falling below this threshold (maximum deviation being approx. 25\%). 
\\
\indent While $V/\sigma$ can capture the contest between the rotation and the random motion of a galaxy in describing its stellar kinematics, it lacks the ability to identify its kinematic features and characterise the true nature of the galactic rotation. $\lambda_{\rm R_e}$, by definition, identifies the complex kinematic structures present in the stellar component and thus traces the true global rotation in the galaxy. In the bottom row of Figure \ref{fig:vbys_spin_ellipticity}, we study the $\lambda_{\rm R_e}$-values of our galaxy samples with respect to their ellipticity, $\epsilon$. The UDGs and the dwarfs are predominantly composed of slowly rotating galaxies, with nearly 50\% of these populations exhibiting $\lambda_{\rm R_e}$ < 0.2. In contrast, the LSBs are more strongly represented at higher values of $\lambda_{\rm R_e}$. The HSBs, on the other hand, display a broad range of $\lambda_{\rm R_e}$ values, suggesting the presence of both FRs and SRs within the population. Similar to the previous case, the distributions of the isolated and tidally bound galaxies are comparable in this space, except for the isolated LSBs. The result obtained in this section is in line with what we observed in our previous study - the isolated HI-rich UDGs and the dwarf irregular galaxies are relatively SRs compared to the LSBs which are SRs \citep{2025NandiPaper1}. A similar conclusion was made by \cite{2020CardonaBarrero_UDG_ShapeTensor} by constructing projected line-of-sight velocity and velocity dispersion maps suggesting that nearly half of the field UDGs in the NIHAO simulation are FRs, the remaining half being SRs. Our results are also supported by the range of rotation properties spectroscopically observed in UDGs. 
\cite{2019Emsellem_UDG_LOSVD} conducted the first spectroscopic analysis of NGC1052-DF2 using the VLT/MUSE integral-field spectrograph and suggested it to be a prolate-rotator. \cite{2019vanDokkum_UDG_LOSVD} analysed the spatially resolved stellar kinematics of the Coma UDG DF44 using the Keck/DEIMOS spectrograph and observed almost no rotation. \cite{2023Iodice_UDG_LEWIS} reported the presence of a UDG in the Hydra I cluster, named UDG11, that shows no rotation. Additionally, \cite{2025Buttitta_UDG_LEWIS} investigated the stellar kinematics of 18 Hydra I UDGs and observed a range of kinematic signatures - 5 showing no rotation, 3 showing photometric major-axis rotation and 4 rotating along the intermediate axis. Therefore, we infer that the UDGs and the dwarfs are distinguishable by their SR-nature as opposed to the LSBs. However, the distinction between the UDGs and the HSBs is still vague and, thus, we move one step further to investigate their higher-order velocity moments.
\subsubsection{Stellar kinematics from the higher order GH moments}\label{results:h3_h4}
The correlation between $h_3$ and V/$\sigma$ varies based on the presence of rotating component in galaxies, while the presence of anisotropy in the rotation is reflected on $h_4$. On the top row of figure \ref{fig:h3_h4_vbys}, we show the variation in $h_3$ vs $\langle V/\sigma\rangle$. Here, $\langle V/\sigma\rangle$ denotes the ratio of flux-weighted average of velocity and velocity dispersion obtained by using Eq. \ref{eq:average}. 
We interpret the correlation between $h_3$ and $\langle V/\sigma\rangle$ as follows: (1) fast rotators are characterised by a strong anti-correlation between $h_3$ and $\langle V/\sigma\rangle$; (2) non-regular or slowly rotating galaxies are distributed almost vertically in this parameter space; and (3) no clear conclusions can be drawn for galaxies that are distributed horizontally or show no particular pattern, possibly due to the coexistence of both correlation and anti-correlation arising from individual kinematic components \citep{2017vandeSande_LOSVD, 2014Naab_LOSVD}. On visual inspection, a clear anti-correlation can be seen between $h_3$ and $V/\sigma$ for the LSBs and HSBs; in comparison, the distributions of the UDG and dwarf galaxies are more vertical in nature. Therefore, we may infer that the UDGs and dwarfs can be classified as slow or non-regular rotators, while the LSBs and HSBs are regular rotators, suggestive of the presence of prominent disc components. Finally, the distribution of $h_4$ as a function of $\langle V/\sigma\rangle$ is shown in the bottom row of Figure \ref{fig:h3_h4_vbys}. A rounder, heart-shaped distribution in the $h_4$–$\langle V/\sigma\rangle$ space is observed for the regular rotators; in contrast, no such correlation is evident for the quasi- or non-regular rotators. Among our galaxy samples, the LSBs and HSBs predominantly form rounder distributions with positive $h_4$ values, indicative of stronger rotation. Conversely, the UDGs and dwarfs show a more vertical distribution with two distinct clusters: (i) galaxies with positive $h_4$ values spanning a range roughly half that of the LSBs and HSBs, and (ii) galaxies with negative $h_4$ values, similar to non-rotating ellipticals \citep{1994Bender_LOSVD}. In this context, we note that \citet{2019vanDokkum_UDG_LOSVD} observed $h_3 = -0.03 \pm 0.04$ for DF44, which falls within the same range as the UDGs in our sample. However, they obtained $h_4$ = 0.13 $\pm$ 0.05 for DF44, which is considerably higher than that of the UDGs in our sample, comparable to the HSBs. Overall, we may infer that the UDGs and the dwarfs are characterised by non-regular or slowly rotating stellar components, in contrast to the more regularly, fast-rotating LSBs and HSBs.
\section{Conclusions}\label{sec:conclusions}
In this work, we studied the morphological and kinematical properties of the UDGs and compared with those of the LSB, HSB and dwarf galaxy samples.
We incorporated a set of UDG, LSB, HSB and dwarf galaxies from the TNG50-1 box for this classification study. We further divided the galaxy samples in two subsamples of isolated galaxies and tidally bound galaxies based on their local environments. At first, we examined the correlations between (a) stellar-to-gas mass ratio and gas mass, (b) stellar-to-gas mass ratio and total dynamical mass, and (c) total baryonic mass and total dynamical mass of our galaxy samples and studied their scaling relations in terms of these pairs of parameters. We observe that in most of the cases, the UDGs and the dwarfs share very similar regression relations as compared to the other galaxy populations. Next, we studied the intrinsic shape of the DM and stellar components of these galaxies by analysing their TNG50-1 galaxy cutouts. All the galaxy classes are observed to have nearly spherical DM shapes, however, the isolated UDGs and dwarfs may follow statistically similar distributions in terms of the DM triaxiality parameter. Moreover, The isolated UDGs and the dwarf galaxies exhibit prolate shapes in comparison to the LSBs and the HSBs which are mostly oblate-triaxials in shape. On the other hand, the UDGs in tidally bound condition may display a range of stellar morphology - from prolate to oblate. To investigate the orbital properties, we studied the DM and stellar velocity anisotropy of our galaxy samples. The variations in the DM and stellar velocity anisotropy are consistent with the differences in their dynamical mass implying that UDGs reside in dwarf-like dark halos. Furthermore, the shapes of their SVE suggest that the UDGs and the dwarfs are ETGs while the LSBs and the HSBs can be characterised as LTGs. Finally, we employed the publicly-available R-software \texttt{SimSpin} to construct the stellar kinematic moment maps of the galaxies considered in our study. Based on the flux-averaged values of the rotation-to-dispersion ratio ($V/\sigma$) and the spin parameter proxy ($\lambda_{\rm R_e}$) as functions of projected ellipticity, we may classify the UDGs and dwarfs as slow rotators, whereas the LSBs and HSBs may be identified as fast rotators. A similar inference is supported by their higher-order Gauss–Hermite moments ($h_3$ and $h_4$), which indicate that the UDGs and dwarfs belong to non-regular or slowly rotating galaxy classes, in contrast to the regularly rotating LSBs and HSBs. Therefore, we may conclude that the kinematical properties of the UDGs and the dwarfs are almost alike suggesting their common dynamical origin. We may also infer that the differences in the stellar morphologies of the UDGs and the dwarfs may be probed further to derive important constraints in explaining the evolution of the UDGs. 
\begin{acknowledgements}
We acknowledge that the project is funded by the Prime
Minister’s Research Fellowship with ID: 0902007. We thank the anonymous referee for his/her detailed, constructive suggestions which have improved the paper.
\end{acknowledgements}
\bibliographystyle{aa_url.bst}
\bibliography{bib}{}

@ARTICLE{2025NandiPaper1,
       author = {{Nandi}, Nilanjana and {Banerjee}, Arunima and {Narayanan}, Ganesh},
        title = "{The dynamical lineage of isolated, HI-rich ultra-diffuse galaxies}",
      journal = {\aap},
         year = 2025,
        month = jan,
       volume = {693},
          eid = {A207},
        pages = {A207},
          doi = {10.1051/0004-6361/202450813},
archivePrefix = {arXiv},
       eprint = {2310.08925},
 primaryClass = {astro-ph.GA},
       adsurl = {https://ui.adsabs.harvard.edu/abs/2025A&A...693A.207N}
}

@ARTICLE{1993Gerhard_LOSVD,
       author = {{Gerhard}, O.~E.},
        title = "{Line-of-sight velocity profiles in spherical galaxies: breaking the degeneracy between anisotropy and mass.}",
      journal = {\mnras},
         year = 1993,
        month = nov,
       volume = {265},
        pages = {213},
          doi = {10.1093/mnras/265.1.213},
       adsurl = {https://ui.adsabs.harvard.edu/abs/1993MNRAS.265..213G}
}

@ARTICLE{1993vanderMarel_Franx_LOSVD,
       author = {{van der Marel}, Roeland P. and {Franx}, Marijn},
        title = "{A New Method for the Identification of Non-Gaussian Line Profiles in Elliptical Galaxies}",
      journal = {\apj},
         year = 1993,
        month = apr,
       volume = {407},
        pages = {525},
          doi = {10.1086/172534},
       adsurl = {https://ui.adsabs.harvard.edu/abs/1993ApJ...407..525V}
}

@ARTICLE{2018TNG_Pillepich_b,
       author = {{Pillepich}, Annalisa and {Springel}, Volker and {Nelson}, Dylan and {Genel}, Shy and {Naiman}, Jill and {Pakmor}, R{\"u}diger and {Hernquist}, Lars and {Torrey}, Paul and {Vogelsberger}, Mark and {Weinberger}, Rainer and {Marinacci}, Federico},
        title = "{Simulating galaxy formation with the IllustrisTNG model}",
      journal = {\mnras},
         year = 2018,
        month = jan,
       volume = {473},
       number = {3},
        pages = {4077-4106},
          doi = {10.1093/mnras/stx2656},
archivePrefix = {arXiv},
       eprint = {1703.02970},
 primaryClass = {astro-ph.GA},
       adsurl = {https://ui.adsabs.harvard.edu/abs/2018MNRAS.473.4077P}
}

@ARTICLE{2018TNG_Pillepich_a,
       author = {{Pillepich}, Annalisa and {Nelson}, Dylan and {Hernquist}, Lars and {Springel}, Volker and {Pakmor}, R{\"u}diger and {Torrey}, Paul and {Weinberger}, Rainer and {Genel}, Shy and {Naiman}, Jill P. and {Marinacci}, Federico and {Vogelsberger}, Mark},
        title = "{First results from the IllustrisTNG simulations: the stellar mass content of groups and clusters of galaxies}",
      journal = {\mnras},
         year = 2018,
        month = mar,
       volume = {475},
       number = {1},
        pages = {648-675},
          doi = {10.1093/mnras/stx3112},
archivePrefix = {arXiv},
       eprint = {1707.03406},
 primaryClass = {astro-ph.GA},
       adsurl = {https://ui.adsabs.harvard.edu/abs/2018MNRAS.475..648P}
}

@ARTICLE{2018TNG_Marinacci,
       author = {{Marinacci}, Federico and {Vogelsberger}, Mark and {Pakmor}, R{\"u}diger and {Torrey}, Paul and {Springel}, Volker and {Hernquist}, Lars and {Nelson}, Dylan and {Weinberger}, Rainer and {Pillepich}, Annalisa and {Naiman}, Jill and {Genel}, Shy},
        title = "{First results from the IllustrisTNG simulations: radio haloes and magnetic fields}",
      journal = {\mnras},
         year = 2018,
        month = nov,
       volume = {480},
       number = {4},
        pages = {5113-5139},
          doi = {10.1093/mnras/sty2206},
archivePrefix = {arXiv},
       eprint = {1707.03396},
 primaryClass = {astro-ph.CO},
       adsurl = {https://ui.adsabs.harvard.edu/abs/2018MNRAS.480.5113M}
}

@ARTICLE{2018TNG_Nelson,
       author = {{Nelson}, Dylan and {Pillepich}, Annalisa and {Springel}, Volker and {Weinberger}, Rainer and {Hernquist}, Lars and {Pakmor}, R{\"u}diger and {Genel}, Shy and {Torrey}, Paul and {Vogelsberger}, Mark and {Kauffmann}, Guinevere and {Marinacci}, Federico and {Naiman}, Jill},
        title = "{First results from the IllustrisTNG simulations: the galaxy colour bimodality}",
      journal = {\mnras},
         year = 2018,
        month = mar,
       volume = {475},
       number = {1},
        pages = {624-647},
          doi = {10.1093/mnras/stx3040},
archivePrefix = {arXiv},
       eprint = {1707.03395},
 primaryClass = {astro-ph.GA},
       adsurl = {https://ui.adsabs.harvard.edu/abs/2018MNRAS.475..624N}
}

@ARTICLE{2019TNG_Nelson_a,
       author = {{Nelson}, Dylan and {Springel}, Volker and {Pillepich}, Annalisa and {Rodriguez-Gomez}, Vicente and {Torrey}, Paul and {Genel}, Shy and {Vogelsberger}, Mark and {Pakmor}, Ruediger and {Marinacci}, Federico and {Weinberger}, Rainer and {Kelley}, Luke and {Lovell}, Mark and {Diemer}, Benedikt and {Hernquist}, Lars},
        title = "{The IllustrisTNG simulations: public data release}",
      journal = {Computational Astrophysics and Cosmology},
         year = 2019,
        month = may,
       volume = {6},
       number = {1},
          eid = {2},
        pages = {2},
          doi = {10.1186/s40668-019-0028-x},
archivePrefix = {arXiv},
       eprint = {1812.05609},
 primaryClass = {astro-ph.GA},
       adsurl = {https://ui.adsabs.harvard.edu/abs/2019ComAC...6....2N}
}

@ARTICLE{2018TNG_Springel,
       author = {{Springel}, Volker and {Pakmor}, R{\"u}diger and {Pillepich}, Annalisa and {Weinberger}, Rainer and {Nelson}, Dylan and {Hernquist}, Lars and {Vogelsberger}, Mark and {Genel}, Shy and {Torrey}, Paul and {Marinacci}, Federico and {Naiman}, Jill},
        title = "{First results from the IllustrisTNG simulations: matter and galaxy clustering}",
      journal = {\mnras},
         year = 2018,
        month = mar,
       volume = {475},
       number = {1},
        pages = {676-698},
          doi = {10.1093/mnras/stx3304},
archivePrefix = {arXiv},
       eprint = {1707.03397},
 primaryClass = {astro-ph.GA},
       adsurl = {https://ui.adsabs.harvard.edu/abs/2018MNRAS.475..676S}
}

@ARTICLE{2018TNG_Naiman,
       author = {{Naiman}, Jill P. and {Pillepich}, Annalisa and {Springel}, Volker and {Ramirez-Ruiz}, Enrico and {Torrey}, Paul and {Vogelsberger}, Mark and {Pakmor}, R{\"u}diger and {Nelson}, Dylan and {Marinacci}, Federico and {Hernquist}, Lars and {Weinberger}, Rainer and {Genel}, Shy},
        title = "{First results from the IllustrisTNG simulations: a tale of two elements - chemical evolution of magnesium and europium}",
      journal = {\mnras},
         year = 2018,
        month = jun,
       volume = {477},
       number = {1},
        pages = {1206-1224},
          doi = {10.1093/mnras/sty618},
archivePrefix = {arXiv},
       eprint = {1707.03401},
 primaryClass = {astro-ph.GA},
       adsurl = {https://ui.adsabs.harvard.edu/abs/2018MNRAS.477.1206N}
}

@ARTICLE{2001Springel_AREPO_GADGET,
       author = {{Springel}, Volker and {Yoshida}, Naoki and {White}, Simon D.~M.},
        title = "{GADGET: a code for collisionless and gasdynamical cosmological simulations}",
      journal = {\na},
         year = 2001,
        month = apr,
       volume = {6},
       number = {2},
        pages = {79-117},
          doi = {10.1016/S1384-1076(01)00042-2},
archivePrefix = {arXiv},
       eprint = {astro-ph/0003162},
 primaryClass = {astro-ph},
       adsurl = {https://ui.adsabs.harvard.edu/abs/2001NewA....6...79S}
}

@ARTICLE{2016PlanckColab,
       author = {{Planck Collaboration} and {Ade}, P.~A.~R. and {Aghanim}, N. and {Arnaud}, M. and {Ashdown}, M. and {Aumont}, J. and {Baccigalupi}, C. and {Banday}, A.~J. and {Barreiro}, R.~B. and {Bartlett}, J.~G. and {Bartolo}, N. and {Battaner}, E. and {Battye}, R. and {Benabed}, K. and {Beno{\^\i}t}, A. and {Benoit-L{\'e}vy}, A. and {Bernard}, J. -P. and {Bersanelli}, M. and {Bielewicz}, P. and {Bock}, J.~J. and {Bonaldi}, A. and {Bonavera}, L. and {Bond}, J.~R. and {Borrill}, J. and {Bouchet}, F.~R. and {Boulanger}, F. and {Bucher}, M. and {Burigana}, C. and {Butler}, R.~C. and {Calabrese}, E. and {Cardoso}, J. -F. and {Catalano}, A. and {Challinor}, A. and {Chamballu}, A. and {Chary}, R. -R. and {Chiang}, H.~C. and {Chluba}, J. and {Christensen}, P.~R. and {Church}, S. and {Clements}, D.~L. and {Colombi}, S. and {Colombo}, L.~P.~L. and {Combet}, C. and {Coulais}, A. and {Crill}, B.~P. and {Curto}, A. and {Cuttaia}, F. and {Danese}, L. and {Davies}, R.~D. and {Davis}, R.~J. and {de Bernardis}, P. and {de Rosa}, A. and {de Zotti}, G. and {Delabrouille}, J. and {D{\'e}sert}, F. -X. and {Di Valentino}, E. and {Dickinson}, C. and {Diego}, J.~M. and {Dolag}, K. and {Dole}, H. and {Donzelli}, S. and {Dor{\'e}}, O. and {Douspis}, M. and {Ducout}, A. and {Dunkley}, J. and {Dupac}, X. and {Efstathiou}, G. and {Elsner}, F. and {En{\ss}lin}, T.~A. and {Eriksen}, H.~K. and {Farhang}, M. and {Fergusson}, J. and {Finelli}, F. and {Forni}, O. and {Frailis}, M. and {Fraisse}, A.~A. and {Franceschi}, E. and {Frejsel}, A. and {Galeotta}, S. and {Galli}, S. and {Ganga}, K. and {Gauthier}, C. and {Gerbino}, M. and {Ghosh}, T. and {Giard}, M. and {Giraud-H{\'e}raud}, Y. and {Giusarma}, E. and {Gjerl{\o}w}, E. and {Gonz{\'a}lez-Nuevo}, J. and {G{\'o}rski}, K.~M. and {Gratton}, S. and {Gregorio}, A. and {Gruppuso}, A. and {Gudmundsson}, J.~E. and {Hamann}, J. and {Hansen}, F.~K. and {Hanson}, D. and {Harrison}, D.~L. and {Helou}, G. and {Henrot-Versill{\'e}}, S. and {Hern{\'a}ndez-Monteagudo}, C. and {Herranz}, D. and {Hildebrandt}, S.~R. and {Hivon}, E. and {Hobson}, M. and {Holmes}, W.~A. and {Hornstrup}, A. and {Hovest}, W. and {Huang}, Z. and {Huffenberger}, K.~M. and {Hurier}, G. and {Jaffe}, A.~H. and {Jaffe}, T.~R. and {Jones}, W.~C. and {Juvela}, M. and {Keih{\"a}nen}, E. and {Keskitalo}, R. and {Kisner}, T.~S. and {Kneissl}, R. and {Knoche}, J. and {Knox}, L. and {Kunz}, M. and {Kurki-Suonio}, H. and {Lagache}, G. and {L{\"a}hteenm{\"a}ki}, A. and {Lamarre}, J. -M. and {Lasenby}, A. and {Lattanzi}, M. and {Lawrence}, C.~R. and {Leahy}, J.~P. and {Leonardi}, R. and {Lesgourgues}, J. and {Levrier}, F. and {Lewis}, A. and {Liguori}, M. and {Lilje}, P.~B. and {Linden-V{\o}rnle}, M. and {L{\'o}pez-Caniego}, M. and {Lubin}, P.~M. and {Mac{\'\i}as-P{\'e}rez}, J.~F. and {Maggio}, G. and {Maino}, D. and {Mandolesi}, N. and {Mangilli}, A. and {Marchini}, A. and {Maris}, M. and {Martin}, P.~G. and {Martinelli}, M. and {Mart{\'\i}nez-Gonz{\'a}lez}, E. and {Masi}, S. and {Matarrese}, S. and {McGehee}, P. and {Meinhold}, P.~R. and {Melchiorri}, A. and {Melin}, J. -B. and {Mendes}, L. and {Mennella}, A. and {Migliaccio}, M. and {Millea}, M. and {Mitra}, S. and {Miville-Desch{\^e}nes}, M. -A. and {Moneti}, A. and {Montier}, L. and {Morgante}, G. and {Mortlock}, D. and {Moss}, A. and {Munshi}, D. and {Murphy}, J.~A. and {Naselsky}, P. and {Nati}, F. and {Natoli}, P. and {Netterfield}, C.~B. and {N{\o}rgaard-Nielsen}, H.~U. and {Noviello}, F. and {Novikov}, D. and {Novikov}, I. and {Oxborrow}, C.~A. and {Paci}, F. and {Pagano}, L. and {Pajot}, F. and {Paladini}, R. and {Paoletti}, D. and {Partridge}, B. and {Pasian}, F. and {Patanchon}, G. and {Pearson}, T.~J. and {Perdereau}, O. and {Perotto}, L. and {Perrotta}, F. and {Pettorino}, V. and {Piacentini}, F. and {Piat}, M. and {Pierpaoli}, E. and {Pietrobon}, D. and {Plaszczynski}, S. and {Pointecouteau}, E. and {Polenta}, G. and {Popa}, L. and {Pratt}, G.~W. and {Pr{\'e}zeau}, G.},
        title = "{Planck 2015 results. XIII. Cosmological parameters}",
      journal = {\aap},
         year = 2016,
        month = sep,
       volume = {594},
          eid = {A13},
        pages = {A13},
          doi = {10.1051/0004-6361/201525830},
archivePrefix = {arXiv},
       eprint = {1502.01589},
 primaryClass = {astro-ph.CO},
       adsurl = {https://ui.adsabs.harvard.edu/abs/2016A&A...594A..13P}
}

@ARTICLE{1985DavisFoF,
       author = {{Davis}, M. and {Efstathiou}, G. and {Frenk}, C.~S. and {White}, S.~D.~M.},
        title = "{The evolution of large-scale structure in a universe dominated by cold dark matter}",
      journal = {\apj},
         year = 1985,
        month = may,
       volume = {292},
        pages = {371-394},
          doi = {10.1086/163168},
       adsurl = {https://ui.adsabs.harvard.edu/abs/1985ApJ...292..371D}
}

@ARTICLE{2001SpringelSUBFIND,
       author = {{Springel}, Volker and {White}, Simon D.~M. and {Tormen}, Giuseppe and {Kauffmann}, Guinevere},
        title = "{Populating a cluster of galaxies - I. Results at z=0}",
      journal = {\mnras},
         year = 2001,
        month = dec,
       volume = {328},
       number = {3},
        pages = {726-750},
          doi = {10.1046/j.1365-8711.2001.04912.x},
archivePrefix = {arXiv},
       eprint = {astro-ph/0012055},
 primaryClass = {astro-ph},
       adsurl = {https://ui.adsabs.harvard.edu/abs/2001MNRAS.328..726S}
}

@ARTICLE{2009DolagSUBFIND,
       author = {{Dolag}, K. and {Borgani}, S. and {Murante}, G. and {Springel}, V.},
        title = "{Substructures in hydrodynamical cluster simulations}",
      journal = {\mnras},
         year = 2009,
        month = oct,
       volume = {399},
       number = {2},
        pages = {497-514},
          doi = {10.1111/j.1365-2966.2009.15034.x},
archivePrefix = {arXiv},
       eprint = {0808.3401},
 primaryClass = {astro-ph},
       adsurl = {https://ui.adsabs.harvard.edu/abs/2009MNRAS.399..497D}
}

@ARTICLE{2019TNG50_Pillepich,
       author = {{Pillepich}, Annalisa and {Nelson}, Dylan and {Springel}, Volker and {Pakmor}, R{\"u}diger and {Torrey}, Paul and {Weinberger}, Rainer and {Vogelsberger}, Mark and {Marinacci}, Federico and {Genel}, Shy and {van der Wel}, Arjen and {Hernquist}, Lars},
        title = "{First results from the TNG50 simulation: the evolution of stellar and gaseous discs across cosmic time}",
      journal = {\mnras},
         year = 2019,
        month = dec,
       volume = {490},
       number = {3},
        pages = {3196-3233},
          doi = {10.1093/mnras/stz2338},
archivePrefix = {arXiv},
       eprint = {1902.05553},
 primaryClass = {astro-ph.GA},
       adsurl = {https://ui.adsabs.harvard.edu/abs/2019MNRAS.490.3196P}
}

@ARTICLE{2019TNG50_Nelson_b,
       author = {{Nelson}, Dylan and {Pillepich}, Annalisa and {Springel}, Volker and {Pakmor}, R{\"u}diger and {Weinberger}, Rainer and {Genel}, Shy and {Torrey}, Paul and {Vogelsberger}, Mark and {Marinacci}, Federico and {Hernquist}, Lars},
        title = "{First results from the TNG50 simulation: galactic outflows driven by supernovae and black hole feedback}",
      journal = {\mnras},
         year = 2019,
        month = dec,
       volume = {490},
       number = {3},
        pages = {3234-3261},
          doi = {10.1093/mnras/stz2306},
archivePrefix = {arXiv},
       eprint = {1902.05554},
 primaryClass = {astro-ph.GA},
       adsurl = {https://ui.adsabs.harvard.edu/abs/2019MNRAS.490.3234N}
}

@ARTICLE{2017Leisman_ALFALFA,
       author = {{Leisman}, Lukas and {Haynes}, Martha P. and {Janowiecki}, Steven and {Hallenbeck}, Gregory and {J{\'o}zsa}, Gyula and {Giovanelli}, Riccardo and {Adams}, Elizabeth A.~K. and {Bernal Neira}, David and {Cannon}, John M. and {Janesh}, William F. and {Rhode}, Katherine L. and {Salzer}, John J.},
        title = "{(Almost) Dark Galaxies in the ALFALFA Survey: Isolated H I-bearing Ultra-diffuse Galaxies}",
      journal = {\apj},
         year = 2017,
        month = jun,
       volume = {842},
       number = {2},
          eid = {133},
        pages = {133},
          doi = {10.3847/1538-4357/aa7575},
archivePrefix = {arXiv},
       eprint = {1703.05293},
 primaryClass = {astro-ph.GA},
       adsurl = {https://ui.adsabs.harvard.edu/abs/2017ApJ...842..133L}
}

@ARTICLE{2017RomanTrujillo_UDG,
       author = {{Rom{\'a}n}, Javier and {Trujillo}, Ignacio},
        title = "{Ultra-diffuse galaxies outside clusters: clues to their formation and evolution}",
      journal = {\mnras},
         year = 2017,
        month = jul,
       volume = {468},
       number = {4},
        pages = {4039-4047},
          doi = {10.1093/mnras/stx694},
archivePrefix = {arXiv},
       eprint = {1610.08980},
 primaryClass = {astro-ph.GA},
       adsurl = {https://ui.adsabs.harvard.edu/abs/2017MNRAS.468.4039R}
}

@ARTICLE{2019Janowiecki_UDG,
       author = {{Janowiecki}, Steven and {Jones}, Michael G. and {Leisman}, Lukas and {Webb}, Andrew},
        title = "{The environment of H I-bearing ultra-diffuse galaxies in the ALFALFA survey}",
      journal = {\mnras},
         year = 2019,
        month = nov,
       volume = {490},
       number = {1},
        pages = {566-577},
          doi = {10.1093/mnras/stz1868},
archivePrefix = {arXiv},
       eprint = {1906.11543},
 primaryClass = {astro-ph.GA},
       adsurl = {https://ui.adsabs.harvard.edu/abs/2019MNRAS.490..566J}
}

@ARTICLE{2021Marleau_UDG,
       author = {{Marleau}, Francine R. and {Habas}, Rebecca and {Poulain}, M{\'e}lina and {Duc}, Pierre-Alain and {M{\"u}ller}, Oliver and {Lim}, Sungsoon and {Durrell}, Patrick R. and {S{\'a}nchez-Janssen}, Rub{\'e}n and {Paudel}, Sanjaya and {Ahad}, Syeda Lammim and {Chougule}, Abhishek and {B{\'\i}lek}, Michal and {Fensch}, J{\'e}r{\'e}my},
        title = "{Ultra diffuse galaxies in the MATLAS low-to-moderate density fields}",
      journal = {\aap},
         year = 2021,
        month = oct,
       volume = {654},
          eid = {A105},
        pages = {A105},
          doi = {10.1051/0004-6361/202141432},
archivePrefix = {arXiv},
       eprint = {2109.13173},
 primaryClass = {astro-ph.GA},
       adsurl = {https://ui.adsabs.harvard.edu/abs/2021A&A...654A.105M}
}

@ARTICLE{2021Kadowaki_UDG,
       author = {{Kadowaki}, Jennifer and {Zaritsky}, Dennis and {Donnerstein}, R.~L. and {RS}, Pranjal and {Karunakaran}, Ananthan and {Spekkens}, Kristine},
        title = "{On the Properties of Spectroscopically Confirmed Ultra-diffuse Galaxies across Environments}",
      journal = {\apj},
         year = 2021,
        month = dec,
       volume = {923},
       number = {2},
          eid = {257},
        pages = {257},
          doi = {10.3847/1538-4357/ac2948},
archivePrefix = {arXiv},
       eprint = {2110.00015},
 primaryClass = {astro-ph.GA},
       adsurl = {https://ui.adsabs.harvard.edu/abs/2021ApJ...923..257K}
}

@ARTICLE{2017Shi_UDG,
       author = {{Shi}, Dong Dong and {Zheng}, Xian Zhong and {Zhao}, Hai Bin and {Pan}, Zhi Zheng and {Li}, Bin and {Zou}, Hu and {Zhou}, Xu and {Guo}, KeXin and {An}, Fang Xia and {Li}, Yu Bin},
        title = "{Deep Imaging of the HCG 95 Field. I. Ultra-diffuse Galaxies}",
      journal = {\apj},
         year = 2017,
        month = sep,
       volume = {846},
       number = {1},
          eid = {26},
        pages = {26},
          doi = {10.3847/1538-4357/aa8327},
archivePrefix = {arXiv},
       eprint = {1708.00013},
 primaryClass = {astro-ph.GA},
       adsurl = {https://ui.adsabs.harvard.edu/abs/2017ApJ...846...26S}
}

@ARTICLE{2023Jones_UDG,
       author = {{Jones}, Michael G. and {Karunakaran}, Ananthan and {Bennet}, Paul and {Sand}, David J. and {Spekkens}, Kristine and {Mutlu-Pakdil}, Bur{\c{c}}in and {Crnojevi{\'c}}, Denija and {Janowiecki}, Steven and {Leisman}, Lukas and {Fielder}, Catherine E.},
        title = "{Gas-rich, Field Ultra-diffuse Galaxies Host Few Globular Clusters}",
      journal = {\apjl},
         year = 2023,
        month = jan,
       volume = {942},
       number = {1},
          eid = {L5},
        pages = {L5},
          doi = {10.3847/2041-8213/acaaab},
archivePrefix = {arXiv},
       eprint = {2211.00651},
 primaryClass = {astro-ph.GA},
       adsurl = {https://ui.adsabs.harvard.edu/abs/2023ApJ...942L...5J}
}

@ARTICLE{2024BeasTNGSKIRT_wavelengthRe,
       author = {{Baes}, Maarten and {Mosenkov}, Aleksandr and {Kelly}, Raymond and {Abdurro'uf} and {Andreadis}, Nick and {Bokona Tulu}, Sena and {Camps}, Peter and {Tassama Emana}, Abdissa and {Fritz}, Jacopo and {Gebek}, Andrea and {Kova{\v{c}}i{\'c}}, Inja and {La Marca}, Antonio and {Martorano}, Marco and {Nersesian}, Angelos and {Rodriguez-Gomez}, Vicente and {Tortora}, Crescenzo and {Tr{\v{c}}ka}, Ana and {Vander Meulen}, Bert and {van der Wel}, Arjen and {Wang}, Lingyu},
        title = "{The TNG50-SKIRT Atlas: Wavelength dependence of the effective radius}",
      journal = {\aap},
         year = 2024,
        month = mar,
       volume = {683},
          eid = {A182},
        pages = {A182},
          doi = {10.1051/0004-6361/202348419},
archivePrefix = {arXiv},
       eprint = {2401.04225},
 primaryClass = {astro-ph.GA},
       adsurl = {https://ui.adsabs.harvard.edu/abs/2024A&A...683A.182B}
}

@ARTICLE{2020Guo_UDG,
       author = {{Guo}, Qi and {Hu}, Huijie and {Zheng}, Zheng and {Liao}, Shihong and {Du}, Wei and {Mao}, Shude and {Jiang}, Linhua and {Wang}, Jing and {Peng}, Yingjie and {Gao}, Liang and {Wang}, Jie and {Wu}, Hong},
        title = "{Further evidence for a population of dark-matter-deficient dwarf galaxies}",
      journal = {Nature Astronomy},
         year = 2020,
        month = jan,
       volume = {4},
        pages = {246-251},
          doi = {10.1038/s41550-019-0930-9},
archivePrefix = {arXiv},
       eprint = {1908.00046},
 primaryClass = {astro-ph.GA},
       adsurl = {https://ui.adsabs.harvard.edu/abs/2020NatAs...4..246G}
}

@ARTICLE{2022Kong_UDG,
       author = {{Kong}, Demao and {Kaplinghat}, Manoj and {Yu}, Hai-Bo and {Fraternali}, Filippo and {Mancera Pi{\~n}a}, Pavel E.},
        title = "{The Odd Dark Matter Halos of Isolated Gas-rich Ultradiffuse Galaxies}",
      journal = {\apj},
         year = 2022,
        month = sep,
       volume = {936},
       number = {2},
          eid = {166},
        pages = {166},
          doi = {10.3847/1538-4357/ac8875},
archivePrefix = {arXiv},
       eprint = {2204.05981},
 primaryClass = {astro-ph.GA},
       adsurl = {https://ui.adsabs.harvard.edu/abs/2022ApJ...936..166K}
}

@ARTICLE{2025Joy_LocalOverdensity,
       author = {{Bhattacharyya}, Joy and {Peter}, Annika H.~G. and {Leauthaud}, Alexie},
        title = "{Dwarf Galaxies in the TNG50 Field: connecting their Star-formation Rates with their Environments}",
      journal = {arXiv e-prints},
         year = 2025,
        month = jan,
          eid = {arXiv:2501.01946},
        pages = {arXiv:2501.01946},
          doi = {10.48550/arXiv.2501.01946},
archivePrefix = {arXiv},
       eprint = {2501.01946},
 primaryClass = {astro-ph.GA},
       adsurl = {https://ui.adsabs.harvard.edu/abs/2025arXiv250101946B}
}

@ARTICLE{2013AKarachentsev_TidalIndex,
       author = {{Karachentsev}, Igor D. and {Makarov}, Dmitry I. and {Kaisina}, Elena I.},
        title = "{Updated Nearby Galaxy Catalog}",
      journal = {\aj},
         year = 2013,
        month = apr,
       volume = {145},
       number = {4},
          eid = {101},
        pages = {101},
          doi = {10.1088/0004-6256/145/4/101},
archivePrefix = {arXiv},
       eprint = {1303.5328},
 primaryClass = {astro-ph.CO},
       adsurl = {https://ui.adsabs.harvard.edu/abs/2013AJ....145..101K}
}

@ARTICLE{2004Karachentsev_TidalIndex,
       author = {{Karachentsev}, Igor D. and {Karachentseva}, Valentina E. and {Huchtmeier}, Walter K. and {Makarov}, Dmitry I.},
        title = "{A Catalog of Neighboring Galaxies}",
      journal = {\aj},
         year = 2004,
        month = apr,
       volume = {127},
       number = {4},
        pages = {2031-2068},
          doi = {10.1086/382905},
       adsurl = {https://ui.adsabs.harvard.edu/abs/2004AJ....127.2031K}
}

@ARTICLE{2018Karachentsev_TidalIndex,
       author = {{Karachentsev}, I.~D. and {Kaisina}, E.~I. and {Makarov}, D.~I.},
        title = "{Morphological properties of galaxies in different Local Volume environments}",
      journal = {\mnras},
         year = 2018,
        month = sep,
       volume = {479},
       number = {3},
        pages = {4136-4152},
          doi = {10.1093/mnras/sty1774},
archivePrefix = {arXiv},
       eprint = {1806.09822},
 primaryClass = {astro-ph.GA},
       adsurl = {https://ui.adsabs.harvard.edu/abs/2018MNRAS.479.4136K}
}

@ARTICLE{2018Besla_TidalIndex,
       author = {{Besla}, Gurtina and {Patton}, David R. and {Stierwalt}, Sabrina and {Rodriguez-Gomez}, Vicente and {Patel}, Ekta and {Kallivayalil}, Nitya J. and {Johnson}, Kelsey E. and {Pearson}, Sarah and {Privon}, George C. and {Putman}, Mary E.},
        title = "{The frequency of dwarf galaxy multiples at low redshift in SDSS versus cosmological expectations}",
      journal = {\mnras},
         year = 2018,
        month = nov,
       volume = {480},
       number = {3},
        pages = {3376-3396},
          doi = {10.1093/mnras/sty2041},
archivePrefix = {arXiv},
       eprint = {1807.06673},
 primaryClass = {astro-ph.GA},
       adsurl = {https://ui.adsabs.harvard.edu/abs/2018MNRAS.480.3376B}
}

@ARTICLE{2024Mutlu-Pakdil_TidalIndex,
       author = {{Mutlu-Pakdil}, Bur{\c{c}}in and {Sand}, David J. and {Crnojevi{\'c}}, Denija and {Bennet}, Paul and {Jones}, Michael G. and {Spekkens}, Kristine and {Karunakaran}, Ananthan and {Zaritsky}, Dennis and {Caldwell}, Nelson and {Fielder}, Catherine E. and {Guhathakurta}, Puragra and {Seth}, Anil C. and {Simon}, Joshua D. and {Strader}, Jay and {Toloba}, Elisa},
        title = "{The Faint Satellite System of NGC 253: Insights into Low-density Environments and No Satellite Plane}",
      journal = {\apj},
         year = 2024,
        month = may,
       volume = {966},
       number = {2},
          eid = {188},
        pages = {188},
          doi = {10.3847/1538-4357/ad36c4},
archivePrefix = {arXiv},
       eprint = {2401.14457},
 primaryClass = {astro-ph.GA},
       adsurl = {https://ui.adsabs.harvard.edu/abs/2024ApJ...966..188M}
}

@INPROCEEDINGS{1999Karachentsev_TidalIndex,
       author = {{Karachentsev}, I.~D. and {Makarov}, D.~I.},
        title = "{Galaxy Interactions in the Local Volume}",
    booktitle = {Galaxy Interactions at Low and High Redshift},
         year = 1999,
       editor = {{Barnes}, J.~E. and {Sanders}, D.~B.},
       series = {IAU Symposium},
       volume = {186},
        month = jan,
        pages = {109},
       adsurl = {https://ui.adsabs.harvard.edu/abs/1999IAUS..186..109K}
}

@ARTICLE{2006Allgood_ShapeTensor,
       author = {{Allgood}, Brandon and {Flores}, Ricardo A. and {Primack}, Joel R. and {Kravtsov}, Andrey V. and {Wechsler}, Risa H. and {Faltenbacher}, Andreas and {Bullock}, James S.},
        title = "{The shape of dark matter haloes: dependence on mass, redshift, radius and formation}",
      journal = {\mnras},
         year = 2006,
        month = apr,
       volume = {367},
       number = {4},
        pages = {1781-1796},
          doi = {10.1111/j.1365-2966.2006.10094.x},
archivePrefix = {arXiv},
       eprint = {astro-ph/0508497},
 primaryClass = {astro-ph},
       adsurl = {https://ui.adsabs.harvard.edu/abs/2006MNRAS.367.1781A}
}

@ARTICLE{2016Tomassetti_ShapeTensor,
       author = {{Tomassetti}, Matteo and {Dekel}, Avishai and {Mandelker}, Nir and {Ceverino}, Daniel and {Lapiner}, Sharon and {Faber}, Sandra and {Kneller}, Omer and {Primack}, Joel and {Sai}, Tanmayi},
        title = "{Evolution of galaxy shapes from prolate to oblate through compaction events}",
      journal = {\mnras},
         year = 2016,
        month = jun,
       volume = {458},
       number = {4},
        pages = {4477-4497},
          doi = {10.1093/mnras/stw606},
archivePrefix = {arXiv},
       eprint = {1512.06268},
 primaryClass = {astro-ph.GA},
       adsurl = {https://ui.adsabs.harvard.edu/abs/2016MNRAS.458.4477T}
}

@ARTICLE{2014VeraCiro_ShapeTensor,
       author = {{Vera-Ciro}, Carlos A. and {Sales}, Laura V. and {Helmi}, Amina and {Navarro}, Julio F.},
        title = "{The shape of dark matter subhaloes in the Aquarius simulations}",
      journal = {\mnras},
         year = 2014,
        month = apr,
       volume = {439},
       number = {3},
        pages = {2863-2872},
          doi = {10.1093/mnras/stu153},
archivePrefix = {arXiv},
       eprint = {1402.0903},
 primaryClass = {astro-ph.CO},
       adsurl = {https://ui.adsabs.harvard.edu/abs/2014MNRAS.439.2863V}
}

@ARTICLE{2017Weinberger_TNGFeedbackProcess,
       author = {{Weinberger}, Rainer and {Springel}, Volker and {Hernquist}, Lars and {Pillepich}, Annalisa and {Marinacci}, Federico and {Pakmor}, R{\"u}diger and {Nelson}, Dylan and {Genel}, Shy and {Vogelsberger}, Mark and {Naiman}, Jill and {Torrey}, Paul},
        title = "{Simulating galaxy formation with black hole driven thermal and kinetic feedback}",
      journal = {\mnras},
         year = 2017,
        month = mar,
       volume = {465},
       number = {3},
        pages = {3291-3308},
          doi = {10.1093/mnras/stw2944},
archivePrefix = {arXiv},
       eprint = {1607.03486},
 primaryClass = {astro-ph.GA},
       adsurl = {https://ui.adsabs.harvard.edu/abs/2017MNRAS.465.3291W}
}

@ARTICLE{2020PinaHI_UDG,
       author = {{Mancera Pi{\~n}a}, Pavel E. and {Fraternali}, Filippo and {Oman}, Kyle A. and {Adams}, Elizabeth A.~K. and {Bacchini}, Cecilia and {Marasco}, Antonino and {Oosterloo}, Tom and {Pezzulli}, Gabriele and {Posti}, Lorenzo and {Leisman}, Lukas and {Cannon}, John M. and {di Teodoro}, Enrico M. and {Gault}, Lexi and {Haynes}, Martha P. and {Reiter}, Kameron and {Rhode}, Katherine L. and {Salzer}, John J. and {Smith}, Nicholas J.},
        title = "{Robust H I kinematics of gas-rich ultra-diffuse galaxies: hints of a weak-feedback formation scenario}",
      journal = {\mnras},
         year = 2020,
        month = jul,
       volume = {495},
       number = {4},
        pages = {3636-3655},
          doi = {10.1093/mnras/staa1256},
archivePrefix = {arXiv},
       eprint = {2004.14392},
 primaryClass = {astro-ph.GA},
       adsurl = {https://ui.adsabs.harvard.edu/abs/2020MNRAS.495.3636M}
}

@ARTICLE{2020KarunakaranHI_UDG,
       author = {{Karunakaran}, Ananthan and {Spekkens}, Kristine and {Zaritsky}, Dennis and {Donnerstein}, Richard L. and {Kadowaki}, Jennifer and {Dey}, Arjun},
        title = "{Systematically Measuring Ultradiffuse Galaxies in H I: Results from the Pilot Survey}",
      journal = {\apj},
         year = 2020,
        month = oct,
       volume = {902},
       number = {1},
          eid = {39},
        pages = {39},
          doi = {10.3847/1538-4357/abb464},
archivePrefix = {arXiv},
       eprint = {2005.14202},
 primaryClass = {astro-ph.GA},
       adsurl = {https://ui.adsabs.harvard.edu/abs/2020ApJ...902...39K}
}

@ARTICLE{2005Pizzella_HSB_Selection,
       author = {{Pizzella}, A. and {Corsini}, E.~M. and {Dalla Bont{\`a}}, E. and {Sarzi}, M. and {Coccato}, L. and {Bertola}, F.},
        title = "{On the Relation between Circular Velocity and Central Velocity Dispersion in High and Low Surface Brightness Galaxies}",
      journal = {\apj},
         year = 2005,
        month = oct,
       volume = {631},
       number = {2},
        pages = {785-791},
          doi = {10.1086/430513},
archivePrefix = {arXiv},
       eprint = {astro-ph/0503649},
 primaryClass = {astro-ph},
       adsurl = {https://ui.adsabs.harvard.edu/abs/2005ApJ...631..785P}
}

@ARTICLE{2005Cooray_HSB_Selection,
       author = {{Cooray}, Asantha and {Milosavljevi{\'c}}, Milo{\v{s}}},
        title = "{What is L$_{*}$? Anatomy of the Galaxy Luminosity Function}",
      journal = {\apjl},
         year = 2005,
        month = jul,
       volume = {627},
       number = {2},
        pages = {L89-L92},
          doi = {10.1086/432259},
archivePrefix = {arXiv},
       eprint = {astro-ph/0504580},
 primaryClass = {astro-ph},
       adsurl = {https://ui.adsabs.harvard.edu/abs/2005ApJ...627L..89C}
}

@ARTICLE{2013Robotham_HSB_Selection,
       author = {{Robotham}, A.~S.~G. and {Liske}, J. and {Driver}, S.~P. and {Sansom}, A.~E. and {Baldry}, I.~K. and {Bauer}, A.~E. and {Bland-Hawthorn}, J. and {Brough}, S. and {Brown}, M.~J.~I. and {Colless}, M. and {Christodoulou}, L. and {Drinkwater}, M.~J. and {Grootes}, M.~W. and {Hopkins}, A.~M. and {Kelvin}, L.~S. and {Norberg}, P. and {Loveday}, J. and {Phillipps}, S. and {Sharp}, R. and {Taylor}, E.~N. and {Tuffs}, R.~J.},
        title = "{Galaxy And Mass Assembly (GAMA): the life and times of L★ galaxies}",
      journal = {\mnras},
         year = 2013,
        month = may,
       volume = {431},
       number = {1},
        pages = {167-193},
          doi = {10.1093/mnras/stt156},
archivePrefix = {arXiv},
       eprint = {1301.7129},
 primaryClass = {astro-ph.CO},
       adsurl = {https://ui.adsabs.harvard.edu/abs/2013MNRAS.431..167R}
}

@ARTICLE{2004Kniazev_LSB_Selection,
       author = {{Kniazev}, Alexei Y. and {Grebel}, Eva K. and {Pustilnik}, Simon A. and {Pramskij}, Alexander G. and {Kniazeva}, Tamara F. and {Prada}, Francisco and {Harbeck}, Daniel},
        title = "{Low Surface Brightness Galaxies in the Sloan Digital Sky Survey. I. Search Method and Test Sample}",
      journal = {\aj},
         year = 2004,
        month = feb,
       volume = {127},
       number = {2},
        pages = {704-727},
          doi = {10.1086/381061},
archivePrefix = {arXiv},
       eprint = {astro-ph/0310644},
 primaryClass = {astro-ph},
       adsurl = {https://ui.adsabs.harvard.edu/abs/2004AJ....127..704K}
}

@ARTICLE{1996Impey_LSB_Selection,
       author = {{Impey}, C.~D. and {Sprayberry}, D. and {Irwin}, M.~J. and {Bothun}, G.~D.},
        title = "{Low Surface Brightness Galaxies in the Local Universe. I. The Catalog}",
      journal = {\apjs},
         year = 1996,
        month = aug,
       volume = {105},
        pages = {209},
          doi = {10.1086/192313},
       adsurl = {https://ui.adsabs.harvard.edu/abs/1996ApJS..105..209I}
}

@ARTICLE{2001Mathews_LSB_Selection,
       author = {{Matthews}, L.~D. and {van Driel}, W. and {Monnier-Ragaigne}, D.},
        title = "{H I observations of giant low surface brightness galaxies}",
      journal = {\aap},
         year = 2001,
        month = jan,
       volume = {365},
        pages = {1-10},
          doi = {10.1051/0004-6361:20000002},
archivePrefix = {arXiv},
       eprint = {astro-ph/0010075},
 primaryClass = {astro-ph},
       adsurl = {https://ui.adsabs.harvard.edu/abs/2001A&A...365....1M}
}

@ARTICLE{2022Poulain_UDG+Dwarf_Selection,
       author = {{Poulain}, M{\'e}lina and {Marleau}, Francine R. and {Habas}, Rebecca and {Duc}, Pierre-Alain and {S{\'a}nchez-Janssen}, Rub{\'e}n and {Durrell}, Patrick R. and {Paudel}, Sanjaya and {M{\"u}ller}, Oliver and {Lim}, Sungsoon and {B{\'\i}lek}, Michal and {Fensch}, J{\'e}r{\'e}my},
        title = "{HI observations of the MATLAS dwarf and ultra-diffuse galaxies}",
      journal = {\aap},
         year = 2022,
        month = mar,
       volume = {659},
          eid = {A14},
        pages = {A14},
          doi = {10.1051/0004-6361/202142012},
archivePrefix = {arXiv},
       eprint = {2111.14491},
 primaryClass = {astro-ph.GA},
       adsurl = {https://ui.adsabs.harvard.edu/abs/2022A&A...659A..14P}
}

@ARTICLE{2018Revaz_Dwarf_Selection,
       author = {{Revaz}, Yves and {Jablonka}, Pascale},
        title = "{Pushing back the limits: detailed properties of dwarf galaxies in a {\ensuremath{\Lambda}}CDM universe}",
      journal = {\aap},
         year = 2018,
        month = aug,
       volume = {616},
          eid = {A96},
        pages = {A96},
          doi = {10.1051/0004-6361/201832669},
archivePrefix = {arXiv},
       eprint = {1801.06222},
 primaryClass = {astro-ph.GA},
       adsurl = {https://ui.adsabs.harvard.edu/abs/2018A&A...616A..96R}
}

@ARTICLE{2019Chua_ShapeTensor,
       author = {{Chua}, Kun Ting Eddie and {Pillepich}, Annalisa and {Vogelsberger}, Mark and {Hernquist}, Lars},
        title = "{Shape of dark matter haloes in the Illustris simulation: effects of baryons}",
      journal = {\mnras},
         year = 2019,
        month = mar,
       volume = {484},
       number = {1},
        pages = {476-493},
          doi = {10.1093/mnras/sty3531},
archivePrefix = {arXiv},
       eprint = {1809.07255},
 primaryClass = {astro-ph.GA},
       adsurl = {https://ui.adsabs.harvard.edu/abs/2019MNRAS.484..476C}
}

@ARTICLE{2021Cataldi_ShapeTensor,
       author = {{Cataldi}, P. and {Pedrosa}, S.~E. and {Tissera}, P.~B. and {Artale}, M.~C.},
        title = "{Baryons shaping dark matter haloes}",
      journal = {\mnras},
         year = 2021,
        month = mar,
       volume = {501},
       number = {4},
        pages = {5679-5691},
          doi = {10.1093/mnras/staa3988},
archivePrefix = {arXiv},
       eprint = {2008.02404},
 primaryClass = {astro-ph.GA},
       adsurl = {https://ui.adsabs.harvard.edu/abs/2021MNRAS.501.5679C}
}

@ARTICLE{2021Emami_ShapeTensor_a,
       author = {{Emami}, Razieh and {Genel}, Shy and {Hernquist}, Lars and {Alcock}, Charles and {Bose}, Sownak and {Weinberger}, Rainer and {Vogelsberger}, Mark and {Marinacci}, Federico and {Loeb}, Abraham and {Torrey}, Paul and {Forbes}, John C.},
        title = "{Morphological Types of DM Halos in Milky Way-like Galaxies in the TNG50 Simulation: Simple, Twisted, or Stretched}",
      journal = {\apj},
         year = 2021,
        month = may,
       volume = {913},
       number = {1},
          eid = {36},
        pages = {36},
          doi = {10.3847/1538-4357/abf147},
archivePrefix = {arXiv},
       eprint = {2009.09220},
 primaryClass = {astro-ph.GA},
       adsurl = {https://ui.adsabs.harvard.edu/abs/2021ApJ...913...36E}
}

@BOOK{2008BinneyTremaine,
       author = {{Binney}, James and {Tremaine}, Scott},
        title = "{Galactic Dynamics: Second Edition}",
         year = 2008,
       adsurl = {https://ui.adsabs.harvard.edu/abs/2008gady.book.....B},
      publisher = {Princeton University Press}
}

@ARTICLE{2022Emami_VelocityDispersion,
       author = {{Emami}, Razieh and {Hernquist}, Lars and {Vogelsberger}, Mark and {Shen}, Xuejian and {Speagle}, Joshua S. and {Moreno}, Jorge and {Alcock}, Charles and {Genel}, Shy and {Forbes}, John C. and {Marinacci}, Federico and {Torrey}, Paul},
        title = "{On the Robustness of the Velocity Anisotropy Parameter in Probing the Stellar Kinematics in Milky Way-Like Galaxies: Takeaway from TNG50 Simulation}",
      journal = {\apj},
         year = 2022,
        month = sep,
       volume = {937},
       number = {1},
          eid = {20},
        pages = {20},
          doi = {10.3847/1538-4357/ac86c7},
archivePrefix = {arXiv},
       eprint = {2202.07162},
 primaryClass = {astro-ph.GA},
       adsurl = {https://ui.adsabs.harvard.edu/abs/2022ApJ...937...20E}
}

@ARTICLE{1980Binney_anisotropy,
       author = {{Binney}, J.},
        title = "{The radius-dependence of velocity dispersion in elliptical galaxies}",
      journal = {\mnras},
         year = 1980,
        month = mar,
       volume = {190},
        pages = {873-880},
          doi = {10.1093/mnras/190.4.873},
       adsurl = {https://ui.adsabs.harvard.edu/abs/1980MNRAS.190..873B}
}

@ARTICLE{2001Halliday_LOSVD,
       author = {{Halliday}, C. and {Davies}, Roger L. and {Kuntschner}, Harald and {Birkinshaw}, M. and {Bender}, Ralf and {Saglia}, R.~P. and {Baggley}, Glenn},
        title = "{Line-of-sight velocity distributions of low-luminosity elliptical galaxies}",
      journal = {\mnras},
         year = 2001,
        month = sep,
       volume = {326},
       number = {2},
        pages = {473-489},
          doi = {10.1046/j.1365-8711.2001.04492.x},
archivePrefix = {arXiv},
       eprint = {astro-ph/0103295},
 primaryClass = {astro-ph},
       adsurl = {https://ui.adsabs.harvard.edu/abs/2001MNRAS.326..473H}
}

@ARTICLE{2020SimSpin_v1,
       author = {{Harborne}, Katherine E. and {Power}, Chris and {Robotham}, Aaron S.~G.},
        title = "{SIMSPIN - Constructing mock IFS kinematic data cubes}",
      journal = {\pasa},
         year = 2020,
        month = may,
       volume = {37},
          eid = {e016},
        pages = {e016},
          doi = {10.1017/pasa.2020.8},
archivePrefix = {arXiv},
       eprint = {2003.07641},
 primaryClass = {astro-ph.GA},
       adsurl = {https://ui.adsabs.harvard.edu/abs/2020PASA...37...16H}
}

@ARTICLE{2023SimSpin_v2,
       author = {{Harborne}, K.~E. and {Serene}, A. and {Davies}, E.~J.~A. and {Derkenne}, C. and {Vaughan}, S. and {Burdon}, A.~I. and {Lagos}, C. del P. and {McDermid}, R. and {O'Toole}, S. and {Power}, C. and {Robotham}, A.~S.~G. and {Santucci}, G. and {Tobar}, R.},
        title = "{SimSpin v2.6.0{\textemdash}constructing synthetic spectral IFU cubes for comparison with observational surveys}",
      journal = {\pasa},
         year = 2023,
        month = oct,
       volume = {40},
          eid = {e048},
        pages = {e048},
          doi = {10.1017/pasa.2023.47},
archivePrefix = {arXiv},
       eprint = {2307.02618},
 primaryClass = {astro-ph.GA},
       adsurl = {https://ui.adsabs.harvard.edu/abs/2023PASA...40...48H}
}

@ARTICLE{2024SimSpin_VorbinLimit,
       author = {{Harborne}, K.~E. and {Lagos}, C. del P. and {Croom}, S.~M. and {van de Sande}, J. and {Ludlow}, A. and {Remus}, R.~S. and {Kimmig}, L.~C. and {Power}, C.},
        title = "{From particles to pixels: how many particles do I really need to construct stellar kinematic mock observational measurements?}",
      journal = {\mnras},
         year = 2024,
        month = dec,
       volume = {535},
       number = {3},
        pages = {2844-2862},
          doi = {10.1093/mnras/stae2526},
archivePrefix = {arXiv},
       eprint = {2411.03791},
 primaryClass = {astro-ph.GA},
       adsurl = {https://ui.adsabs.harvard.edu/abs/2024MNRAS.535.2844H}
}

@ARTICLE{1963Sersic,
       author = {{S{\'e}rsic}, J.~L.},
        title = "{Influence of the atmospheric and instrumental dispersion on the brightness distribution in a galaxy}",
      journal = {Boletin de la Asociacion Argentina de Astronomia La Plata Argentina},
         year = 1963,
        month = feb,
       volume = {6},
        pages = {41-43},
       adsurl = {https://ui.adsabs.harvard.edu/abs/1963BAAA....6...41S}
}

@BOOK{1968Sersic,
       author = {{Sersic}, Jose Luis},
        title = "{Atlas de Galaxias Australes}",
         year = 1968,
       adsurl = {https://ui.adsabs.harvard.edu/abs/1968adga.book.....S},
      publisher = {Observatorio Astronomico, Universidad Nacional de Cordoba}
}

@ARTICLE{2007Cappellari_LOSVD,
       author = {{Cappellari}, Michele and {Emsellem}, Eric and {Bacon}, R. and {Bureau}, M. and {Davies}, Roger L. and {de Zeeuw}, P.~T. and {Falc{\'o}n-Barroso}, Jes{\'u}s and {Krajnovi{\'c}}, Davor and {Kuntschner}, Harald and {McDermid}, Richard M. and {Peletier}, Reynier F. and {Sarzi}, Marc and {van den Bosch}, Remco C.~E. and {van de Ven}, Glenn},
        title = "{The SAURON project - X. The orbital anisotropy of elliptical and lenticular galaxies: revisiting the (V/{\ensuremath{\sigma}}, ɛ) diagram with integral-field stellar kinematics}",
      journal = {\mnras},
         year = 2007,
        month = aug,
       volume = {379},
       number = {2},
        pages = {418-444},
          doi = {10.1111/j.1365-2966.2007.11963.x},
archivePrefix = {arXiv},
       eprint = {astro-ph/0703533},
 primaryClass = {astro-ph},
       adsurl = {https://ui.adsabs.harvard.edu/abs/2007MNRAS.379..418C}
}

@ARTICLE{2007Emsellem_LOSVD,
       author = {{Emsellem}, Eric and {Cappellari}, Michele and {Krajnovi{\'c}}, Davor and {van de Ven}, Glenn and {Bacon}, R. and {Bureau}, M. and {Davies}, Roger L. and {de Zeeuw}, P.~T. and {Falc{\'o}n-Barroso}, Jes{\'u}s and {Kuntschner}, Harald and {McDermid}, Richard and {Peletier}, Reynier F. and {Sarzi}, Marc},
        title = "{The SAURON project - IX. A kinematic classification for early-type galaxies}",
      journal = {\mnras},
         year = 2007,
        month = aug,
       volume = {379},
       number = {2},
        pages = {401-417},
          doi = {10.1111/j.1365-2966.2007.11752.x},
archivePrefix = {arXiv},
       eprint = {astro-ph/0703531},
 primaryClass = {astro-ph},
       adsurl = {https://ui.adsabs.harvard.edu/abs/2007MNRAS.379..401E}
}

@ARTICLE{2018vandeSande_LOSVD_VbyS,
       author = {{van de Sande}, Jesse and {Scott}, Nicholas and {Bland-Hawthorn}, Joss and {Brough}, Sarah and {Bryant}, Julia J. and {Colless}, Matthew and {Cortese}, Luca and {Croom}, Scott M. and {d'Eugenio}, Francesco and {Foster}, Caroline and {Goodwin}, Michael and {Konstantopoulos}, Iraklis S. and {Lawrence}, Jon S. and {McDermid}, Richard M. and {Medling}, Anne M. and {Owers}, Matt S. and {Richards}, Samuel N. and {Sharp}, Rob},
        title = "{A relation between the characteristic stellar ages of galaxies and their intrinsic shapes}",
      journal = {Nature Astronomy},
         year = 2018,
        month = apr,
       volume = {2},
        pages = {483-488},
          doi = {10.1038/s41550-018-0436-x},
archivePrefix = {arXiv},
       eprint = {1804.07769},
 primaryClass = {astro-ph.GA},
       adsurl = {https://ui.adsabs.harvard.edu/abs/2018NatAs...2..483V}
}

@ARTICLE{2021Pinna_VbyS_SAMI,
       author = {{Pinna}, Francesca and {Neumayer}, Nadine and {Seth}, Anil and {Emsellem}, Eric and {Nguyen}, Dieu D. and {B{\"o}ker}, Torsten and {Cappellari}, Michele and {McDermid}, Richard M. and {Voggel}, Karina and {Walcher}, C. Jakob},
        title = "{Resolved Nuclear Kinematics Link the Formation and Growth of Nuclear Star Clusters with the Evolution of Their Early- and Late-type Hosts}",
      journal = {\apj},
         year = 2021,
        month = nov,
       volume = {921},
       number = {1},
          eid = {8},
        pages = {8},
          doi = {10.3847/1538-4357/ac158f},
archivePrefix = {arXiv},
       eprint = {2107.08903},
 primaryClass = {astro-ph.GA},
       adsurl = {https://ui.adsabs.harvard.edu/abs/2021ApJ...921....8P}
}

@ARTICLE{2021vandeSande_VbyS_SAMI,
       author = {{van de Sande}, Jesse and {Croom}, Scott M. and {Bland-Hawthorn}, Joss and {Cortese}, Luca and {Scott}, Nicholas and {Lagos}, Claudia D.~P. and {D'Eugenio}, Francesco and {Bryant}, Julia J. and {Brough}, Sarah and {Catinella}, Barbara and {Foster}, Caroline and {Groves}, Brent and {Harborne}, Katherine E. and {L{\'o}pez-S{\'a}nchez}, {\'A}ngel R. and {McDermid}, Richard and {Medling}, Anne and {Owers}, Matt S. and {Richards}, Samuel N. and {Sweet}, Sarah M. and {Vaughan}, Sam P.},
        title = "{The SAMI galaxy survey: Mass and environment as independent drivers of galaxy dynamics}",
      journal = {\mnras},
         year = 2021,
        month = dec,
       volume = {508},
       number = {2},
        pages = {2307-2328},
          doi = {10.1093/mnras/stab2647},
archivePrefix = {arXiv},
       eprint = {2109.06189},
 primaryClass = {astro-ph.GA},
       adsurl = {https://ui.adsabs.harvard.edu/abs/2021MNRAS.508.2307V}
}

@ARTICLE{2011Emsellem_vbys_ATLAS,
       author = {{Emsellem}, Eric and {Cappellari}, Michele and {Krajnovi{\'c}}, Davor and {Alatalo}, Katherine and {Blitz}, Leo and {Bois}, Maxime and {Bournaud}, Fr{\'e}d{\'e}ric and {Bureau}, Martin and {Davies}, Roger L. and {Davis}, Timothy A. and {de Zeeuw}, P.~T. and {Khochfar}, Sadegh and {Kuntschner}, Harald and {Lablanche}, Pierre-Yves and {McDermid}, Richard M. and {Morganti}, Raffaella and {Naab}, Thorsten and {Oosterloo}, Tom and {Sarzi}, Marc and {Scott}, Nicholas and {Serra}, Paolo and {van de Ven}, Glenn and {Weijmans}, Anne-Marie and {Young}, Lisa M.},
        title = "{The ATLAS$^{3D}$ project - III. A census of the stellar angular momentum within the effective radius of early-type galaxies: unveiling the distribution of fast and slow rotators}",
      journal = {\mnras},
         year = 2011,
        month = jun,
       volume = {414},
       number = {2},
        pages = {888-912},
          doi = {10.1111/j.1365-2966.2011.18496.x},
archivePrefix = {arXiv},
       eprint = {1102.4444},
 primaryClass = {astro-ph.CO},
       adsurl = {https://ui.adsabs.harvard.edu/abs/2011MNRAS.414..888E}
}

@ARTICLE{2005Binney,
       author = {{Binney}, James},
        title = "{Rotation and anisotropy of galaxies revisited}",
      journal = {\mnras},
         year = 2005,
        month = nov,
       volume = {363},
       number = {3},
        pages = {937-942},
          doi = {10.1111/j.1365-2966.2005.09495.x},
archivePrefix = {arXiv},
       eprint = {astro-ph/0504387},
 primaryClass = {astro-ph},
       adsurl = {https://ui.adsabs.harvard.edu/abs/2005MNRAS.363..937B}
}

@ARTICLE{2006Cappellari_FR_SR_Formation,
       author = {{Cappellari}, Michele and {Bacon}, R. and {Bureau}, M. and {Damen}, M.~C. and {Davies}, Roger L. and {de Zeeuw}, P.~T. and {Emsellem}, Eric and {Falc{\'o}n-Barroso}, Jes{\'u}s and {Krajnovi{\'c}}, Davor and {Kuntschner}, Harald and {McDermid}, Richard M. and {Peletier}, Reynier F. and {Sarzi}, Marc and {van den Bosch}, Remco C.~E. and {van de Ven}, Glenn},
        title = "{The SAURON project - IV. The mass-to-light ratio, the virial mass estimator and the Fundamental Plane of elliptical and lenticular galaxies}",
      journal = {\mnras},
         year = 2006,
        month = mar,
       volume = {366},
       number = {4},
        pages = {1126-1150},
          doi = {10.1111/j.1365-2966.2005.09981.x},
archivePrefix = {arXiv},
       eprint = {astro-ph/0505042},
 primaryClass = {astro-ph},
       adsurl = {https://ui.adsabs.harvard.edu/abs/2006MNRAS.366.1126C}
}

@ARTICLE{2022Zhang_ShapeTensor,
       author = {{Zhang}, Junkai and {Wuyts}, Stijn and {Witten}, Callum and {Avery}, Charlotte R. and {Hao}, Lei and {Sharma}, Raman and {Shen}, Juntai and {Toshikawa}, Jun and {Villforth}, Carolin},
        title = "{3D intrinsic shapes of quiescent galaxies in observations and simulations}",
      journal = {\mnras},
         year = 2022,
        month = jul,
       volume = {513},
       number = {4},
        pages = {4814-4832},
          doi = {10.1093/mnras/stac1083},
archivePrefix = {arXiv},
       eprint = {2204.10867},
 primaryClass = {astro-ph.GA},
       adsurl = {https://ui.adsabs.harvard.edu/abs/2022MNRAS.513.4814Z}
}

@ARTICLE{2014vanderWel_ShapeTensor,
       author = {{van der Wel}, A. and {Chang}, Yu-Yen and {Bell}, E.~F. and {Holden}, B.~P. and {Ferguson}, H.~C. and {Giavalisco}, M. and {Rix}, H. -W. and {Skelton}, R. and {Whitaker}, K. and {Momcheva}, I. and {Brammer}, G. and {Kassin}, S.~A. and {Martig}, M. and {Dekel}, A. and {Ceverino}, D. and {Koo}, D.~C. and {Mozena}, M. and {van Dokkum}, P.~G. and {Franx}, M. and {Faber}, S.~M. and {Primack}, J.},
        title = "{Geometry of Star-forming Galaxies from SDSS, 3D-HST, and CANDELS}",
      journal = {\apjl},
         year = 2014,
        month = sep,
       volume = {792},
       number = {1},
          eid = {L6},
        pages = {L6},
          doi = {10.1088/2041-8205/792/1/L6},
archivePrefix = {arXiv},
       eprint = {1407.4233},
 primaryClass = {astro-ph.GA},
       adsurl = {https://ui.adsabs.harvard.edu/abs/2014ApJ...792L...6V}
}

@ARTICLE{2019Bernardi_MANGA,
       author = {{Bernardi}, M. and {Dom{\'\i}nguez S{\'a}nchez}, H. and {Brownstein}, J.~R. and {Drory}, N. and {Sheth}, R.~K.},
        title = "{Galaxy properties as revealed by MaNGA - II. Differences in stellar populations of slow and fast rotator ellipticals and dependence on environment}",
      journal = {\mnras},
         year = 2019,
        month = nov,
       volume = {489},
       number = {4},
        pages = {5633-5652},
          doi = {10.1093/mnras/stz2413},
archivePrefix = {arXiv},
       eprint = {1904.11996},
 primaryClass = {astro-ph.GA},
       adsurl = {https://ui.adsabs.harvard.edu/abs/2019MNRAS.489.5633B}
}

@ARTICLE{2022Chua_ShapeTensor,
       author = {{Chua}, Kun Ting Eddie and {Vogelsberger}, Mark and {Pillepich}, Annalisa and {Hernquist}, Lars},
        title = "{The impact of galactic feedback on the shapes of dark matter haloes}",
      journal = {\mnras},
         year = 2022,
        month = sep,
       volume = {515},
       number = {2},
        pages = {2681-2697},
          doi = {10.1093/mnras/stac1897},
archivePrefix = {arXiv},
       eprint = {2109.00012},
 primaryClass = {astro-ph.GA},
       adsurl = {https://ui.adsabs.harvard.edu/abs/2022MNRAS.515.2681C}
}

@ARTICLE{2020Rong_UDG_ShapeTensor,
       author = {{Rong}, Yu and {Dong}, Xiao-Yu and {Puzia}, Thomas H. and {Galaz}, Gaspar and {S{\'a}nchez-Janssen}, Ruben and {Cao}, Tianwen and {van der Burg}, Remco F.~J. and {Sif{\'o}n}, Crist{\'o}bal and {Mancera Pi{\~n}a}, Pavel E. and {Marcelo}, Mora and {D'Ago}, Giuseppe and {Zhang}, Hong-Xin and {Johnston}, Evelyn J. and {Eigenthaler}, Paul},
        title = "{Intrinsic Morphology of Ultra-diffuse Galaxies}",
      journal = {\apj},
         year = 2020,
        month = aug,
       volume = {899},
       number = {1},
          eid = {78},
        pages = {78},
          doi = {10.3847/1538-4357/aba74a},
archivePrefix = {arXiv},
       eprint = {1907.10079},
 primaryClass = {astro-ph.GA},
       adsurl = {https://ui.adsabs.harvard.edu/abs/2020ApJ...899...78R}
}

@ARTICLE{2017Burkert_UDG_ShapeTensor,
       author = {{Burkert}, A.},
        title = "{The Geometry and Origin of Ultra-diffuse Ghost Galaxies}",
      journal = {\apj},
         year = 2017,
        month = apr,
       volume = {838},
       number = {2},
          eid = {93},
        pages = {93},
          doi = {10.3847/1538-4357/aa671c},
archivePrefix = {arXiv},
       eprint = {1609.00052},
 primaryClass = {astro-ph.GA},
       adsurl = {https://ui.adsabs.harvard.edu/abs/2017ApJ...838...93B}
}

@ARTICLE{2019Jiang_UDG_ShapeTensor,
       author = {{Jiang}, Fangzhou and {Dekel}, Avishai and {Freundlich}, Jonathan and {Romanowsky}, Aaron J. and {Dutton}, Aaron A. and {Macci{\`o}}, Andrea V. and {Di Cintio}, Arianna},
        title = "{Formation of ultra-diffuse galaxies in the field and in galaxy groups}",
      journal = {\mnras},
         year = 2019,
        month = aug,
       volume = {487},
       number = {4},
        pages = {5272-5290},
          doi = {10.1093/mnras/stz1499},
archivePrefix = {arXiv},
       eprint = {1811.10607},
 primaryClass = {astro-ph.GA},
       adsurl = {https://ui.adsabs.harvard.edu/abs/2019MNRAS.487.5272J}
}

@ARTICLE{2020CardonaBarrero_UDG_ShapeTensor,
       author = {{Cardona-Barrero}, Salvador and {Di Cintio}, Arianna and {Brook}, Christopher B.~A. and {Ruiz-Lara}, Tomas and {Beasley}, Michael A. and {Falc{\'o}n-Barroso}, Jesus and {Macci{\`o}}, Andrea V.},
        title = "{NIHAO XXIV: rotation- or pressure-supported systems? Simulated Ultra Diffuse Galaxies show a broad distribution in their stellar kinematics}",
      journal = {\mnras},
         year = 2020,
        month = oct,
       volume = {497},
       number = {4},
        pages = {4282-4292},
          doi = {10.1093/mnras/staa2094},
archivePrefix = {arXiv},
       eprint = {2004.09535},
 primaryClass = {astro-ph.GA},
       adsurl = {https://ui.adsabs.harvard.edu/abs/2020MNRAS.497.4282C}
}

@ARTICLE{2021KadoFong_UDG_ShapeTensor,
       author = {{Kado-Fong}, Erin and {Petrescu}, Mihai and {Mohammad}, Majid and {Greco}, Johnny and {Greene}, Jenny E. and {Adams}, Elizabeth A.~K. and {Huang}, Song and {Leisman}, Lukas and {Munshi}, Ferah and {Tanoglidis}, Dimitrios and {Van Nest}, Jordan},
        title = "{The Intrinsic Shapes of Low Surface Brightness Galaxies (LSBGs): A Discriminant of LSBG Galaxy Formation Mechanisms}",
      journal = {\apj},
         year = 2021,
        month = oct,
       volume = {920},
       number = {2},
          eid = {72},
        pages = {72},
          doi = {10.3847/1538-4357/ac15f0},
archivePrefix = {arXiv},
       eprint = {2106.05288},
 primaryClass = {astro-ph.GA},
       adsurl = {https://ui.adsabs.harvard.edu/abs/2021ApJ...920...72K}
}

@ARTICLE{2003Shapiro_SVE,
       author = {{Shapiro}, Kristen L. and {Gerssen}, Joris and {van der Marel}, Roeland P.},
        title = "{Observational Constraints on Disk Heating as a Function of Hubble Type}",
      journal = {\aj},
         year = 2003,
        month = dec,
       volume = {126},
       number = {6},
        pages = {2707-2716},
          doi = {10.1086/379306},
archivePrefix = {arXiv},
       eprint = {astro-ph/0308489},
 primaryClass = {astro-ph},
       adsurl = {https://ui.adsabs.harvard.edu/abs/2003AJ....126.2707S}
}

@ARTICLE{2012Gerssen_SVE,
       author = {{Gerssen}, J. and {Shapiro Griffin}, K.},
        title = "{Disc heating agents across the Hubble sequence}",
      journal = {\mnras},
         year = 2012,
        month = jul,
       volume = {423},
       number = {3},
        pages = {2726-2735},
          doi = {10.1111/j.1365-2966.2012.21078.x},
archivePrefix = {arXiv},
       eprint = {1204.3430},
 primaryClass = {astro-ph.CO},
       adsurl = {https://ui.adsabs.harvard.edu/abs/2012MNRAS.423.2726G}
}

@ARTICLE{1997Gerseen_SVE,
       author = {{Gerssen}, Joris and {Kuijken}, Konrad and {Merrifield}, Michael R.},
        title = "{The shape of the velocity ellipsoid in NGC 488}",
      journal = {\mnras},
         year = 1997,
        month = jul,
       volume = {288},
       number = {3},
        pages = {618-622},
          doi = {10.1093/mnras/288.3.618},
archivePrefix = {arXiv},
       eprint = {astro-ph/9702128},
 primaryClass = {astro-ph},
       adsurl = {https://ui.adsabs.harvard.edu/abs/1997MNRAS.288..618G}
}

@ARTICLE{2018Pinna_SVE,
       author = {{Pinna}, F. and {Falc{\'o}n-Barroso}, J. and {Martig}, M. and {Mart{\'\i}nez-Valpuesta}, I. and {M{\'e}ndez-Abreu}, J. and {van de Ven}, G. and {Leaman}, R. and {Lyubenova}, M.},
        title = "{Revisiting the stellar velocity ellipsoid-Hubble-type relation: observations versus simulations}",
      journal = {\mnras},
         year = 2018,
        month = apr,
       volume = {475},
       number = {2},
        pages = {2697-2712},
          doi = {10.1093/mnras/stx3331},
archivePrefix = {arXiv},
       eprint = {1801.03352},
 primaryClass = {astro-ph.GA},
       adsurl = {https://ui.adsabs.harvard.edu/abs/2018MNRAS.475.2697P}
}

@ARTICLE{2022WaloMartin_SVE,
       author = {{Walo-Mart{\'\i}n}, Daniel and {Pinna}, Francesca and {Grand}, Robert J.~J. and {P{\'e}rez}, Isabel and {Falc{\'o}n-Barroso}, Jes{\'u}s and {Fragkoudi}, Francesca and {Martig}, Marie},
        title = "{Local variations of the stellar velocity ellipsoid - II. The effect of the bar in the inner regions of Auriga galaxies}",
      journal = {\mnras},
         year = 2022,
        month = jul,
       volume = {513},
       number = {3},
        pages = {4587-4604},
          doi = {10.1093/mnras/stac769},
archivePrefix = {arXiv},
       eprint = {2203.07288},
 primaryClass = {astro-ph.GA},
       adsurl = {https://ui.adsabs.harvard.edu/abs/2022MNRAS.513.4587W}
}

@ARTICLE{1995AMcGaugh_SVE_LSB,
       author = {{McGaugh}, Stacy S. and {Schombert}, James M. and {Bothun}, Gregory D.},
        title = "{The Morphology of Low Surface Brightness Disk Galaxies}",
      journal = {\aj},
         year = 1995,
        month = may,
       volume = {109},
        pages = {2019},
          doi = {10.1086/117427},
archivePrefix = {arXiv},
       eprint = {astro-ph/9501085},
 primaryClass = {astro-ph},
       adsurl = {https://ui.adsabs.harvard.edu/abs/1995AJ....109.2019M}
}

@ARTICLE{1996deBlok_SVE_LSB,
       author = {{de Blok}, W.~J.~G. and {McGaugh}, S.~S. and {van der Hulst}, J.~M.},
        title = "{HI observations of low surface brightness galaxies: probing low-density galaxies}",
      journal = {\mnras},
         year = 1996,
        month = nov,
       volume = {283},
       number = {1},
        pages = {18-54},
          doi = {10.1093/mnras/283.1.18},
archivePrefix = {arXiv},
       eprint = {astro-ph/9605069},
 primaryClass = {astro-ph},
       adsurl = {https://ui.adsabs.harvard.edu/abs/1996MNRAS.283...18D}
}

@ARTICLE{2014Adams_SVE_Dwarfs,
       author = {{Adams}, Joshua J. and {Simon}, Joshua D. and {Fabricius}, Maximilian H. and {van den Bosch}, Remco C.~E. and {Barentine}, John C. and {Bender}, Ralf and {Gebhardt}, Karl and {Hill}, Gary J. and {Murphy}, Jeremy D. and {Swaters}, R.~A. and {Thomas}, Jens and {van de Ven}, Glenn},
        title = "{Dwarf Galaxy Dark Matter Density Profiles Inferred from Stellar and Gas Kinematics}",
      journal = {\apj},
         year = 2014,
        month = jul,
       volume = {789},
       number = {1},
          eid = {63},
        pages = {63},
          doi = {10.1088/0004-637X/789/1/63},
archivePrefix = {arXiv},
       eprint = {1405.4854},
 primaryClass = {astro-ph.GA},
       adsurl = {https://ui.adsabs.harvard.edu/abs/2014ApJ...789...63A}
}

@ARTICLE{2018Chemin_SVE,
       author = {{Chemin}, Laurent},
        title = "{A mass-velocity anisotropy relation in galactic stellar disks}",
      journal = {\aap},
         year = 2018,
        month = oct,
       volume = {618},
          eid = {A121},
        pages = {A121},
          doi = {10.1051/0004-6361/201832573},
archivePrefix = {arXiv},
       eprint = {1807.09833},
 primaryClass = {astro-ph.GA},
       adsurl = {https://ui.adsabs.harvard.edu/abs/2018A&A...618A.121C}
}

@ARTICLE{2019Mogotsi_SVE,
       author = {{Mogotsi}, Keoikantse Moses and {Romeo}, Alessandro B.},
        title = "{The stellar velocity dispersion in nearby spirals: radial profiles and correlations}",
      journal = {\mnras},
         year = 2019,
        month = nov,
       volume = {489},
       number = {3},
        pages = {3797-3809},
          doi = {10.1093/mnras/stz2370},
archivePrefix = {arXiv},
       eprint = {1804.10119},
 primaryClass = {astro-ph.GA},
       adsurl = {https://ui.adsabs.harvard.edu/abs/2019MNRAS.489.3797M}
}

@ARTICLE{2024He_anisotropy,
       author = {{He}, Jiaxin and {Wang}, Wenting and {Li}, Zhaozhou and {Han}, Jiaxin and {Rodriguez-Gomez}, Vicente and {Zhao}, Donghai and {Meng}, Xianguang and {Jing}, Yipeng and {Shao}, Shi and {Shi}, Rui and {Tan}, Zhenlin},
        title = "{How Do the Velocity Anisotropies of Halo Stars, Dark Matter, and Satellite Galaxies Depend on Host Halo Properties?}",
      journal = {\apj},
         year = 2024,
        month = dec,
       volume = {976},
       number = {2},
          eid = {187},
        pages = {187},
          doi = {10.3847/1538-4357/ad8882},
archivePrefix = {arXiv},
       eprint = {2407.14827},
 primaryClass = {astro-ph.GA},
       adsurl = {https://ui.adsabs.harvard.edu/abs/2024ApJ...976..187H}
}

@ARTICLE{2017GargBanerjee_plotC,
       author = {{Garg}, Prerak and {Banerjee}, Arunima},
        title = "{Origin of low surface brightness galaxies: a dynamical study}",
      journal = {\mnras},
         year = 2017,
        month = nov,
       volume = {472},
       number = {1},
        pages = {166-173},
          doi = {10.1093/mnras/stx1918},
archivePrefix = {arXiv},
       eprint = {1707.08085},
 primaryClass = {astro-ph.GA},
       adsurl = {https://ui.adsabs.harvard.edu/abs/2017MNRAS.472..166G}
}

@ARTICLE{2025Dou_plotD,
       author = {{Dou}, Jing and {Peng}, Yingjie and {Gu}, Qiusheng and {Ho}, Luis C. and {Renzini}, Alvio and {Shi}, Yong and {Daddi}, Emanuele and {Zhao}, Dingyi and {Zhang}, Chengpeng and {Gao}, Zeyu and {Li}, Di and {Lyu}, Cheqiu and {Mannucci}, Filippo and {Maiolino}, Roberto and {Wang}, Tao and {Yuan}, Feng},
        title = "{The Critical Role of Dark Matter Halos in Driving Star Formation}",
      journal = {\apjl},
         year = 2025,
        month = mar,
       volume = {982},
       number = {1},
          eid = {L12},
        pages = {L12},
          doi = {10.3847/2041-8213/adb95c},
archivePrefix = {arXiv},
       eprint = {2503.04243},
 primaryClass = {astro-ph.GA},
       adsurl = {https://ui.adsabs.harvard.edu/abs/2025ApJ...982L..12D}
}

@ARTICLE{2016AmoriscoLoeb_UDGformation,
       author = {{Amorisco}, N.~C. and {Loeb}, A.},
        title = "{Ultradiffuse galaxies: the high-spin tail of the abundant dwarf galaxy population}",
      journal = {\mnras},
         year = 2016,
        month = jun,
       volume = {459},
       number = {1},
        pages = {L51-L55},
          doi = {10.1093/mnrasl/slw055},
archivePrefix = {arXiv},
       eprint = {1603.00463},
 primaryClass = {astro-ph.GA},
       adsurl = {https://ui.adsabs.harvard.edu/abs/2016MNRAS.459L..51A}
}

@ARTICLE{2017DiCintio_UDGFormation,
       author = {{Di Cintio}, Arianna and {Brook}, Chris B. and {Dutton}, Aaron A. and {Macci{\`o}}, Andrea V. and {Obreja}, Aura and {Dekel}, Avishai},
        title = "{NIHAO - XI. Formation of ultra-diffuse galaxies by outflows}",
      journal = {\mnras},
         year = 2017,
        month = mar,
       volume = {466},
       number = {1},
        pages = {L1-L6},
          doi = {10.1093/mnrasl/slw210},
archivePrefix = {arXiv},
       eprint = {1608.01327},
 primaryClass = {astro-ph.GA},
       adsurl = {https://ui.adsabs.harvard.edu/abs/2017MNRAS.466L...1D}
}

@ARTICLE{2017vandeSande_LOSVD,
       author = {{van de Sande}, Jesse and {Bland-Hawthorn}, Joss and {Fogarty}, Lisa M.~R. and {Cortese}, Luca and {d'Eugenio}, Francesco and {Croom}, Scott M. and {Scott}, Nicholas and {Allen}, James T. and {Brough}, Sarah and {Bryant}, Julia J. and {Cecil}, Gerald and {Colless}, Matthew and {Couch}, Warrick J. and {Davies}, Roger and {Elahi}, Pascal J. and {Foster}, Caroline and {Goldstein}, Gregory and {Goodwin}, Michael and {Groves}, Brent and {Ho}, I. -Ting and {Jeong}, Hyunjin and {Jones}, D. Heath and {Konstantopoulos}, Iraklis S. and {Lawrence}, Jon S. and {Leslie}, Sarah K. and {L{\'o}pez-S{\'a}nchez}, {\'A}ngel R. and {McDermid}, Richard M. and {McElroy}, Rebecca and {Medling}, Anne M. and {Oh}, Sree and {Owers}, Matt S. and {Richards}, Samuel N. and {Schaefer}, Adam L. and {Sharp}, Rob and {Sweet}, Sarah M. and {Taranu}, Dan and {Tonini}, Chiara and {Walcher}, C. Jakob and {Yi}, Sukyoung K.},
        title = "{The SAMI Galaxy Survey: Revisiting Galaxy Classification through High-order Stellar Kinematics}",
      journal = {\apj},
         year = 2017,
        month = jan,
       volume = {835},
       number = {1},
          eid = {104},
        pages = {104},
          doi = {10.3847/1538-4357/835/1/104},
archivePrefix = {arXiv},
       eprint = {1611.07039},
 primaryClass = {astro-ph.GA},
       adsurl = {https://ui.adsabs.harvard.edu/abs/2017ApJ...835..104V}
}

@ARTICLE{2017Veale_LOSVD,
       author = {{Veale}, Melanie and {Ma}, Chung-Pei and {Greene}, Jenny E. and {Thomas}, Jens and {Blakeslee}, John P. and {McConnell}, Nicholas and {Walsh}, Jonelle L. and {Ito}, Jennifer},
        title = "{The MASSIVE Survey - VII. The relationship of angular momentum, stellar mass and environment of early-type galaxies}",
      journal = {\mnras},
         year = 2017,
        month = oct,
       volume = {471},
       number = {2},
        pages = {1428-1445},
          doi = {10.1093/mnras/stx1639},
archivePrefix = {arXiv},
       eprint = {1703.08573},
 primaryClass = {astro-ph.GA},
       adsurl = {https://ui.adsabs.harvard.edu/abs/2017MNRAS.471.1428V}
}

@ARTICLE{2019Liang_UDG_ShapeTensor,
       author = {{Liao}, Shihong and {Gao}, Liang and {Frenk}, Carlos S. and {Grand}, Robert J.~J. and {Guo}, Qi and {G{\'o}mez}, Facundo A. and {Marinacci}, Federico and {Pakmor}, R{\"u}diger and {Shao}, Shi and {Springel}, Volker},
        title = "{Ultra-diffuse galaxies in the Auriga simulations}",
      journal = {\mnras},
         year = 2019,
        month = dec,
       volume = {490},
       number = {4},
        pages = {5182-5195},
          doi = {10.1093/mnras/stz2969},
archivePrefix = {arXiv},
       eprint = {1904.06356},
 primaryClass = {astro-ph.GA},
       adsurl = {https://ui.adsabs.harvard.edu/abs/2019MNRAS.490.5182L}
}

@ARTICLE{2022vanNest_UDG_morphology,
       author = {{Van Nest}, Jordan D. and {Munshi}, F. and {Wright}, A.~C. and {Tremmel}, M. and {Brooks}, A.~M. and {Nagai}, D. and {Quinn}, T.},
        title = "{What's in a Name? Quantifying the Interplay between the Definition, Orientation, and Shape of Ultra-diffuse Galaxies Using the Romulus Simulations}",
      journal = {\apj},
         year = 2022,
        month = feb,
       volume = {926},
       number = {1},
          eid = {92},
        pages = {92},
          doi = {10.3847/1538-4357/ac43b7},
archivePrefix = {arXiv},
       eprint = {2108.12985},
 primaryClass = {astro-ph.GA},
       adsurl = {https://ui.adsabs.harvard.edu/abs/2022ApJ...926...92V}
}

@BOOK{1936Hubble,
       author = {{Hubble}, E.~P.},
        title = "{Realm of the Nebulae}",
         year = 1936,
       adsurl = {https://ui.adsabs.harvard.edu/abs/1936rene.book.....H},
        publisher = {Yale University Press}
}

@ARTICLE{1926Hubble,
       author = {{Hubble}, E.~P.},
        title = "{Extragalactic nebulae.}",
      journal = {\apj},
         year = 1926,
        month = dec,
       volume = {64},
        pages = {321-369},
          doi = {10.1086/143018},
       adsurl = {https://ui.adsabs.harvard.edu/abs/1926ApJ....64..321H}
}

@ARTICLE{1966Liller,
       author = {{Liller}, Martha H.},
        title = "{The Distribution of Intensity in Elliptical Galaxies of the Virgo Cluster. II}",
      journal = {\apj},
         year = 1966,
        month = oct,
       volume = {146},
        pages = {28},
          doi = {10.1086/148857},
       adsurl = {https://ui.adsabs.harvard.edu/abs/1966ApJ...146...28L}
}

@ARTICLE{1996KormendyBender,
       author = {{Kormendy}, John and {Bender}, Ralf},
        title = "{A Proposed Revision of the Hubble Sequence for Elliptical Galaxies}",
      journal = {\apjl},
         year = 1996,
        month = jun,
       volume = {464},
        pages = {L119},
          doi = {10.1086/310095},
       adsurl = {https://ui.adsabs.harvard.edu/abs/1996ApJ...464L.119K}
}

@ARTICLE{2017RodriguezGomez_morphology,
       author = {{Rodriguez-Gomez}, Vicente and {Sales}, Laura V. and {Genel}, Shy and {Pillepich}, Annalisa and {Zjupa}, Jolanta and {Nelson}, Dylan and {Griffen}, Brendan and {Torrey}, Paul and {Snyder}, Gregory F. and {Vogelsberger}, Mark and {Springel}, Volker and {Ma}, Chung-Pei and {Hernquist}, Lars},
        title = "{The role of mergers and halo spin in shaping galaxy morphology}",
      journal = {\mnras},
         year = 2017,
        month = may,
       volume = {467},
       number = {3},
        pages = {3083-3098},
          doi = {10.1093/mnras/stx305},
archivePrefix = {arXiv},
       eprint = {1609.09498},
 primaryClass = {astro-ph.GA},
       adsurl = {https://ui.adsabs.harvard.edu/abs/2017MNRAS.467.3083R}
}

@ARTICLE{2025Kolesnikov_morphology,
       author = {{Kolesnikov}, I. and {Sampaio}, V.~M. and {de Carvalho}, R.~R. and {Conselice}, C.},
        title = "{Galaxy morphology in CANDELS: addressing evolutionary changes across 0.2 {\ensuremath{\leqslant}} z {\ensuremath{\leqslant}} 2.4 with hybrid classification approach}",
      journal = {\mnras},
         year = 2025,
        month = may,
       volume = {539},
       number = {3},
        pages = {2765-2779},
          doi = {10.1093/mnras/staf625},
archivePrefix = {arXiv},
       eprint = {2412.03778},
 primaryClass = {astro-ph.GA},
       adsurl = {https://ui.adsabs.harvard.edu/abs/2025MNRAS.539.2765K}
}

@ARTICLE{2016Yagi_UDG_morphology,
       author = {{Yagi}, Masafumi and {Koda}, Jin and {Komiyama}, Yutaka and {Yamanoi}, Hitomo},
        title = "{Catalog of Ultra-diffuse Galaxies in the Coma Clusters from Subaru Imaging Data}",
      journal = {\apjs},
         year = 2016,
        month = jul,
       volume = {225},
       number = {1},
          eid = {11},
        pages = {11},
          doi = {10.3847/0067-0049/225/1/11},
       adsurl = {https://ui.adsabs.harvard.edu/abs/2016ApJS..225...11Y}
}

@ARTICLE{2021Wright_UDG_formation,
       author = {{Wright}, Anna C. and {Tremmel}, Michael and {Brooks}, Alyson M. and {Munshi}, Ferah and {Nagai}, Daisuke and {Sharma}, Ray S. and {Quinn}, Thomas R.},
        title = "{The formation of isolated ultradiffuse galaxies in ROMULUS25}",
      journal = {\mnras},
         year = 2021,
        month = apr,
       volume = {502},
       number = {4},
        pages = {5370-5389},
          doi = {10.1093/mnras/stab081},
archivePrefix = {arXiv},
       eprint = {2005.07634},
 primaryClass = {astro-ph.GA},
       adsurl = {https://ui.adsabs.harvard.edu/abs/2021MNRAS.502.5370W}
}

@ARTICLE{2015vanDokkum_UDG_formation_a,
       author = {{van Dokkum}, Pieter G. and {Abraham}, Roberto and {Merritt}, Allison and {Zhang}, Jielai and {Geha}, Marla and {Conroy}, Charlie},
        title = "{Forty-seven Milky Way-sized, Extremely Diffuse Galaxies in the Coma Cluster}",
      journal = {\apjl},
         year = 2015,
        month = jan,
       volume = {798},
       number = {2},
          eid = {L45},
        pages = {L45},
          doi = {10.1088/2041-8205/798/2/L45},
archivePrefix = {arXiv},
       eprint = {1410.8141},
 primaryClass = {astro-ph.GA},
       adsurl = {https://ui.adsabs.harvard.edu/abs/2015ApJ...798L..45V}
}

@ARTICLE{2015vanDokkum_UDG_formation_b,
       author = {{van Dokkum}, Pieter G. and {Romanowsky}, Aaron J. and {Abraham}, Roberto and {Brodie}, Jean P. and {Conroy}, Charlie and {Geha}, Marla and {Merritt}, Allison and {Villaume}, Alexa and {Zhang}, Jielai},
        title = "{Spectroscopic Confirmation of the Existence of Large, Diffuse Galaxies in the Coma Cluster}",
      journal = {\apjl},
         year = 2015,
        month = may,
       volume = {804},
       number = {1},
          eid = {L26},
        pages = {L26},
          doi = {10.1088/2041-8205/804/1/L26},
archivePrefix = {arXiv},
       eprint = {1504.03320},
 primaryClass = {astro-ph.GA},
       adsurl = {https://ui.adsabs.harvard.edu/abs/2015ApJ...804L..26V}
}

@ARTICLE{2017FalconBarroso_kine,
       author = {{Falc{\'o}n-Barroso}, J. and {Lyubenova}, M. and {van de Ven}, G. and {Mendez-Abreu}, J. and {Aguerri}, J.~A.~L. and {Garc{\'\i}a-Lorenzo}, B. and {Bekerait{\'e}}, S. and {S{\'a}nchez}, S.~F. and {Husemann}, B. and {Garc{\'\i}a-Benito}, R. and {Mast}, D. and {Walcher}, C.~J. and {Zibetti}, S. and {Barrera-Ballesteros}, J.~K. and {Galbany}, L. and {S{\'a}nchez-Bl{\'a}zquez}, P. and {Singh}, R. and {van den Bosch}, R.~C.~E. and {Wild}, V. and {Zhu}, L. and {Bland-Hawthorn}, J. and {Cid Fernandes}, R. and {de Lorenzo-C{\'a}ceres}, A. and {Gallazzi}, A. and {Gonz{\'a}lez Delgado}, R.~M. and {Marino}, R.~A. and {M{\'a}rquez}, I. and {P{\'e}rez}, E. and {P{\'e}rez}, I. and {Roth}, M.~M. and {Rosales-Ortega}, F.~F. and {Ruiz-Lara}, T. and {Wisotzki}, L. and {Ziegler}, B. and {CALIFA Collaboration}},
        title = "{Stellar kinematics across the Hubble sequence in the CALIFA survey: general properties and aperture corrections}",
      journal = {\aap},
         year = 2017,
        month = jan,
       volume = {597},
          eid = {A48},
        pages = {A48},
          doi = {10.1051/0004-6361/201628625},
archivePrefix = {arXiv},
       eprint = {1609.06446},
 primaryClass = {astro-ph.GA},
       adsurl = {https://ui.adsabs.harvard.edu/abs/2017A&A...597A..48F}
}

@ARTICLE{2019FalconBarroso_kine,
       author = {{Falc{\'o}n-Barroso}, J. and {van de Ven}, G. and {Lyubenova}, M. and {Mendez-Abreu}, J. and {Aguerri}, J.~A.~L. and {Garc{\'\i}a-Lorenzo}, B. and {Bekerait{\'e}}, S. and {S{\'a}nchez}, S.~F. and {Husemann}, B. and {Garc{\'\i}a-Benito}, R. and {Gonz{\'a}lez Delgado}, R.~M. and {Mast}, D. and {Walcher}, C.~J. and {Zibetti}, S. and {Zhu}, L. and {Barrera-Ballesteros}, J.~K. and {Galbany}, L. and {S{\'a}nchez-Bl{\'a}zquez}, P. and {Singh}, R. and {van den Bosch}, R.~C.~E. and {Wild}, V. and {Bland-Hawthorn}, J. and {Cid Fernandes}, R. and {de Lorenzo-C{\'a}ceres}, A. and {Gallazzi}, A. and {Marino}, R.~A. and {M{\'a}rquez}, I. and {Peletier}, R.~F. and {P{\'e}rez}, E. and {P{\'e}rez}, I. and {Roth}, M.~M. and {Rosales-Ortega}, F.~F. and {Ruiz-Lara}, T. and {Wisotzki}, L. and {Ziegler}, B.},
        title = "{The CALIFA view on stellar angular momentum across the Hubble sequence}",
      journal = {\aap},
         year = 2019,
        month = dec,
       volume = {632},
          eid = {A59},
        pages = {A59},
          doi = {10.1051/0004-6361/201936413},
archivePrefix = {arXiv},
       eprint = {1910.06236},
 primaryClass = {astro-ph.GA},
       adsurl = {https://ui.adsabs.harvard.edu/abs/2019A&A...632A..59F}
}

@ARTICLE{1907Schwarzschild,
       author = {{Schwarzschild}, Karl},
        title = "{Ueber die Eigenbewegungen der Fixsterne}",
      journal = {Nachrichten von der Gesellschaft der Wissenschaften zu Goettingen, Mathematisch-Physikalische Klasse},
         year = 1907,
        month = jan,
       volume = {5},
        pages = {614},
       adsurl = {https://ui.adsabs.harvard.edu/abs/1907NWGot...5..614S}
}

@ARTICLE{2014Agnello_anisotropy,
       author = {{Agnello}, A. and {Evans}, N.~W. and {Romanowsky}, A.~J.},
        title = "{Dynamical models of elliptical galaxies - I. Simple methods}",
      journal = {\mnras},
         year = 2014,
        month = aug,
       volume = {442},
       number = {4},
        pages = {3284-3298},
          doi = {10.1093/mnras/stu959},
archivePrefix = {arXiv},
       eprint = {1401.4462},
 primaryClass = {astro-ph.GA},
       adsurl = {https://ui.adsabs.harvard.edu/abs/2014MNRAS.442.3284A}
}

@ARTICLE{2001Bacon_LOSVD,
       author = {{Bacon}, R. and {Copin}, Y. and {Monnet}, G. and {Miller}, Bryan W. and {Allington-Smith}, J.~R. and {Bureau}, M. and {Carollo}, C.~M. and {Davies}, Roger L. and {Emsellem}, Eric and {Kuntschner}, Harald and {Peletier}, Reynier F. and {Verolme}, E.~K. and {de Zeeuw}, P. Tim},
        title = "{The SAURON project - I. The panoramic integral-field spectrograph}",
      journal = {\mnras},
         year = 2001,
        month = sep,
       volume = {326},
       number = {1},
        pages = {23-35},
          doi = {10.1046/j.1365-8711.2001.04612.x},
archivePrefix = {arXiv},
       eprint = {astro-ph/0103451},
 primaryClass = {astro-ph},
       adsurl = {https://ui.adsabs.harvard.edu/abs/2001MNRAS.326...23B}
}

@ARTICLE{2004Emsellem_LOSVD,
       author = {{Emsellem}, Eric and {Cappellari}, Michele and {Peletier}, Reynier F. and {McDermid}, Richard M. and {Bacon}, R. and {Bureau}, M. and {Copin}, Y. and {Davies}, Roger L. and {Krajnovi{\'c}}, Davor and {Kuntschner}, Harald and {Miller}, Bryan W. and {de Zeeuw}, P. Tim},
        title = "{The SAURON project - III. Integral-field absorption-line kinematics of 48 elliptical and lenticular galaxies}",
      journal = {\mnras},
         year = 2004,
        month = aug,
       volume = {352},
       number = {3},
        pages = {721-743},
          doi = {10.1111/j.1365-2966.2004.07948.x},
archivePrefix = {arXiv},
       eprint = {astro-ph/0404034},
 primaryClass = {astro-ph},
       adsurl = {https://ui.adsabs.harvard.edu/abs/2004MNRAS.352..721E}
}

@ARTICLE{2002deZeeuw_LOSVD,
       author = {{de Zeeuw}, P.~T. and {Bureau}, M. and {Emsellem}, Eric and {Bacon}, R. and {Carollo}, C.~M. and {Copin}, Y. and {Davies}, Roger L. and {Kuntschner}, Harald and {Miller}, Bryan W. and {Monnet}, G. and {Peletier}, Reynier F. and {Verolme}, E.~K.},
        title = "{The SAURON project - II. Sample and early results}",
      journal = {\mnras},
         year = 2002,
        month = jan,
       volume = {329},
       number = {3},
        pages = {513-530},
          doi = {10.1046/j.1365-8711.2002.05059.x},
archivePrefix = {arXiv},
       eprint = {astro-ph/0109511},
 primaryClass = {astro-ph},
       adsurl = {https://ui.adsabs.harvard.edu/abs/2002MNRAS.329..513D}
}

@ARTICLE{2019vanDokkum_UDG_LOSVD,
       author = {{van Dokkum}, Pieter and {Wasserman}, Asher and {Danieli}, Shany and {Abraham}, Roberto and {Brodie}, Jean and {Conroy}, Charlie and {Forbes}, Duncan A. and {Martin}, Christopher and {Matuszewski}, Matt and {Romanowsky}, Aaron J. and {Villaume}, Alexa},
        title = "{Spatially Resolved Stellar Kinematics of the Ultra-diffuse Galaxy Dragonfly 44. I. Observations, Kinematics, and Cold Dark Matter Halo Fits}",
      journal = {\apj},
         year = 2019,
        month = aug,
       volume = {880},
       number = {2},
          eid = {91},
        pages = {91},
          doi = {10.3847/1538-4357/ab2914},
archivePrefix = {arXiv},
       eprint = {1904.04838},
 primaryClass = {astro-ph.GA},
       adsurl = {https://ui.adsabs.harvard.edu/abs/2019ApJ...880...91V}
}

@ARTICLE{2019Pina_UDG_LOSVD_offBTFR,
       author = {{Mancera Pi{\~n}a}, Pavel E. and {Fraternali}, Filippo and {Adams}, Elizabeth A.~K. and {Marasco}, Antonino and {Oosterloo}, Tom and {Oman}, Kyle A. and {Leisman}, Lukas and {di Teodoro}, Enrico M. and {Posti}, Lorenzo and {Battipaglia}, Michael and {Cannon}, John M. and {Gault}, Lexi and {Haynes}, Martha P. and {Janowiecki}, Steven and {McAllan}, Elizabeth and {Pagel}, Hannah J. and {Reiter}, Kameron and {Rhode}, Katherine L. and {Salzer}, John J. and {Smith}, Nicholas J.},
        title = "{Off the Baryonic Tully-Fisher Relation: A Population of Baryon-dominated Ultra-diffuse Galaxies}",
      journal = {\apjl},
         year = 2019,
        month = oct,
       volume = {883},
       number = {2},
          eid = {L33},
        pages = {L33},
          doi = {10.3847/2041-8213/ab40c7},
archivePrefix = {arXiv},
       eprint = {1909.01363},
 primaryClass = {astro-ph.GA},
       adsurl = {https://ui.adsabs.harvard.edu/abs/2019ApJ...883L..33M}
}

@ARTICLE{2019Sengupta_UDG_LOSVD,
       author = {{Sengupta}, Chandreyee and {Scott}, T.~C. and {Chung}, Aeree and {Wong}, O. Ivy},
        title = "{Dark matter and H I in ultra-diffuse galaxy UGC 2162}",
      journal = {\mnras},
         year = 2019,
        month = sep,
       volume = {488},
       number = {3},
        pages = {3222-3230},
          doi = {10.1093/mnras/stz1884},
archivePrefix = {arXiv},
       eprint = {1907.10240},
 primaryClass = {astro-ph.GA},
       adsurl = {https://ui.adsabs.harvard.edu/abs/2019MNRAS.488.3222S}
}

@ARTICLE{2019Chilingarian_UDG_LOSVD,
       author = {{Chilingarian}, Igor V. and {Afanasiev}, Anton V. and {Grishin}, Kirill A. and {Fabricant}, Daniel and {Moran}, Sean},
        title = "{Internal Dynamics and Stellar Content of Nine Ultra-diffuse Galaxies in the Coma Cluster Prove Their Evolutionary Link with Dwarf Early-type Galaxies}",
      journal = {\apj},
         year = 2019,
        month = oct,
       volume = {884},
       number = {1},
          eid = {79},
        pages = {79},
          doi = {10.3847/1538-4357/ab4205},
archivePrefix = {arXiv},
       eprint = {1901.05489},
 primaryClass = {astro-ph.GA},
       adsurl = {https://ui.adsabs.harvard.edu/abs/2019ApJ...884...79C}
}

@ARTICLE{2019Emsellem_UDG_LOSVD,
       author = {{Emsellem}, Eric and {van der Burg}, Remco F.~J. and {Fensch}, J{\'e}r{\'e}my and {Je{\v{r}}{\'a}bkov{\'a}}, Tereza and {Zanella}, Anita and {Agnello}, Adriano and {Hilker}, Michael and {M{\"u}ller}, Oliver and {Rejkuba}, Marina and {Duc}, Pierre-Alain and {Durrell}, Patrick and {Habas}, Rebecca and {Lelli}, Federico and {Lim}, Sungsoon and {Marleau}, Francine R. and {Peng}, Eric and {S{\'a}nchez-Janssen}, Rub{\'e}n},
        title = "{The ultra-diffuse galaxy NGC 1052-DF2 with MUSE. I. Kinematics of the stellar body}",
      journal = {\aap},
         year = 2019,
        month = may,
       volume = {625},
          eid = {A76},
        pages = {A76},
          doi = {10.1051/0004-6361/201834909},
archivePrefix = {arXiv},
       eprint = {1812.07345},
 primaryClass = {astro-ph.GA},
       adsurl = {https://ui.adsabs.harvard.edu/abs/2019A&A...625A..76E}
}

@ARTICLE{2023Iodice_UDG_LEWIS,
       author = {{Iodice}, Enrichetta and {Hilker}, Michael and {Doll}, Goran and {Mirabile}, Marco and {Buttitta}, Chiara and {Hartke}, Johanna and {Mieske}, Steffen and {Cantiello}, Michele and {D'Ago}, Giuseppe and {Forbes}, Duncan A. and {Gullieuszik}, Marco and {Rejkuba}, Marina and {Spavone}, Marilena and {Spiniello}, Chiara and {Arnaboldi}, Magda and {Corsini}, Enrico M. and {Greggio}, Laura and {Falc{\'o}n-Barroso}, Jesus and {Fahrion}, Katja and {Fritz}, Jacopo and {La Marca}, Antonio and {Paolillo}, Maurizio and {Angela Raj}, Maria and {Rampazzo}, Roberto and {Sarzi}, Marc and {Capasso}, Giulio},
        title = "{Looking into the faintEst WIth MUSE (LEWIS): Exploring the nature of ultra-diffuse galaxies in the Hydra-I cluster. I. Project description and preliminary results}",
      journal = {\aap},
         year = 2023,
        month = nov,
       volume = {679},
          eid = {A69},
        pages = {A69},
          doi = {10.1051/0004-6361/202347129},
archivePrefix = {arXiv},
       eprint = {2308.11493},
 primaryClass = {astro-ph.GA},
       adsurl = {https://ui.adsabs.harvard.edu/abs/2023A&A...679A..69I}
}

@ARTICLE{2025Buttitta_UDG_LEWIS,
       author = {{Buttitta}, Chiara and {Iodice}, Enrichetta and {Doll}, Goran and {Hartke}, Johanna and {Hilker}, Michael and {Forbes}, Duncan A. and {Corsini}, Enrico M. and {Rossi}, Luca and {Arnaboldi}, Magda and {Cantiello}, Michele and {D'Ago}, Giuseppe and {Falc{\'o}n-Barroso}, Jesus and {Gullieuszik}, Marco and {La Marca}, Antonio and {Mieske}, Steffen and {Mirabile}, Marco and {Paolillo}, Maurizio and {Rejkuba}, Marina and {Spavone}, Marilena and {Spiniello}, Chiara and {Sarzi}, Marc},
        title = "{Looking into the faintEst WIth MUSE (LEWIS): Exploring the nature of ultra-diffuse galaxies in the Hydra-I cluster: II. Stellar kinematics and dynamical masses}",
      journal = {\aap},
         year = 2025,
        month = feb,
       volume = {694},
          eid = {A276},
        pages = {A276},
          doi = {10.1051/0004-6361/202452915},
archivePrefix = {arXiv},
       eprint = {2501.16190},
 primaryClass = {astro-ph.GA},
       adsurl = {https://ui.adsabs.harvard.edu/abs/2025A&A...694A.276B}
}

@ARTICLE{2016MartinezDelgado_UDGENV,
       author = {{Mart{\'\i}nez-Delgado}, David and {L{\"a}sker}, Ronald and {Sharina}, Margarita and {Toloba}, Elisa and {Fliri}, J{\"u}rgen and {Beaton}, Rachael and {Valls-Gabaud}, David and {Karachentsev}, Igor D. and {Chonis}, Taylor S. and {Grebel}, Eva K. and {Forbes}, Duncan A. and {Romanowsky}, Aaron J. and {Gallego-Laborda}, J. and {Teuwen}, Karel and {G{\'o}mez-Flechoso}, M.~A. and {Wang}, Jie and {Guhathakurta}, Puragra and {Kaisin}, Serafim and {Ho}, Nhung},
        title = "{Discovery of an Ultra-diffuse Galaxy in the Pisces--Perseus Supercluster}",
      journal = {\aj},
         year = 2016,
        month = apr,
       volume = {151},
       number = {4},
          eid = {96},
        pages = {96},
          doi = {10.3847/0004-6256/151/4/96},
archivePrefix = {arXiv},
       eprint = {1601.06960},
 primaryClass = {astro-ph.GA},
       adsurl = {https://ui.adsabs.harvard.edu/abs/2016AJ....151...96M}
}

@ARTICLE{2017vanderBurg_UDGENV,
       author = {{van der Burg}, Remco F.~J. and {Hoekstra}, Henk and {Muzzin}, Adam and {Sif{\'o}n}, Crist{\'o}bal and {Viola}, Massimo and {Bremer}, Malcolm N. and {Brough}, Sarah and {Driver}, Simon P. and {Erben}, Thomas and {Heymans}, Catherine and {Hildebrandt}, Hendrik and {Holwerda}, Benne W. and {Klaes}, Dominik and {Kuijken}, Konrad and {McGee}, Sean and {Nakajima}, Reiko and {Napolitano}, Nicola and {Norberg}, Peder and {Taylor}, Edward N. and {Valentijn}, Edwin},
        title = "{The abundance of ultra-diffuse galaxies from groups to clusters. UDGs are relatively more common in more massive haloes}",
      journal = {\aap},
         year = 2017,
        month = nov,
       volume = {607},
          eid = {A79},
        pages = {A79},
          doi = {10.1051/0004-6361/201731335},
archivePrefix = {arXiv},
       eprint = {1706.02704},
 primaryClass = {astro-ph.GA},
       adsurl = {https://ui.adsabs.harvard.edu/abs/2017A&A...607A..79V}
}

@ARTICLE{2019Roman_UDGENV,
       author = {{Rom{\'a}n}, Javier and {Beasley}, Michael A. and {Ruiz-Lara}, Tom{\'a}s and {Valls-Gabaud}, David},
        title = "{Discovery of a red ultra-diffuse galaxy in a nearby void based on its globular cluster luminosity function}",
      journal = {\mnras},
         year = 2019,
        month = jun,
       volume = {486},
       number = {1},
        pages = {823-835},
          doi = {10.1093/mnras/stz835},
archivePrefix = {arXiv},
       eprint = {1903.08168},
 primaryClass = {astro-ph.GA},
       adsurl = {https://ui.adsabs.harvard.edu/abs/2019MNRAS.486..823R}
}

@ARTICLE{2020Forbes_UDGENVb,
       author = {{Forbes}, Duncan A. and {Alabi}, Adebusola and {Romanowsky}, Aaron J. and {Brodie}, Jean P. and {Arimoto}, Nobuo},
        title = "{Globular clusters in Coma cluster ultra-diffuse galaxies (UDGs): evidence for two types of UDG?}",
      journal = {\mnras},
         year = 2020,
        month = mar,
       volume = {492},
       number = {4},
        pages = {4874-4883},
          doi = {10.1093/mnras/staa180},
archivePrefix = {arXiv},
       eprint = {2001.10031},
 primaryClass = {astro-ph.GA},
       adsurl = {https://ui.adsabs.harvard.edu/abs/2020MNRAS.492.4874F}
}

@ARTICLE{2019Forbes_UDGENV,
       author = {{Forbes}, Duncan A. and {Gannon}, Jonah and {Couch}, Warrick J. and {Iodice}, Enrichetta and {Spavone}, Marilena and {Cantiello}, Michele and {Napolitano}, Nicola and {Schipani}, Pietro},
        title = "{An ultra diffuse galaxy in the NGC 5846 group from the VEGAS survey}",
      journal = {\aap},
         year = 2019,
        month = jun,
       volume = {626},
          eid = {A66},
        pages = {A66},
          doi = {10.1051/0004-6361/201935499},
archivePrefix = {arXiv},
       eprint = {1905.06415},
 primaryClass = {astro-ph.GA},
       adsurl = {https://ui.adsabs.harvard.edu/abs/2019A&A...626A..66F}
}

@ARTICLE{2018Muller_UDGENV,
       author = {{M{\"u}ller}, Oliver and {Jerjen}, Helmut and {Binggeli}, Bruno},
        title = "{The Leo-I group: new dwarf galaxy and ultra diffuse galaxy candidates}",
      journal = {\aap},
         year = 2018,
        month = jul,
       volume = {615},
          eid = {A105},
        pages = {A105},
          doi = {10.1051/0004-6361/201832897},
archivePrefix = {arXiv},
       eprint = {1802.08657},
 primaryClass = {astro-ph.GA},
       adsurl = {https://ui.adsabs.harvard.edu/abs/2018A&A...615A.105M}
}

@ARTICLE{2019Janssens_UDGENV,
       author = {{Janssens}, Steven R. and {Abraham}, Roberto and {Brodie}, Jean and {Forbes}, Duncan A. and {Romanowsky}, Aaron J.},
        title = "{The Distribution of Ultra-diffuse and Ultra-compact Galaxies in the Frontier Fields}",
      journal = {\apj},
         year = 2019,
        month = dec,
       volume = {887},
       number = {1},
          eid = {92},
        pages = {92},
          doi = {10.3847/1538-4357/ab536c},
archivePrefix = {arXiv},
       eprint = {1911.00011},
 primaryClass = {astro-ph.GA},
       adsurl = {https://ui.adsabs.harvard.edu/abs/2019ApJ...887...92J}
}

@ARTICLE{2022Janssens_UDGENV,
       author = {{Janssens}, Steven R. and {Romanowsky}, Aaron J. and {Abraham}, Roberto and {Brodie}, Jean P. and {Couch}, Warrick J. and {Forbes}, Duncan A. and {Laine}, Seppo and {Mart{\'\i}nez-Delgado}, David and {van Dokkum}, Pieter G.},
        title = "{The globular clusters and star formation history of the isolated, quiescent ultra-diffuse galaxy DGSAT I}",
      journal = {\mnras},
         year = 2022,
        month = nov,
       volume = {517},
       number = {1},
        pages = {858-871},
          doi = {10.1093/mnras/stac2717},
archivePrefix = {arXiv},
       eprint = {2209.09910},
 primaryClass = {astro-ph.GA},
       adsurl = {https://ui.adsabs.harvard.edu/abs/2022MNRAS.517..858J}
}

@ARTICLE{2020Barbosa_UDGENV,
       author = {{Barbosa}, C.~E. and {Zaritsky}, D. and {Donnerstein}, R. and {Zhang}, H. and {Dey}, A. and {Mendes de Oliveira}, C. and {Sampedro}, L. and {Molino}, A. and {Costa-Duarte}, M.~V. and {Coelho}, P. and {Cortesi}, A. and {Herpich}, F.~R. and {Hernandez-Jimenez}, J.~A. and {Santos-Silva}, T. and {Pereira}, E. and {Werle}, A. and {Overzier}, R.~A. and {Cid Fernandes}, R. and {Smith Castelli}, A.~V. and {Ribeiro}, T. and {Schoenell}, W. and {Kanaan}, A.},
        title = "{One Hundred SMUDGes in S-PLUS: Ultra-diffuse Galaxies Flourish in the Field}",
      journal = {\apjs},
         year = 2020,
        month = apr,
       volume = {247},
       number = {2},
          eid = {46},
        pages = {46},
          doi = {10.3847/1538-4365/ab7660},
archivePrefix = {arXiv},
       eprint = {2002.05171},
 primaryClass = {astro-ph.GA},
       adsurl = {https://ui.adsabs.harvard.edu/abs/2020ApJS..247...46B}
}

@ARTICLE{2017Janssens_UDGENV,
       author = {{Janssens}, Steven and {Abraham}, Roberto and {Brodie}, Jean and {Forbes}, Duncan and {Romanowsky}, Aaron J. and {van Dokkum}, Pieter},
        title = "{Ultra-diffuse and Ultra-compact Galaxies in the Frontier Fields Cluster Abell 2744}",
      journal = {\apjl},
         year = 2017,
        month = apr,
       volume = {839},
       number = {1},
          eid = {L17},
        pages = {L17},
          doi = {10.3847/2041-8213/aa667d},
archivePrefix = {arXiv},
       eprint = {1701.00011},
 primaryClass = {astro-ph.GA},
       adsurl = {https://ui.adsabs.harvard.edu/abs/2017ApJ...839L..17J}
}

@ARTICLE{2021Prole_UDGENV,
       author = {{Prole}, D.~J. and {van der Burg}, R.~F.~J. and {Hilker}, M. and {Spitler}, L.~R.},
        title = "{The quiescent fraction of isolated low surface brightness galaxies: observational constraints}",
      journal = {\mnras},
         year = 2021,
        month = jan,
       volume = {500},
       number = {2},
        pages = {2049-2062},
          doi = {10.1093/mnras/staa3296},
archivePrefix = {arXiv},
       eprint = {2010.11210},
 primaryClass = {astro-ph.GA},
       adsurl = {https://ui.adsabs.harvard.edu/abs/2021MNRAS.500.2049P}
}

@ARTICLE{2019Zaritsky_UDGENV,
       author = {{Zaritsky}, Dennis and {Donnerstein}, Richard and {Dey}, Arjun and {Kadowaki}, Jennifer and {Zhang}, Huanian and {Karunakaran}, Ananthan and {Mart{\'\i}nez-Delgado}, David and {Rahman}, Mubdi and {Spekkens}, Kristine},
        title = "{Systematically Measuring Ultra-diffuse Galaxies (SMUDGes). I. Survey Description and First Results in the Coma Galaxy Cluster and Environs}",
      journal = {\apjs},
         year = 2019,
        month = jan,
       volume = {240},
       number = {1},
          eid = {1},
        pages = {1},
          doi = {10.3847/1538-4365/aaefe9},
archivePrefix = {arXiv},
       eprint = {1811.04098},
 primaryClass = {astro-ph.GA},
       adsurl = {https://ui.adsabs.harvard.edu/abs/2019ApJS..240....1Z}
}

@ARTICLE{2015Koda_UDGENV,
       author = {{Koda}, Jin and {Yagi}, Masafumi and {Yamanoi}, Hitomi and {Komiyama}, Yutaka},
        title = "{Approximately a Thousand Ultra-diffuse Galaxies in the Coma Cluster}",
      journal = {\apjl},
         year = 2015,
        month = jul,
       volume = {807},
       number = {1},
          eid = {L2},
        pages = {L2},
          doi = {10.1088/2041-8205/807/1/L2},
archivePrefix = {arXiv},
       eprint = {1506.01712},
 primaryClass = {astro-ph.GA},
       adsurl = {https://ui.adsabs.harvard.edu/abs/2015ApJ...807L...2K}
}

@ARTICLE{2015Mihos_UDGENV,
       author = {{Mihos}, J. Christopher and {Durrell}, Patrick R. and {Ferrarese}, Laura and {Feldmeier}, John J. and {C{\^o}t{\'e}}, Patrick and {Peng}, Eric W. and {Harding}, Paul and {Liu}, Chengze and {Gwyn}, Stephen and {Cuillandre}, Jean-Charles},
        title = "{Galaxies at the Extremes: Ultra-diffuse Galaxies in the Virgo Cluster}",
      journal = {\apjl},
         year = 2015,
        month = aug,
       volume = {809},
       number = {2},
          eid = {L21},
        pages = {L21},
          doi = {10.1088/2041-8205/809/2/L21},
archivePrefix = {arXiv},
       eprint = {1507.02270},
 primaryClass = {astro-ph.GA},
       adsurl = {https://ui.adsabs.harvard.edu/abs/2015ApJ...809L..21M}
}

@BOOK{2002Combes,
       author = {{Combes}, Francoise and {Boisse}, Patrick and {Mazure}, Alain and {Blanchard}, Alain and {Seymour}, M.},
        title = "{Galaxies and cosmology}",
         year = 2002,
        publisher = {Springer Science \& Business Media},
       adsurl = {https://ui.adsabs.harvard.edu/abs/2002gaco.book.....C}
}

\begin{appendix}

\section{Tidal indices for the ‘central’ and ‘satellite’ galaxies:}\label{appendix1}
\begin{figure}
    \centering
    \includegraphics[width=1.\linewidth]{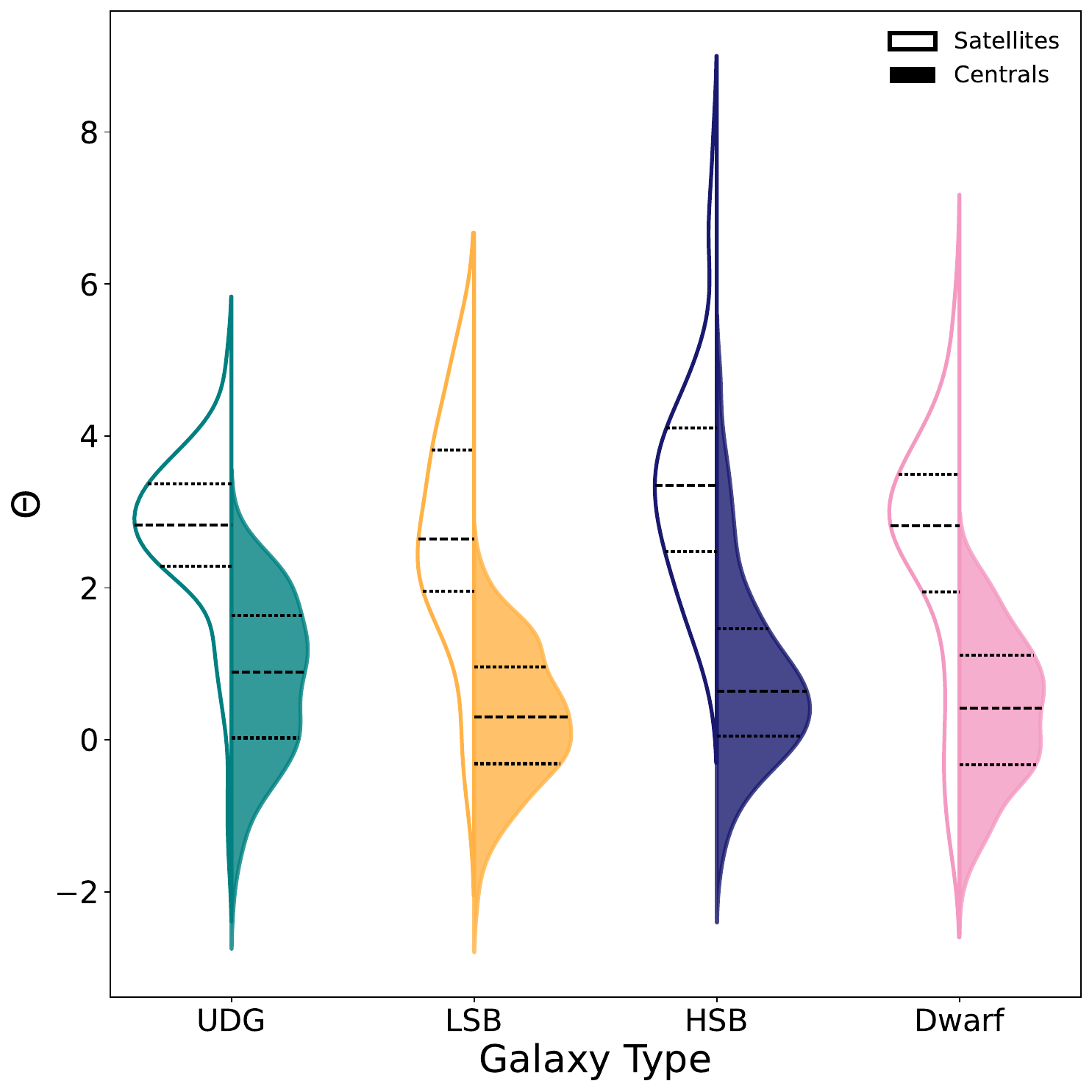}
    \caption{Distribution of the tidal index, $\Theta$, for our galaxy samples divided into two subsamples: ‘central’ on the right side (colour-filled) and ‘satellite’ on the left side (empty) of each violin plot.}
    \label{fig:primary_flag}
\end{figure}
Based on the \texttt{FoF} algorithm implemented in TNG50, each dark matter halo is assumed to host exactly one central galaxy, located at the position of the most-bound particle within the halo. In contrast, galaxies associated with any other particle within the same halo (i.e. not the most-bound one) are identified as satellite galaxies. Within this framework, ‘central’ galaxies are regarded as field galaxies that may eventually merge with a larger halo to become ‘satellites’ \citep{2001SpringelSUBFIND}. In the TNG50 data archive, the \texttt{primary\_flag} parameter can be used for this classification, taking a value of 1 for centrals and 0 for satellites.
\\
\indent The main motivation behind introducing the tidal index $\Theta$ in our study (discussed in Sect. \ref{sec:tidal_index}) is to differentiate isolated galaxies from tidally influenced ones. To compare this classification scheme with the conventional technique used in TNG50, we obtained the distribution of $\Theta$ values for all our galaxy samples, subdividing them into centrals (right-hand side of each violin plot) and satellites (left-hand side), as shown in Figure \ref{fig:primary_flag}. For each distribution, the median is indicated by a dashed line, while the dotted lines denote the 25th and 75th percentiles. We note that, based on the $\Theta$ distributions, the central and satellite populations are distinguishable from each other. However, the median $\Theta$ values are greater than zero in both cases (theoretically expecting, $\Theta < 0$ for central galaxies and $\Theta > 0$ for satellite galaxies). In addition, there is a significant overlap between the distributions. Therefore, we argue that \texttt{primary\_flag} alone may not provide a robust distinction between central and satellite galaxies, highlighting the need for a more reliable determination of the local tidal field, as implemented in this work.  However, instead of a rather rigid selection criterion, we relaxed the boundaries such that (1) galaxies with $\Theta < 0.25$ were classified as isolated, and (2) galaxies with $\Theta > 1.5$ were classified as tidally bound, while choosing $\Theta \sim 2$ as the indicator of significant tidal influence (following the distribution of $\Theta$ in Figure \ref{fig:primary_flag}).
\section{Distribution of galaxies in the TNG50-1 box:}\label{appendix2}
\begin{figure}
    \centering
    \includegraphics[width=1.\linewidth]{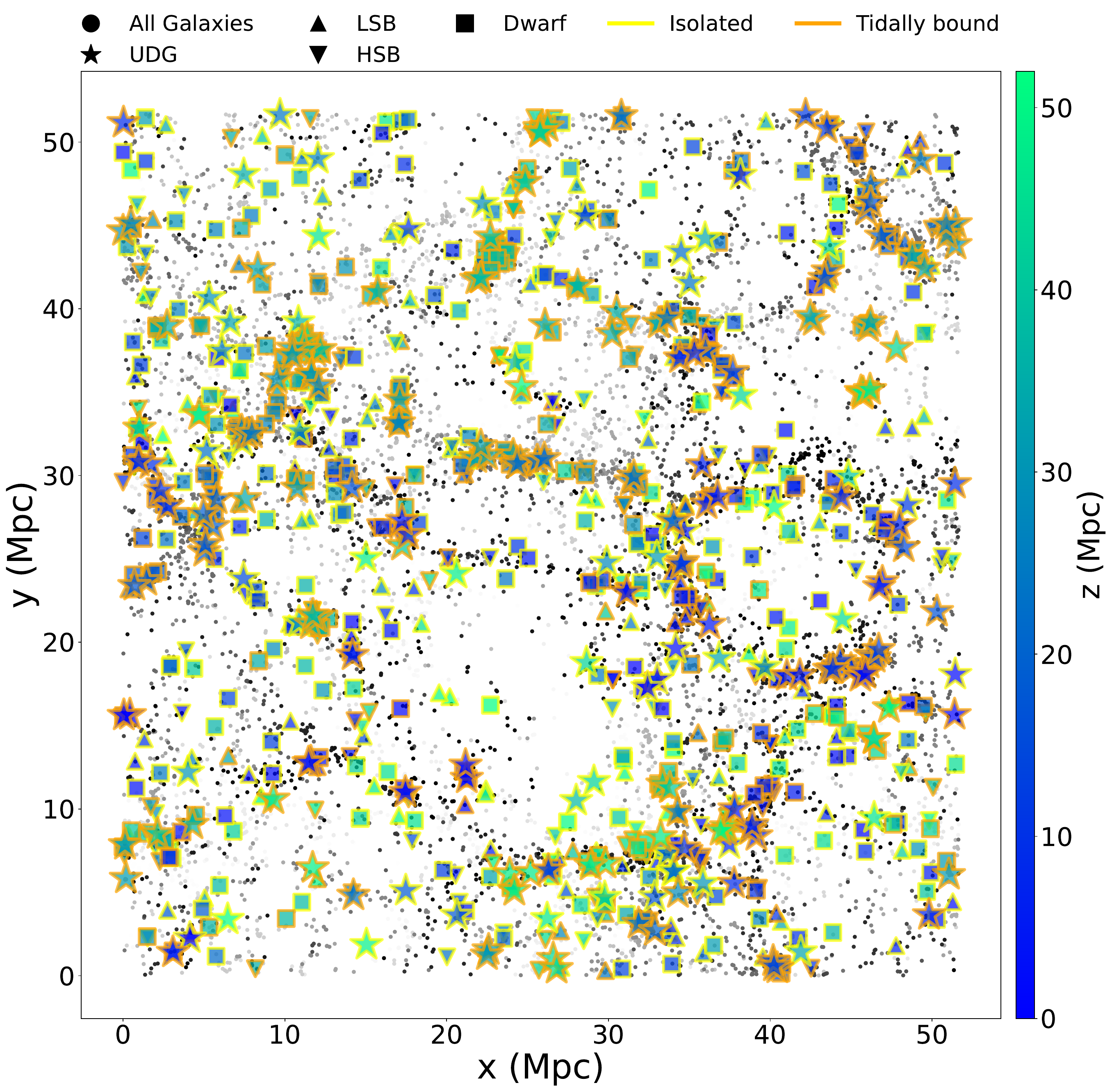}
    \caption{Distribution of the UDGs ($\star$), LSBs ($\blacktriangle$), HSBs ($\blacktriangledown$), and the dwarfs ($\blacksquare$) in the $x$-$y$ plane of the TNG50-1 box. The isolated and tidally bound galaxies are represented with yellow and orange borders, respectively. The dots presented in grey scale represent the neighbouring galaxies in the TNG50-1 box. The variation in marker colour denotes the location of the galaxy on the $z$-axis (direction along the plane of the paper) - blue (black) denoting the nearest ($\sim$ 0 Mpc) and green (fainter grey) denoting the farthest ($\sim$ 50 Mpc comoving) galaxies of our sample (remaining in the box). }
    \label{fig:environment}
\end{figure}

The distribution of our samples in the TNG50-1 box is shown in Figure \ref{fig:environment}. The UDGs, LSBs, HSBs, and the dwarfs are represented with $\star$, $\blacktriangle$, $\blacktriangledown$, and $\blacksquare$, respectively, following the same colour scheme as discussed in Sect. \ref{sec:tidal_index}. Each of our samples are bordered with either yellow or orange colours - yellow for the isolated and orange for the tidally bound galaxies.
\section{Regression fitting coefficients:}\label{appendix3}
Slopes ($m$), intercepts ($c$), coefficients of determination (R$^2$) and Spearman's rank correlation coefficients (S) corresponding to each of the possible scaling relations for isolated (Iso) and tidally bound (Bound) subsets of the UDG, LSB, HSB and the dwarf samples obtained in Sect. \ref{results:regression} are enlisted in Table \ref{tab:regression}. The $p$-values associated with R$^2$ and S-coefficients are smaller than 0.05 in all cases except for the isolated UDGs and the isolated HSBs (denoted with $\textbf{*}$).
\begin{table*}
\begin{center}
\caption{$m$, $c$, R$^2$ and S-values obtained in Sect. \ref{results:regression} for the UDGs, LSBs, HSBs and dwarfs divided in isolated and tidally-bound subsamples.}
\begin{tabular}{c c l l c c c c c c c c c c c}
\hline
Regression &&& Galaxy && Environment && $m$ && $c$ && R$^2$ && S &\\
\hline
\multirow{8}{*}{log$_{10}$ (M$_{*}$/M$_{\rm gas}$) vs. log$_{10}$ M$_{\rm gas}$} 
&&& \multirow{2}{*}{UDGs}  && Iso   && -0.90$\pm$0.09 && 7.41$\pm$0.87 && 0.62 && -0.61 &\\
  &&&                      && Bound && -1.03$\pm$0.03 && 8.73$\pm$0.26 && 0.86 && -0.93 &\\
&&& \multirow{2}{*}{LSBs}  && Iso   && -0.48$\pm$0.06 && 4.16$\pm$0.57 && 0.41 && -0.64 &\\
  &&&                      && Bound && -0.94$\pm$0.04 && 8.73$\pm$0.38 && 0.88 && -0.82 &\\
&&& \multirow{2}{*}{HSBs}  && Iso   && -0.70$\pm$0.06 && 7.31$\pm$0.67 && 0.64 && -0.64 &\\
  &&&                      && Bound && -0.91$\pm$0.01 && 9.61$\pm$0.15 && 0.96 && -0.94 &\\
&&& \multirow{2}{*}{Dwarfs}&& Iso   && -0.93$\pm$0.03 && 7.73$\pm$0.26 && 0.85 && -0.82 &\\
  &&&                      && Bound && -1.06$\pm$0.02 && 8.93$\pm$0.15 && 0.95 && -0.96 &\\
\hline
\multirow{8}{*}{log$_{10}$ M$_*$/M$_{\rm gas}$ vs. log$_{10}$ M$_{\rm dyn}$} 
&&& \multirow{2}{*}{UDGs}  && Iso$^\textbf{*}$   && -0.39$\pm$0.14 && 3.28$\pm$1.52 && 0.12 && -0.26 &\\
  &&&                      && Bound && -0.83$\pm$0.11 && 8.25$\pm$1.13 && 0.23 && -0.48 &\\
&&& \multirow{2}{*}{LSBs}  && Iso   && -0.26$\pm$0.08 && 2.24$\pm$0.92 && 0.09 && -0.35 &\\
  &&&                      && Bound && -0.96$\pm$0.16 && 10.22$\pm$1.75 && 0.32 && -0.53 &\\
&&& \multirow{2}{*}{HSBs}  &&Iso$^\textbf{*}$&& -0.08$\pm$0.22 && 0.90$\pm$2.58 && 0.00 && 0.03 &\\
  &&&                      && Bound && -1.39$\pm$0.12 && 16.56$\pm$1.42 && 0.46 && -0.79 &\\
&&& \multirow{2}{*}{Dwarfs}&& Iso   && -0.96$\pm$0.07 && 9.16$\pm$0.70 && 0.52 && -0.50 &\\
  &&&                      && Bound && -1.51$\pm$0.11 && 15.26$\pm$1.14 && 0.48 && -0.66 &\\
\hline
\multirow{8}{*}{log$_{10}$ M$_{\rm b}$ vs. log$_{10}$ M$_{\rm dyn}$} 
&&& \multirow{2}{*}{UDGs}  && Iso   && 0.50$\pm$0.11 && 4.02$\pm$1.13 && 0.24 && 0.60 &\\
  &&&                      && Bound && 0.60$\pm$0.05 && 2.86$\pm$0.50 && 0.36 && 0.61 &\\
&&& \multirow{2}{*}{LSBs}  && Iso   && 0.93$\pm$0.04 && -0.24$\pm$0.49 && 0.82 && 0.89 &\\
  &&&                      && Bound && 0.83$\pm$0.03 && 0.86$\pm$0.37 && 0.78 && 0.87 &\\
&&& \multirow{2}{*}{HSBs}  && Iso   && 0.88$\pm$0.08 && 0.41$\pm$0.90 && 0.75 && 0.84 &\\
  &&&                      && Bound && 0.59$\pm$0.02 && 3.87$\pm$0.25 && 0.83 && 0.92 &\\
&&& \multirow{2}{*}{Dwarfs}&& Iso   && 0.76$\pm$0.03 && 1.44$\pm$0.37 && 0.63 && 0.76 &\\
  &&&                      && Bound && 0.86$\pm$0.04 && 0.27$\pm$0.45 && 0.44 && 0.72 &\\
\hline
\end{tabular}
\label{tab:regression}
\end{center}
\end{table*}
\end{appendix}

\end{document}